\UseRawInputEncoding 
\documentclass{nature3}
\usepackage{graphicx}
\usepackage{xcolor}
\usepackage{float}
\usepackage{longtable}
\usepackage{tabularx}
\usepackage{url}
\usepackage{amsmath}
\usepackage{amssymb}
\usepackage{gensymb}
\usepackage{subcaption}
\usepackage{courier}
\usepackage{adjustbox}
\usepackage{rotating}
\usepackage{inputenc}

\usepackage[labelfont=bf]{caption}

\usepackage[super]{natbib}

\usepackage[colorlinks=true,citecolor=blue,urlcolor=cyan]{hyperref}
\usepackage{astjnlabbrev}

\usepackage[affil-it]{authblk}

\makeatletter
\def\@maketitle{%
  \newpage
  \begin{center}%
  \let \footnote \thanks
    {\Large \@title \par}%
    \vskip -3em
    {\large
      \begin{tabular}[t]{c}%
        \@author
      \end{tabular}\par}%
  \end{center}%
  \par
  \vskip 0.5em}
\makeatother

\title{Evidence for the volatile-rich composition of a 1.5-$R_\oplus$ planet}

\author[1]{Caroline Piaulet} 

\affil[1]{Department of Physics and Institute for Research on Exoplanets, Universit\'{e} de Montr\'{e}al, Montreal, QC, Canada}

\author[1]{Bj\"{o}rn Benneke}

\author[2]{Jose M. Almenara} 
\affil[2]{Universit\'{e} Grenoble Alpes, CNRS, IPAG, 38000 Grenoble, France}

\author[3]{Diana~Dragomir}%0000-0003-2313-467X
\affil[3]{Department of Physics and Astronomy, University of New Mexico, 210 Yale Blvd NE, Albuquerque, NM 87106, USA}

\author[4]{Heather A. Knutson} 
\affil[4]{Division of Geological and Planetary Sciences, California Institute of Technology, Pasadena, CA 91125, USA}

\author[1,5]{Daniel Thorngren}
\affil[5]{Department of Astronomy \& Astrophysics, University of California, Santa Cruz, CA 95064, USA}

\author[1]{Merrin S. Peterson} 

\author[6]{Ian J.M. Crossfield}
\affil[6]{The University of Kansas, Department of Physics and Astronomy, Malott Room 1082, 1251 Wescoe Hall Drive, Lawrence, KS, 66045, USA}

\author[7]{Eliza M.-R. Kempton}
\affil[7]{Department of Astronomy, University of Maryland, College Park, MD 20742, USA}

\author[8]{Daria Kubyshkina}
\affil[8]{Space Research Institute, Austrian Academy of Sciences, Schmiedlstrasse 6, 8042 Graz, Austria}

\author[9]{Andrew W. Howard}
\affil[9]{Department of Astronomy, California Institute of Technology, Pasadena, CA 91125, USA}

\author[10, 11]{Ruth Angus}
\affil[10]{Department of Astrophysics, American Museum of Natural History, 200 Central Park West, Manhattan, NY 10024, USA}
\affil[11]{Center for Computational Astrophysics, Flatiron Institute, 162 5th Ave, Manhattan, NY 10010, USA}

\author[12]{Howard Isaacson}
\affil[12]{Department of Astronomy, University of California - Berkeley, Berkeley, CA, 94720, USA}

\author[13]{Lauren M. Weiss}
\affil[13]{Department of Physics, University of Notre Dame, Notre Dame, IN 46556, USA}

\author[14]{Charles A. Beichman}
\affil[14]{NASA Exoplanet Science Institute, Caltech/IPAC, Pasadena, CA 91125, USA}

\author[5]{Jonathan J. Fortney}

\author[8]{Luca Fossati}
\author[8]{Helmut Lammer}

\author[15,16]{P. R. McCullough}
\affil[15]{The William H. Miller III Department of Physics and Astronomy, Johns Hopkins University, Baltimore, MD 21218, USA}
\affil[16]{Space Telescope Science Institute, 3700 San Martin Dr., Baltimore, MD 21218, USA}

\author[17]{Caroline V. Morley}
\affil[17]{Department of Astronomy, University of Texas, Austin, TX 78712, USA}

\author[18,19]{Ian Wong}
\affil[18]{NASA Goddard Space Flight Center, 8800 Greenbelt Road, Greenbelt, MD 20771, USA}
\affil[19]{NASA Postdoctoral Program Fellow}

\begin{document}
\maketitle

%% MAIN TEXT FIGURES AND TABLES
\newcommand{\figfullphotodynres}{Figure 1}
\newcommand{\figmassradiussimple}{Figure 2}
\newcommand{\figmasscompare}{Figure 3}
\newcommand{\figstructure}{Figure 4}
\newcommand{\tabtrtimes}{Table 1}
\newcommand{\tabphotodynres}{Table 2}

%% EXTENDED DATA FIGURES
\newcommand{\figthreeplaphotodyn}{Extended Data Figure 1}
\newcommand{\figplanetesearch}{Extended Data Figure 2}
\newcommand{\figRVpdgm}{Extended Data Figure 3}
\newcommand{\figphotGP}{Extended Data Figure 4}
\newcommand{\figRVfit}{Extended Data Figure 5}
\newcommand{\figstrucmodel}{Extended Data Figure 6}
\newcommand{\figcompd}{Extended Data Figure 7}
\newcommand{\figspots}{Extended Data Figure 8}
\newcommand{\figcompc}{Extended Data Figure 9}
\newcommand{\figretrieval}{Extended Data Figure 10}

%% SUPPLEMENTARY FIGURES AND TABLES
\newcommand{\figtimediff}{Supplementary Figure 1}
\newcommand{\fighstlcs}{Supplementary Figure 2}
\newcommand{\figspitzerlcs}{Supplementary Figure 3}
\newcommand{\figfitlcsb}{Supplementary Figure 4}
\newcommand{\figfitlcsc}{Supplementary Figure 5}
\newcommand{\figfitlcsd}{Supplementary Figure 6}
\newcommand{\figfitlcsdspitzerhst}{Supplementary Figure 7}
\newcommand{\figphotodyncorner}{Supplementary Figure 8}
\newcommand{\figRVcorner}{Supplementary Figure 9}
\newcommand{\figmassradius}{Supplementary Figure 10}
\newcommand{\fighhb}{Supplementary Figure 11}
\newcommand{\figstellarage}{Supplementary Figure 12}
\newcommand{\figescape}{Supplementary Figure 13}

\newcommand{\tabstarparams}{Supplementary Table 1}
\newcommand{\tabRVresults}{Supplementary Table 2}
\newcommand{\tabspectrum}{Supplementary Table 3}

\newcommand{\tabRVdata}{Supplementary Dataset 1}

\begin{abstract}

The population of planets smaller than approximately $1.7\,R_\oplus$ is widely interpreted as consisting of rocky worlds, generally referred to as super-Earths. This picture is largely corroborated by radial-velocity (RV) mass measurements for close-in super-Earths but lacks constraints at lower insolations. Here we present the results of a detailed study of the Kepler-138 system using 13 \textit{Hubble} and \textit{Spitzer} transit observations of the warm-temperate $1.51\pm0.04\,R_\oplus$ planet Kepler-138~d ($T_{\mathrm{eq, A_B=0.3}}\approx350\,K$) combined with new \textit{Keck}/HIRES RV measurements of its host star. We find evidence for a volatile-rich “water world” nature of Kepler-138~d, with a large fraction of its mass contained in a thick volatile layer.
This finding is independently supported by transit timing variations, RV observations ($M_\mathrm{d}=2.1_{-0.7}^{+0.6}\,M_\oplus$), as well as the flat optical/IR transmission spectrum. Quantitatively, we infer a composition of $11_{-4}^{+3}$\% volatiles by mass or $\sim 51\%$ by volume, with a 2000 km deep water mantle and atmosphere on top of a core with an Earth-like silicates/iron ratio. Any hypothetical hydrogen layer consistent with the observations ($<0.003\,M_\oplus$) would have swiftly been lost on a $\sim10$ Myr timescale. The bulk composition of Kepler-138~d therefore resembles those of the icy moons rather than the terrestrial planets in the solar system. We conclude that not all super-Earth-sized planets are rocky worlds, but that volatile-rich water worlds exist in an overlapping size regime, especially at lower insolations. Finally, our photodynamical analysis also reveals that Kepler-138~c ($R_\mathrm{c}=1.51 \pm 0.04\,R_\oplus$, $M_\mathrm{c}=2.3_{-0.5}^{+0.6}\,M_\oplus$) is a slightly warmer twin of Kepler-138~d, i.e., another water world in the same system, and we infer the presence of Kepler-138~e, a likely non-transiting planet at the inner edge of the habitable zone.
\end{abstract}

We observed 13 new transits of Kepler-138~d with \textit{HST} and \textit{Spitzer} (\tabtrtimes) as part of the \textit{HST} survey program GO~13665 (PI Benneke, three transits) and the \textit{Spitzer} program GO~11131 (PI Dragomir, five transits at 3.6 $\mu$m and five transits at 4.5 $\mu$m). We chose the Kepler-138 system for this detailed study because the three known transiting planets on near-resonant orbits open a rare opportunity for measuring the masses of low-temperature, super-Earth-sized planets \citep{rowe_validation_2014,kipping_hunt_2014,jontof-hutter_mass_2015,almenara_absolute_2018}. The \textit{HST} and \textit{Spitzer} observations critically extend the baseline for the transit-timing variations (TTV) analysis to over~7~years. 
Therefore, the new transits enable our analysis to cover nearly two super-periods of TTV modulation for the interaction of Kepler-138~c and d, almost doubling the baseline compared to the \textit{Kepler} transit timing measurements alone. We complement this dataset with 28 \textit{Keck}/HIRES RV measurements of Kepler-138 that support the TTV analysis. The \textit{HST}/WFC3 and \textit{Spitzer}/IRAC light curves are extracted using the ExoTEP pipeline. The transit parameters are then constrained for each visit by jointly fitting a set of astrophysical and instrumental model parameters (see Methods). We ensure the robustness of our transit analyses by verifying  the statistical consistency of transit depths from individual visits with the same instrument. We also find that the transit timing constraints are consistent for a transit simultaneously observed with \textit{HST} and \textit{Spitzer} (\tabtrtimes).

\textbf{Results}~~
Our measured \textit{HST} and \textit{Spitzer} transit times for Kepler-138~d (\tabtrtimes) are inconsistent with the forward predictions from the photodynamical fit to the \textit{Kepler} transits (\figtimediff). We therefore revisit the orbital solution using an MCMC analysis of the transit-timing variations (TTV) over the full 7-year dataset combining \textit{Kepler}, \textit{HST} and \textit{Spitzer} transits (see Methods). No three-planet model can simultaneously reproduce the \textit{Kepler}, \textit{HST}, and \textit{Spitzer} transit times of Kepler-138~d (\figfullphotodynres abc, \figthreeplaphotodyn). We therefore explore possible orbits and masses of a fourth planet (see Methods). We infer the presence of a fourth, likely non-transiting planet (\figplanetesearch) exterior to Kepler-138~d near the 5:3 resonance, providing a good match to the observed transit times (\figfullphotodynres). We subsequently analyze the light curves directly using a photodynamical fitting framework, and derive consistent parameters with the TTV analysis for the four planets (see Methods; \tabstarparams and \tabphotodynres).
In parallel with the TTV analysis, we analyze the \textit{Keck}/HIRES RVs of Kepler-138 (see Methods). The  data are reduced following standard data reduction procedures of the California Planet Search \citep{howard_california_2010}. We then use a Gaussian Process model trained on the portion of the long-cadence \textit{Kepler} photometry simultaneous with the HIRES dataset to mitigate stellar contamination (see Methods; \figRVpdgm, \figphotGP) and derive additional independent constraints on the masses of Kepler-138 b,c,d, and e (see \tabRVresults, \figRVfit ~and \figRVcorner).

Accounting for the presence of the newly inferred planet Kepler-138~e has a significant impact on the masses of Kepler-138~c and d (\figfullphotodynres, \tabphotodynres). While previously believed to have drastically-different densities \citep{jontof-hutter_mass_2015,almenara_absolute_2018}, Kepler-138~c and d are revealed to be low-density ``twins'', with consistent masses and radii ($M_\mathrm{c}=2.3_{-0.5}^{+0.6}\,M_\oplus$, $M_\mathrm{d}=2.1_{-0.7}^{+0.6}\,M_\oplus$, \figmassradiussimple). We confirm the Mars-mass of Kepler-138 b ($M_b=0.07 \pm 0.02\,M_\oplus$), and the newly-discovered outer planet Kepler-138~e has a mass of $M_e=0.43_{-0.10}^{+0.21}\,M_\oplus$ (\figmassradius). This uncommon configuration with one small planet, followed by two larger ``twin'' planets and a lighter outer planet resembles a scaled version of the inner solar system (\figfullphotodynres d). 
The temperate Kepler-138~e ($T_\mathrm{eq}\sim 292$\,K assuming an Earth-like Bond albedo of 0.3), lies at the inner edge of the classical 1D habitable zone (\figfullphotodynres e, \tabphotodynres, Ref. \citealp{kopparapu_habitable_2013}). 
Kepler-138~e is, however, likely not amenable to detailed characterization as its orbital solution is consistent with a non-transiting geometry, in line with its non-detection in the \textit{Kepler} light curves (see Methods and \figplanetesearch).

The mass of the 1.51$R_\oplus$ planet Kepler-138~d is lower than the expectation for a rocky planet of its size. For an Earth-like interior composition, the measured mass of Kepler-138~d requires the presence of a volatile envelope with $>99.81$\% confidence.
Even completely iron-free scenarios are disfavored at $>$98.75\% confidence from the combined posterior on the mass of Kepler-138~d from the photodynamical and RV analyses (\figmasscompare). This indicates the presence of either a H$_2$/He envelope, a volatile-rich layer, or a combination of the two. We explore the range of plausible compositions for Kepler-138~d by coupling a four-layer (iron, silicates, water and hydrogen) interior model with a self-consistent non-gray atmosphere model (see Methods). This new coupled full-planet model enables us to account for the contributions of both the interior and the potentially puffy atmosphere to the measured radius (\figstrucmodel).  In the interior model, water serves as a proxy for any composition of similarly-dense ices (e.g. methane, ammonia). We first look into how much H$_2$/He could be present atop Kepler-138~d. For an interior composed of an Earth-like mixture of rock and iron, only a thin H$_2$/He layer of maximum $\sim 0.01$wt\% (percent by mass) would be allowed to match the measured mass and radius of Kepler-138~d.  Hydrogen mass fractions greater than 0.1\% (i.e. $\approx 0.003\,M_\oplus$ of H$_2$/He) are excluded at 99.7\% (3$\sigma$) confidence (\figstructure a). Any water present in the interior of Kepler-138~d underneath the H$_2$/He would further decrease this upper limit on the amount of H$_2$/He. 
The existence of such a light $\lesssim$0.01 wt\% H$_2$/He envelope is, however, fundamentally challenged by its vulnerability to loss to space.
We compute the expected hydrogen envelope lifetime under the influence of hydrodynamic energy-limited escape, as well as using a full 1D hydrodynamic upper atmosphere model (see Methods, Ref. \citep{kubyshkina_grid_2018}). We calculate escape timescales of only tens of Myr, orders of magnitude shorter than the estimated age of the system of 1 to 2.7 Gyr (see Methods).
We therefore regard the survival of any hydrogen-rich atmosphere with a maximum mass of $0.003\,M_\oplus$ on Kepler-138~d as implausible. Fine-tuning would be required for us to observe Kepler-138~d right before the last remains of the H$_2$/He envelope are lost, which is expectantly even more unlikely given that the more highly-irradiated Kepler-138~c would also need to be in the same fine-tuned state.
In addition, beyond thermal escape, non-thermal processes including ion escape could accelerate the atmospheric loss, with loss rates that are harder to quantify but potentially orders of magnitudes larger than what the inner solar system planets experience \citep{lammer_outgassing_2013, dong_atmospheric_2020}. A magnetic field could at best decrease the mass-loss rate by a factor of a few \citep{khodachenko_atmosphere_2015}, while orders of magnitude would be needed for Kepler-138~d to safely retain a hydrogen envelope. Finally, while interior outgassing can in some cases at least temporarily replenish lost primary atmospheres, the resulting atmospheres are volatile-rich, rather than hydrogen-dominated (see Methods, Refs.\citep{kite_exoplanet_2020,bower_retention_2021}).

With the implausibility of a hydrogen-rich envelope composition, Kepler-138~d's low density can instead be explained by a large exposed volatile layer dominated by water or other ices (e.g. methane, ammonia). We investigate this possibility using three-layer models with silicates+iron cores underlying a water layer with a high-metallicity water steam atmosphere \citep{aguichine_mass-radius_2021}, and we explore the full range of water fractions consistent with Kepler-138~d's mass and radius 
using the \texttt{smint} package\citep{piaulet_wasp-107bs_2021}
(see Methods). This analysis reveals that $11^{+3}_{-4}$\% of the mass of Kepler-138~d needs to be composed of water, which corresponds to $\sim51\%$ water by volume, for a planetary interior with Earth-like silicates/iron ratio (\figcompd). This water content is in line with the water contents of the icy moons of the outer solar system (Jupiter's moon Europa has a water content of $\approx 8$wt\%), rather than the terrestrial planets in the inner solar system (\figmasscompare). When additionally considering silicates/iron ratios strongly deviant from Earth-like, all the way from pure silicate interiors to iron-rich interiors, our data indicate a water mass fractions of $14^{+6}_{-5}$\% for Kepler-138~d (\figstructure). We verify that this conclusion cannot be challenged by an overestimated planet radius due to stellar contamination (see Methods, \figspots), hazes (see Methods), planetary rings (see Methods), or a partially molten rock interior \citep{bower_linking_2019}. 
Because of its virtually identical mass and radius, a similar ``warm water world'' composition can explain the structure of Kepler-138~c, with the nuance that planet c receives more intense stellar irradiation, yielding a slightly lower inferred water mass fraction of $9_{-3}^{+2}$\% for an Earth-like core composition (\figcompc). We do not expect our conclusions for Kepler-138~d to be affected by the presence of a magma ocean due to its low instellation \citep{kite_atmosphere_2020}, however the radius of the warmer Kepler-138~c could potentially accommodate even larger water mass fractions if the rock near the rock-water interface is molten \citep{dorn_hidden_2021}. 

Contrary to the fragility of a light H$_2$/He layer atop Kepler-138~d, the high mean molecular weight envelope of a volatile-rich water world is stable against atmosphere stripping. Small initial water reservoirs of a few Earth oceans could be lost due to the runaway greenhouse followed by water photolysis triggered by the pre-main sequence evolution of Kepler-138 \citep{luger_extreme_2015}. However, Kepler-138~d's inferred water layer of $\sim 1000$ modern Earth oceans is sufficiently massive to be robust against complete stripping by the star's early high energy irradiation \citep{lopez_born_2017}. Moreover, large amounts of water can be shielded from early loss within the magma ocean\citep{bower_linking_2019} while the mantle is still molten because of water's high solubility in the melt\citep{dorn_hidden_2021}, which limits early water outgassing\citep{bower_retention_2021}. This at least partially molten mantle stage can last up to Gyrs and has been theorized to play a key role in sustaining\citep{bower_linking_2019} and even fostering\citep{kite_water_2021} long-lived water reservoirs. Therefore, we conclude that a thick layer of water or other volatiles stands out as the most plausible explanation for the low density of Kepler-138~d and c. 

Beyond the constraints on bulk planetary compositions offered by the new transit and RV observations, our spectroscopic near-IR observations with \textit{HST}, and mid-IR broadband measurements with \textit{Spitzer}, enable us to simultaneously obtain first insights into Kepler-138~d's transmission spectrum. The retrieval analysis of the  optical-to-IR transmission spectrum of Kepler-138~d further supports our conclusion on its volatile-rich nature (see Methods, \figretrieval, \tabspectrum). While not conclusive in its own right, we find that the observed transmission spectrum is fully consistent with the high metallicity atmosphere of a volatile-rich water world as large spectral features are not observed (see Methods; \figretrieval). 

\textbf{Discussion}~~
Multiple theories have been proposed to explain the formation of volatile-rich ``water worlds'', based on different volatile supply mechanisms. Volatiles could be delivered by the surrounding gas and solids in the protoplanetary disk during planet formation, assuming the planet formed farther from its host than its present location \citep{kuchner_volatile-rich_2003}. This picture finds support in the  dynamical architecture of the Kepler-138 system, with all adjacent planet pairs being close to first- or second-order mean-motion resonances (4:3 for b and c, 5:3 for c and d, and 5:3 for d and e). Sequential formation near the water ice line and subsequent inwards migration of the four planets through the protoplanetary disk could have locked them in this near-resonant configuration \citep{huang_dynamics_2021}.
Alternatively, solids may have contributed to the water budget of Kepler-138~c and d via the plausible in-situ accretion of volatile-rich bodies or outgassing of meteoritic material (see Methods;\citep{elkins-tanton_ranges_2008, kite_exoplanet_2020}). Recently, an endogenous source of water has also been proposed. The oxygen present in the young planet's magma ocean in the form of iron oxide can react with accreted nebular hydrogen and produce a dissolved water reservoir in the magma of up to a few percent by mass \citep{kite_atmosphere_2020,kite_water_2021}. Significant amounts of water can be shielded in the planetary interior \citep{dorn_hidden_2021,bower_retention_2021}, potentially enabling a long-lived water world stage following the shedding of the hydrogen due to stellar irradiation \citep{luger_habitable_2015}. The inferred amounts of water for Kepler-138~c and d, however, suggest that this endogenous water supply mechanism could at best account for a fraction of the volatiles present.

The inference of a volatile-rich ``water world'' composition for the warm-temperate ($T_\mathrm{eq, A_B=0.3} = 345 \pm 7$\,K), super-Earth-sized (1.51 $R_\oplus$) planet Kepler-138~d reveals that the super-Earth population is not uniform in composition. Our analysis shows that at least some small planets on warm and temperate orbits have compositions akin to the icy moons rather than the terrestrial planets of the solar system, pointing to a distinct origins story compared to the close-in rocky super-Earths. Both mass measurements of individual close-in super-Earths \citep{weiss_mass-radius_2014, lundkvist_hot_2016, otegi_revisited_2020} (see \figmassradius) and theoretical predictions motivated by population studies of close-in planets \citep{gupta_sculpting_2019, lee_breeding_2016,owen_kepler_2013,lopez_role_2013} agree on the rocky nature of short-period super-Earth-sized planets. Kepler-138~d shows that this is not universally true, especially at longer orbital periods and lower equilibrium temperatures. Previous studies had hinted at the lower density of small planets beyond 11 days using TTV mass measurements \citep{mills_planetary_2017}, but not to the extent that a planet as small as Kepler-138~c or d would be expected to have a non-rocky bulk density. 
Future discoveries of small transiting planets at low instellations combined with RV/TTV follow-up and atmospheric characterization have the potential to identify more of these temperate water worlds. This would provide us with an understanding of the relative occurrence of rocky vs. volatile-rich water worlds within the super-Earth size range, and the relative importance of incident irradiation and formation pathway \citep{kite_water_2021, bitsch_rocky_2019} for the planet's internal composition.

%######################################################################################################################

\begin{methods}

\subsection{\textit{HST}/WFC3 observations and light curve extraction.}

The \textit{HST} observed three transits of Kepler-138~d using the G141 grism of the WFC3 instrument as part of a multi-year survey program (GO~13665, PI Benneke; see \tabtrtimes). The first two transit observations consisted of five 96-minute orbits with 46-minute-long inter-orbit gaps in data acquisition due to Earth occultation, with the third and fourth-orbit observations deliberately timed to observe the transit ingress and egress of Kepler-138~d, respectively. The third transit observation, on the other hand, consisted of only four orbits, with the third orbit centered near mid-transit. At the beginning of each transit observation, we first obtained an image using the F130N filter (exposure time: 0.8s) to be used for the wavelength calibration purposes of the subsequent telescopic science observations. 
For each transit observation, these science observations consisted of a time-series of 103s exposures using the G141 grism, providing low-resolution spectrophotometry across the 1.1--1.7~$\mu$m range. To avoid instrumental overheads and allow for longer exposures, we used the spatial scan mode in which the telescope slews in time in the  cross-dispersion direction. We utilize both forward and backward scans of maximal length across a large fraction of the detector $256 \times 256$ pixel subarray, again to optimize the efficiency of the observing strategy. Throughout the observations, the number of electron counts per pixel did not exceed 32,000, or approximately $40$\% of the detector's saturation limit.

The \textit{HST} light curve extraction was performed within the ExoTEP framework following the procedures described in Refs. \citep{benneke_sub-neptune_2019,benneke_water_2019}. Starting from the 15 non-destructive reads stored in each `ima' file from the STScI standard reduction pipeline, we build one background-noise-reduced frame per exposure by subtracting consecutive non-destructive reads and adding only the rows of the detector that were illuminated by Kepler-138 in the time interval between those non-destructive reads \citep{deming_infrared_2013,benneke_sub-neptune_2019}. A wavelength-dependent flat-field image created from the 2D wavelength solution\citep{tsiaras_new_2016} is then used to produce a series of flat-fielded exposures from these frames\citep{benneke_sub-neptune_2019}. Bad pixels are flagged as 6$\sigma$ outliers within a region of 11$\times$11 pixels and replaced with the mean of the pixels within this region. 
In order to build the white light curves, we add all the electron counts within a rectangle covering the illuminated detector area.
For the spectrophotometric light curves used exclusively to build the transmission spectrum, on the other hand, we account for the fact that the grism dispersion is not perfectly uniform as the star's spectrum is scanned across the detector. This results in a near-rectangular but slightly trapezoidal-shaped illuminated region on the detector. To account for this 2-3 pixel shift of the wavelength solution, we sum the flux in trapezoidal wavelength bins, built using the 2D wavelength solution\citep{tsiaras_new_2016}. We thus integrate the flux over trapezoidal bins defined by pre-computed lines of constant wavelength. We perform three distinct extractions using either 30~nm bins, 120~nm bins, or four bins tailored to match the 1.4~$\mu$m absorption feature, spanning the wavelength range from 1.11 to 1.59~$\mu$m. No pre-smoothing is applied to the pixels, and we ensure total flux conservation by adding a fraction of the pixel fluxes that are intersected by the bin boundaries\citep{benneke_sub-neptune_2019}. In the extraction process, small drifts in the star position resulting in $x$ position shifts are accounted for exposure-by-exposure.

\subsection{\textit{Spitzer}/IRAC observations and light curve extraction.}

We observed 10 transits of Kepler-138~d with \textit{Spitzer}, five in each of the 3.6~$\mu$m and 4.5~$\mu$m channels of the IRAC detector (Program GO 11131, PI Dragomir; see \tabtrtimes). 
Each observing sequence was preceded with 30-minute peak-up mode pre-observations (using for positional reference the Pointing Calibration and Reference Sensor) to enable the mitigation of telescope drift and temperature variations during the transition to a new target \citep{grillmair_pointing_2012} prior to the science observations. We chose an exposure time of 2.0~s, to minimize nonlinear detector effects while simultaneously optimizing integration efficiency. Overall, each transit observation consists of 225 individual frames taken over 8 hours.

We used ExoTEP to extract photometric time series from the \textit{Spitzer} observations \citep{benneke_spitzer_2017}, with
the flat-fielded and dark-substracted ``Basic Calibration Data'' (BCD) images from the standard IRAC pipeline as starting point.
The star position was obtained using flux-weighted centroiding with a radius of 3.0 pixels. Background subtraction was performed by fitting a Gaussian function to a histogram of pixel-count values for pixels away from the point spread function of the target star. We ignored all elements within 12 pixels of the star position, as well as those in the 32$^{nd}$ row of the array which are systematically lower than what is observed in the rest of the image. We removed 3$\sigma$ outliers prior to background estimation. 
Finally, the photometric time series were obtained by adding up the flux in a circular aperture centered on the star's position. We tried aperture radii of [1.5, 2.0, 2.5, 3.0] pixels and selected for each visit the aperture radius that minimizes both the RMS in the unbinned residuals, and time-correlated noise in the systematics-corrected data.
The light curve was median-normalized and binned to 80-seconds cadence, to ease the subsequent systematics removal, with BJD UTC mid-exposure times calculated from the time stamp in the headers of the BCD images.

Following standard procedure, we discard the first orbit as well as the first forward and backward scan from each of the subsequent orbits, which are affected by a stronger systematic effect. We also remove the 50$^{\mathrm{th}}$ and 56$^{\mathrm{th}}$ exposures in the first transit observation and the 21$^{\mathrm{st}}$, 33$^{\mathrm{rd}}$ and 55$^{\mathrm{th}}$ exposures in the second transit observation  that are affected by cosmic ray hits.

\subsection{\textit{HST} and \textit{Spitzer} photometric light curve analysis.}

Following the procedures described in Refs. \citep{benneke_spitzer_2017,benneke_sub-neptune_2019,benneke_water_2019}, we analyze each of the \textit{HST}/WFC3 and \textit{Spitzer}/IRAC transit observations by simultaneously fitting the astrophysical transit-light curve model, an instrument-specific systematics model, and the photometric scatter using Affine Invariant Markov chain Monte Carlo (MCMC)\citep{foreman-mackey_emcee_2013}. The transit light curve model $f(t)$ is computed using \texttt{batman} \citep{batman}, and we fit to each transit observation the transit mid-time , the apparent planet-to-star radius ratio, $R_p/R_\star$ in the spectral bandpass at hand, as well as the scaled orbital distance $a/R_\star$ and transit impact parameter $b$. We impose Gaussian correlated priors on $a/R_\star$ and $b$ informed by the tightly-constrained posterior distributions from the fit to the \textit{Kepler} observations of Kepler-138\citep{almenara_absolute_2018}. For stellar limb-darkening, we use uncorrelated Gaussian priors to marginalize over the uncertainties of the parameters in each bandpass.  

The limb-darkening for the \textit{HST} transits is modeled using the LDTK package \citep{parviainen_ldtk_2015}.  We use LDTK to calculate the four coefficients of a four-parameter non-linear law as well as their uncertainties using the MCMC sampling option in LDTK, and provide as inputs to the model the constraints on the stellar $T_\mathrm{eff}$, $\log~g_\star$, and [Fe/H] (\tabstarparams). The choice of a four-parameter non-linear limb-darkening law was motivated by the impact of the choice of limb-darkening law on the retrieved transit depths at these wavelengths. Similar fits to the \textit{HST} transits using a quadratic limb-darkening law resulted in a systematic offset of about 25 ppm in the white light curve transit depths. Meanwhile, for \textit{Spitzer}/IRAC, we adopt a quadratic limb-darkening parametrization after checking using the same method as for \textit{HST}/WFC3 that at these longer wavelengths, the choice of limb-darkening law does not noticeably or systematically impact the transit depths.. We set the prior mean of the priors on both coefficients to the values corresponding to the closest-matching set of stellar parameters in a grid of precomputed coefficients \citep{sing_stellar_2010}. We use the typical difference between parameters at neighboring nodes in the grid in terms of $T_\mathrm{eff}$, $\log~g_\star$, and [Fe/H] as the standard deviation of the Gaussian priors in this case, as the separation between grid nodes is greater than the uncertainty on stellar parameters.

The systematics model for the \textit{HST}/WFC3 analysis accounts for the presence of well-documented instrumental systematics in \textit{HST}/WFC3 observations \citep{deming_infrared_2013,kreidberg_clouds_2014, kreidberg_detection_2015,benneke_water_2019} and captures visit- and orbit-long trends using a parametric model:
\begin{equation}\label{eq:sys_wlc}
S_{\mathrm{WFC3}}(t) = (c s(t) + vt_v) \times (1-e^{-a t_{\mathrm{orb}} - (b + d(t))}).
\end{equation}
\noindent The first term describes the visit-long trend and differences between forward and backward scans. The normalization constant $c s(t)$ is equal to $c$ for forward scans and $c s$ for backward scans, while $v$ is a visit-long slope that multiplies $t_v$, the time since the start of the visit. The second term accounts for systematic variations within each \textit{HST} orbit. We fit for the rate $a$ of the exponential ramp as a function of $t_{\mathrm{orb}}$, the time elapsed since the start of the orbit. The term $b+d(t)$ describe the ramp amplitude and $d(t)$ has a value of 0 for forward scans and $d$ for background scans. This adds up to a total of 6 free parameters ($c$, $s$, $v$, $a$, $b$, $d$) that describe the \textit{HST} instrument model and are fitted jointly with the astrophysical transit model parameters. Fitting the light curves with more complex systematics models where $v$ or $a$ take different values for forward vs. backward scans results in consistent retrieved transit times and transit depths while providing no significant improvement in terms of the quality of the systematics removal.

Equivalently, the \textit{Spitzer}/IRAC systematics model accounts for variations associated with non-uniform intra-pixel sensitivity and a temporal drift. We correct for the intra-pixel sensitivity variations using a pixel-level decorrelation (PLD) method \citep{deming_spitzer_2015,benneke_spitzer_2017} and combine the PLD term with a `ramp' term describing variations in the detector sensitivity over time. We fit successively all 10 \textit{Spitzer}/IRAC light curves using ExoTEP for a variety of analytical forms for the time ramp \citep{stevenson_transit_2012}, and select the ramp description for which the residuals to the best-fit models most closely match the expectation for photon noise-limited precision. The full expression of the systematics model is:
\begin{equation}\label{eq:sys_spitzer}
\begin{split}
    S_\mathrm{Spitzer}(t)& = \left(1 + \frac{\sum_{k=1}^{9} w_k D_k(t)}{\sum_{k=1}^{9} D_k(t)} \right) \\
    &\times \left(1 + r_0 \ln(t-t_0) + r_1 \left[\ln(t-t_0)\right]^2\right).
\end{split}
\end{equation}
\noindent where the $D_k$ are the raw counts on the central $3 \times 3$ pixels of the IRAC detector. 
The \textit{Spitzer} instrument model has 12 free parameters, including 9 PLD weights ($w_k$) describing intra-pixel variations (first term in Eq. \ref{eq:sys_spitzer}) and 3 additional parameters ($r_0$, $t_0$ and $t_1$) that account for time-dependent variations (second term in Eq. \ref{eq:sys_spitzer}). We discard the start of the out-of transit baseline for five of the visits, which were heavily affected by detector systematics. The first 30 minutes of the visits on 2015 Aug 8, 2015 Sep 1 and 2016 Sep 4, and the first hour of the observations taken on 2015 Oct 17 and 2016 Aug 12 were therefore ignored in our analysis.  

Finally, in each light curve analysis, the log-likelihood function optimized for the fit to each visit $V$ takes the form:
\begin{equation}
\begin{split}
\ln \mathcal{L} =  &-n_V \ln \sigma_V -\frac{n_V}{2} \ln 2\pi \\
&- \sum_{i=1}^{n_V}\frac{\left[D_V(t_i)-S_V(t_i)\times f_V(t_i)\right]^2}{\sigma_V^2}
\end{split}
\end{equation} 
\noindent where $n_V$ is the number of points in the visit, $S_V$ and $f_V$ are the instrument and astrophysical models suited to the visit and instrument at hand, and the $D_V(t_i)$ are the datapoints of the broadband light curve. The photometric scatter $\sigma_V$ is fitted alongside with the parameters of the instrument and astrophysical models. We use four times as many walkers as there are free parameters in the fit, run the chains for 10,000 steps and discard the first 60\% as burn-in. We check for convergence by calculating the autocorrelation time $\tau$ of the chains and find that all have run for more than 80$\tau$ past the burn-in phase. The \textit{HST} systematics-corrected white and spectroscopic light curves are shown in \fighstlcs for the main extraction in four equal-width 120nm wavelength bins, and the \textit{Spitzer} light curves are shown in \figspitzerlcs. In the end, we obtain the desired transit times and their uncertainties, as well as the radius ratio $\left(R_p/R_\star\right)$ by marginalizing the posterior distribution from the MCMC over all other parameters.

\subsection{TTV analysis.}

To infer the masses and orbital parameters of the planets in the Kepler-138 system, we combine the observations from our targeted \textit{HST} and \textit{Spitzer} transit campaign with the previously obtained observations from the \textit{Kepler} mission. We perform both an initial exploratory TTV analysis based on the individually inferred transit times (using  literature values for the \textit{Kepler} transit times \citep{jontof-hutter_mass_2015}), as well as a full photodynamical analysis directly leveraging the photometric observations (see next section). The \textit{Kepler} space telescope observed Kepler-138 throughout Quarters Q0-Q17. In total, \textit{Kepler} recorded 121 transits of Kepler-138~b, 85 transits of Kepler-138~c and 51 transits of Kepler-138~d between 2008 and 2013. Adding the \textit{HST} and \textit{Spitzer} observations, our dataset covers Kepler-138 over 7 years, with the \textit{HST} and \textit{Spitzer} critically extending to coverage from 1 to nearly 2 super-periods of the interaction between Kepler-138~c and d (see \figfullphotodynres). A baseline covering more than one full cycle of TTV modulation is essential to ensure that the exploration of the parameter space is not hindered by the presence of disconnected local likelihood maxima \citep{ragozzine_value_2010,agol_refining_2020,jontof-hutter_following_2021}.

We perform the TTV analysis using \texttt{TTVFast} \citep{deck_ttvfast_2014} in combination with the Markov Chain Monte Carlo package \texttt{emcee} \citep{foreman-mackey_emcee_2013}.
For each planet, we adopt as the fitting basis the planet mass $M_\mathrm{p}$ and its orbital elements described by their Jacobi coordinates, consisting of the orbital period $P$, the eccentricity and argument of periastron parametrized as $\sqrt{e} \cos \omega$ and $\sqrt{e} \sin \omega$, the inclination $i$, the mean anomaly $M_\mathrm{0}$ and the longitude of the ascending node, $\Omega$. We use the basis $(\sqrt{e} \cos \omega, \sqrt{e} \sin \omega)$ rather than $(e \cos \omega, e \sin \omega)$ to avoid a bias towards high eccentricities in the MCMC \citep{ford_improving_2006}.
We also fit $M_\mathrm{0} + \omega$ rather than $M_\mathrm{0}$ directly, because transit observations better constrain the planet's position relative to the transit than the planet's position relative to the ascending node.
For a N-planet fit, we only fit $\Omega$ for $N-1$ planets as the quantity of interest is the relative value of the longitudes of the ascending node, and fix $\Omega_\mathrm{c}$ to 180 degrees. \texttt{TTVFast} takes as inputs the masses of the host star and the planets, but the quantity that is constrained by the TTVs is the planet-to-star mass ratio $M_\mathrm{p}/M_\mathrm{\star}$ rather than the planet masses themselves. Therefore, we fit the stellar mass along with the planet parameters to marginalize over its uncertainty when constraining planet masses. We impose a Gaussian prior on the stellar mass with a mean and standard deviation that match the updated Gaia DR2 stellar parameters for Kepler-138 obtained from empirical relations for M dwarfs \citep{mann_how_2015,mann_how_2019,berger_gaia-kepler_2020}. Flat priors are used for the other parameters. Our TTV fit for three planets has 21 free parameters, sampled using \texttt{emcee} \citep{foreman-mackey_emcee_2013}. For the modeling of the TTVs, we integrate the orbital evolution of the  planets from $t_{\mathrm{start}} \mathrm{[BJD_{TDB}]} = 2454955.0$ to $t_{\mathrm{end}} \mathrm{[BJD_{TDB}]} = 2457650.0$ using a time step of 0.5 days.

No three-planet model can provide a satisfactory fit to the combined set of transit times of Kepler-138~d from \textit{HST}, \textit{Spitzer} and \textit{Kepler} (\figfullphotodynres abc). We therefore perform a suite of four-planet fits, raising the number of free parameters to 28. We scan the parameter space of orbits beyond Kepler-138~d for a potential planet e. In particular, a position near a mean-motion resonance is needed to explain the long-term deviation of the transit times of Kepler-138~d from the previously inferred three-planet solution \citep{jontof-hutter_following_2021, almenara_absolute_2018}. Therefore, we specifically investigate the presence of planet e near the first-order (2:1, 3:2, 4:3, 5:4) or second-order (3:1, 5:3) mean-motion resonances with Kepler-138~d, as well as the third-order resonance 5:2.
For this exploratory phase, we adopt Gaussian priors on the planets' eccentricities (mean of 0 and standard deviation of 0.1 on $\sqrt{e} \cos \omega$ and $ \sqrt{e} \sin \omega$), inclinations (mean of 90 degrees, standard deviation of 2 degrees), longitudes of ascending node (mean of 180 degrees, standard deviation of 2 degrees) and on the period ratio $P_\mathrm{e}/P_\mathrm{d}$ (mean at the target mean-motion resonance and standard deviation of 0.1). We test a range of spreads for the initialization of the \texttt{emcee} walkers for each parameter, and run fits where the proposal for the next step of each walker uses either the ``stretch move''\citep{goodman_ensemble_2010} or the ``Differential Evolution'' move\citep{nelson_run_2013}, in order to capture potential local maxima.

We discover that within the set of all explored orbits for Kepler-138~e, only a solution where Kepler-138~e is in a $\sim 38$d orbit (near the 5:3 second-order mean-motion resonance with Kepler-138~d) can simultaneously reproduce well the observed \textit{Kepler}, \textit{HST} and \textit{Spitzer} transit times of planets b, c, and d (\figfullphotodynres abc). 
Once this orbital solution was identified, we use it as an initial guess for a final TTV fit that does not impose a prior on the period ratios, to obtain statistical uncertainties on the planet parameters. 
We use 20 times as many walkers as there are free parameters. The chains run for 200,000 steps and we check for their convergence by ensuring that the number of steps exceeds 50 times the autocorrelation time for each parameter. 
We use this final TTV fit to validate the results from our photodynamical fit (see below).

\subsection{Photodynamical analysis.}

We refine system parameters following the exploratory TTV analysis using a photodynamical model. The photodynamical approach directly couples a dynamical code to a light curve model to leverage the information contained in the transit light curves, rather than only the fitted transit times.

The photodynamical model is parametrized by the stellar density, the planet-to-star mass, radius ratio, and the orbital parameters of each planet at a reference time $t_\mathrm{ref}$. The jump parameters that are sampled with \texttt{emcee} are set following previous work\citep{almenara_absolute_2018} such that correlations are minimized. We define the two jump parameters $P'$ and $T_0'$ as follows:
\begin{equation}
  P' \equiv \sqrt{\frac{3\pi}{G \rho_\star} \left(\frac{a}{R_\star}\right)^3}  
\end{equation}
\begin{equation}
T_0' \equiv t_\mathrm{ref} - \frac{P'}{2 \pi} \left(M_0 - E + e\sin E\right)
\end{equation}
\noindent with
\begin{equation}
E=2 \arctan \left\{\sqrt{\frac{1-e}{1+e}} \tan \left[\frac{1}{2} \left(\frac{\pi}{2}- \omega\right)\right]\right\}.
\end{equation}
Our model neglects light-time and relativistic effects, which should only be of the order of milliseconds for the Kepler-138 system \citep{heyl_using_2007, almenara_absolute_2018}. This timescale is orders of magnitude lower than the precision determined from the transit light curves. We simulate the orbital evolution of the system with \texttt{REBOUND} \citep{rein_rebound_2012} and the \texttt{WHFast} integrator \citep{rein_whfast_2015}, using time steps of 0.01 days. We interpolate the positions of the objects between integration points using a cubic spline for the \textit{Kepler} short-cadence light curves and calculate their positions at 30 evenly spaced points around each observation date for all other transits. We compute transit light curves using the analytic description of the transit shape \citep{mandel_analytic_2002}.

The \textit{HST} and \textit{Spitzer} inputs for the photodynamical fit are the best-fit systematics-corrected light curves. 
For the \textit{Kepler} observations, we use simple aperture photometry (SAP) light curves retrieved from the Mikulski Archive for Space Telescopes, and select short-cadence over long-cadence observations where available (Q6-Q17). We favor the SAP over the Pre-search Data Conditioning SAP (PDCSAP) light curves because the latter are missing 1 transit of planet b and 2 transits of planet d. There are no important differences between the SAP and PDCSAP light curves on the time scale over which the fitted transits occur.
The light curves are processed following previous work and including a correction for flux contamination\citep{almenara_absolute_2018}. We use a window size of three transit durations around each transit for the light curve modeling. The light curve within each transit window was normalized with a second-order polynomial (attempts using a single or third-order polynomial yielded similar results), and corrected for the effect of stellar activity using a spot model from a previous fit to the \textit{Kepler} light curves \citep{almenara_absolute_2018}.

We set the limb-darkening coefficients to the median value used in the Gaussian prior for the ExoTEP fit (fixing or fitting these coefficients did not impact our conclusions; see below), and fit the two parameters of a quadratic limb-darkening law to the \textit{Kepler} transits. 
We account for any detectable offsets in the \textit{Kepler}, \textit{HST}, and \textit{Spitzer} 3.6 and 4.5$\mu$m broadband transit depths by fitting four values of $R_\mathrm{p}/R_\mathrm{\star}$ to the light curves of Kepler-138~d (one per bandpass). 

In total, we perform four MCMC fits using the photodynamical model to ensure that our results are not affected by the choice of prior on the stellar parameters, or the two degenerate solutions for the inclinations of Kepler-138 b and c \citep{almenara_absolute_2018}. For each of the four fits, we use the best-fit parameters yielded by the four-planets TTV analysis as an initial condition for the masses and orbital locations of Kepler-138~b, c, d, and e in our photodynamical analysis.
In two of the fits, a Gaussian prior is imposed on the stellar density based on the most recent literature values for the mass and radius (\tabstarparams, Ref. \citealp{berger_gaia-kepler_2020}), while the other two invoke a flat prior on the stellar density. For a given stellar density prior, we initialize two fits, each with its set of \texttt{emcee} MCMC chains in one of the two degenerate inclination configuration (with $i_b$, $i_c$ both either above or below 90 degrees\citep{almenara_absolute_2018}). Altogether, each of these analyses has 48 free parameters. 
For each of the 4 MCMC runs, chains are run for 320,000 steps to sample the posterior near the solution identified by the TTV fits, with 200 walkers for each combination of inclination configuration and stellar density prior. We check that the chains have reached convergence by computing the autocorrelation timescale for each walker and parameter, and ensuring that all chains have run for over 60 autocorrelation time scales. We combine the results for the two inclination configurations and the same stellar density prior as a postprocessing step to produce the full posterior distribution.
Additionally, we performed another series of four-planet photodynamical fits to test the impact of marginalizing over the systematics in the \textit{HST} and \textit{Spitzer} light curves, as well as fitting the limb-darkening coefficients in these two bandpasses. For this test, we inflated the error bars on the best-fit light curves by adding in quadrature the additional dispersion introduced by different systematics models. The increase in the single-point errors is typically small (below 10 to 15\%) compared to the white light curve error already computed for the best-fit model. The limb-darkening coefficients were fitted using the same laws and Gaussian priors as in the ExoTEP fit. We find that these analyses yield results that are statistically indistinguishable from the previous fits, with comparable parameter values and uncertainties. Finally, we also performed one additional photodynamical fit with only the three known planets Kepler-138~b, c, and d. In this case,  the starting point is set the best-fit solution of the three-planet, instead of four-planet, TTV fit. The results from this fit illustrate how significantly the new transit times of Kepler-138~d deviate from the three-planet prediction (see \figthreeplaphotodyn). The new \textit{HST} and \textit{Spitzer} observations were essential in constraining the timescale over which the transit times of Kepler-138~d are modulated, and reveal at high significance the presence of a fourth planet in the system. 

Our best fitting four-planet photodynamical model reproduces well the \textit{HST}, \textit{Spitzer} and \textit{Kepler} transit observations 
(\figfitlcsb, \figfitlcsc, \figfitlcsd, \figfitlcsdspitzerhst) and provides independent constraints on stellar parameters. We find $\rho_\star=4.9 \pm 0.5$ g/cm$^3$ for a uniform prior on $\rho_\star$, compared to $\rho_\star=4.9 \pm 0.4$ g/cm$^3$ using a Gaussian prior informed by literature values \citep{berger_gaia-kepler_2020}.
We obtain an independent stellar radius estimate of $R_\star=0.535 \pm 0.017\,R_\odot$ from the fit with the uniform prior on $\rho_\star$, using the literature value of the stellar mass (\tabstarparams). The choice of stellar density prior does not impact our inference of the stellar or planetary parameters, and we choose to report the planet parameters inferred using the Gaussian stellar density prior (\figfullphotodynres, \figphotodyncorner, \tabphotodynres, \figmassradius). 

We do not expect dilution to affect the inferred Kepler transit depth of Kepler-138~d. Both the consistency in the transit depths of each planet inferred from different quarters of the Kepler data after their independent correction for flux contamination\citep{almenara_absolute_2018}, and the absence of any detected stellar companion to Kepler-138 from adaptive optics imaging\citep{wang_influence_2015}, suggest that our results are not biased by this potential source of contamination.

We derive from the photodynamical fit transit depths of $672 \pm 16$ ppm, $648 \pm 44$ ppm, $511 \pm 47$ ppm and $565 \pm 57$ ppm in the broadband \textit{Kepler}, \textit{HST}/WFC3, \textit{Spitzer} 3.6$\mu$m and 4.5$\mu$m bandpasses respectively. We validate their consistency with the results from the light curve fits (\tabspectrum), which gives further credence to the fitted orbital solution. Indeed, a poor match to the transit times from the dynamical model would have resulted in smaller inferred planet radii due to the mismatch between the light curve model and the observations. For the spectrum, we keep the ExoTEP results that are more conservative estimates of the transit depth uncertainties, due to the simultaneous fitting of the systematics and astrophysical models.

Our photodynamical modeling also indicates that Kepler-138~e might not be transiting with an impact parameter of $1.8_{-1.2}^{+1.9}$.
We confirm this using the full \textit{Kepler} long-cadence PDCSAP light curve of Kepler-138 to search for the transit of Kepler-138~e. We correct for the modulation associated with the presence of stellar surface inhomogeneities using our spot model \citep{almenara_absolute_2018}, and then fold the light curves around the median transit times from the posterior distribution of the fit to Kepler-138~b, c, d, and e in a window of two days around each transit epoch (the 3-sigma uncertainty on the transit times of Kepler-138~e is of $\sim 50$ hours). Finally, we normalize each segment by the out of transit median value, apply 3$\sigma$ clipping to the folded light curves and superimpose \texttt{batman} transit models corresponding to the median retrieved parameters of the four planets. For Kepler-138~e, we assume an Earth-like composition to estimate the planet's radius and explore several values for the inclination of the planet (\figplanetesearch). The non-detection of Kepler-138~e in the \textit{Kepler} light curves is compatible with the constraint on its inclination. For example, assuming an Earth-like (vs. 90\% iron, 10\% silicates) composition, Kepler-138~e has a 25.4\% (25.3\%) probability of transiting its host star but only a 0.3\% ($<0.01$\%) probability of producing transits deeper than the $\approx$ 300 ppm scatter in the folded \textit{Kepler} light curve.

\subsection{Keck/HIRES Radial Velocity observations and analysis.} \label{sec:rv_analysis}

We collected a total of 28 RV measurements of Kepler-138 over 28 nights between 2011 and 2015 using the High Resolution Echelle Spectrometer~HIRES\citep{vogt_hires_1994} on the Keck I Telescope. The observations were conducted by the California Planet Search. The ``C2'' decker was used for data acquisition and sky subtraction, with median exposure times of 1920 seconds ($\sim 32$ minutes), reaching a median S/N of 91/pixel. Wavelength calibrations were performed using the iodine cell~\citep{butler_attaining_1996}. The HIRES data reduction followed the standard procedures of the California Planet Search \citep{howard_california_2010}. These RVs, and activity indicators, are included in \tabRVdata.

We analyze the \textit{Keck}/HIRES RVs of Kepler-138 to independently constrain the masses of Kepler-138~b, c, d, and e using \texttt{RadVel} \citep{fulton_radvel:_2018}. In the analysis, we account for the effect of stellar activity using a Gaussian process (GP) model trained on the \textit{Kepler} light curve because the expected radial-velocity (RV) semiamplitudes $K$ of the planetary signals are below the typical uncertainties of $\sim2$ m/s on the HIRES RVs. We choose to only fit for $K_\mathrm{b}$, $K_\mathrm{c}$, $K_\mathrm{d}$, and $K_\mathrm{e}$ and leave other orbital parameters fixed to their median values from the photodynamical fit (\tabphotodynres) in the Keplerian orbit model. We use an additional jitter term $\sigma_\mathrm{H}$ to account for residual RV scatter due to stellar activity. 

The measured RVs are correlated with the S-index which acts as a stellar activity tracer (Pearson-r=0.67). We thus proceed to a careful treatment of the stellar activity component. 
We use a GP kernel to model the covariance between observations that are not only close in time, but also close in terms of their phase with respect to stellar rotation. Such analyses have proven more helpful to tease out low-amplitude RV variations than using a single jitter term or even parametric periodic models (see e.g. \citealp{amado_carmenes_2021,ahrer_harps_2021}). 
We build our GP model of the covariance structure of the RV dataset by optimizing the following Gaussian log-likelihood function:

\begin{equation}
    \ln \mathcal{L} = - \frac{1}{2} \left(N \ln 2 \pi + \ln |\Sigma| + \textbf{y}^T \Sigma^{-1}\textbf{y}\right)
\end{equation}
\noindent where $N$ is the number of points in the light curve, $\Sigma$ is the covariance matrix (described below), and $\textbf{y}$ is a vector of all photometric data points. The covariance matrix is constructed such that each of its elements $\Sigma_{ij}$ describes the covariance between observations at times $t_i$ and $t_j$ following a quasi-periodic kernel:
\begin{equation}
    \Sigma_{ij}=a_\mathrm{GP}^2 \exp\left[-\frac{(t_i - t_j)^2}{\lambda^2} - \frac{\sin^2\left(\frac{\pi |t_i-t_j|}{P_{\mathrm{GP}}}\right)}{2\Gamma^2}\right] + \sigma_w^2\delta_{ij}
\end{equation}
\noindent where $a$ is the correlation amplitude, $\lambda$ is the coherence timescale of the stochastic phenomena, $\Gamma$ is the timescale of the periodic variations, $P_{\mathrm{GP}}$ is the stellar rotational period and $\sigma_w$ is a white noise term along the diagonal. The radial term in the exponential encodes the stochastic nature of the variations, while the periodic term describes the rotational modulation of the signal. 

\textit{Kepler} observed Kepler-138 nearly continuously throughout Quarters 0 to 17, recording the star's brightness variations over more than 60 stellar rotation cycles and providing exquisite constraints on the covariance structure of time-correlated signals associated with stellar activity. We therefore fit the same GP model to the \textit{Kepler} light curve to use this ``trained'' GP model as a prior on the covariance structure of the RV time series. We preprocess the \textit{Kepler} PDCSAP light curve prior to training, mostly to reduce the number of data points in the training set while still retaining critical information on the stellar brightness variations on short timescales. First, we discard epochs where transits occur, as the star's brightness is then affected by partial occultation from the planet, and perform 5$\sigma$ clipping. The median-normalized light curve then passes through a median filtering step, followed by a resampling (1 in 20 points are kept). Finally, we retain only observations taken after BJD=2455750, i.e. that were simultaneous with the RV time series. A subset of the last 200 days of observations is shown in the left panel of \figphotGP, which demonstrates that the final time sampling remains sufficient for characterizing the variability in the stellar brightness.

We use the \texttt{george} \citep{ambikasaran_fast_2015} package for the GP model and fit the five kernel parameters to obtain constraints on the light curve's covariance structure within a Bayesian framework. The parameter space is explored using the \texttt{emcee} package via a Gaussian likelihood function and Jeffreys priors on $a_{\mathrm{GP}}$, $\lambda$, $\Gamma$ and $\sigma_w$. For $P_{\mathrm{GP}}$ we adopt a uniform prior from 15 to 22 days, informed by previous studies \citep{almenara_absolute_2018,mcquillan_measuring_2013,mcquillan_stellar_2013} and the periodogram of the light curve which exhibits significant peaks at both the stellar rotation period and its first harmonic (\figRVpdgm). 

In the training step, we run 20 chains for 5,000 iterations, 60\% of which are discarded as burn-in. We confirm that the chains are converged by calculating autocorrelation timescales for all chains and ensuring that they amount to less than 1/50 of the total number of steps. We obtain posterior distributions on all parameters, including the period of the orbital modulations (see \figphotGP) and correlation length scales, and transfer this knowledge via the prior when fitting the RV data.
Three GP parameters have priors informed by this training step: $\lambda$, $\Gamma$ and $P_{\mathrm{GP}}$. The GP amplitude $a_\mathrm{GP}$ and residual white noise term $\sigma_w$ are fitted independently for both time series, as the scatter and amplitude of the stellar activity induced variations are not expected to be shared between photometric and RV datasets. One notable possible caveat is that the \textit{Kepler} light curve extends to BJD=2456424, approximately 2 years prior to the last recorded RV measurements in 2015. Therefore, any change in the covariance structure of the stellar activity signal between the RV and photometric time series will not be captured in our trained model. 

Our final RV fit has 9 free parameters: the four planets' semi-amplitudes, the jitter term as well as $a_\mathrm{GP}$ (the amplitude of the stellar activity component) and the three trained parameters $\lambda$, $\Gamma$ and $P_{\mathrm{GP}}$ for which we used as priors the kernel density estimates from the post-burnin samples of the fit to the \textit{Kepler} photometry. We run 20 chains for 10,000 steps and use the same criterion as above to assess that convergence was achieved. We obtain the joint posterior distribution of the planet parameters and the stellar activity component (\tabRVresults, \figRVfit and \figRVcorner). The addition of the trained GP reduces the residual jitter by $\approx 1$ m/s compared to our initial fit which did not account for the stellar activity contribution. 

We exclude from the RV dataset an outlier measurement at BJD=2457294.89 and attribute it to either stellar activity or instrumental noise: this point has a Mt. Wilson S-value, RV internal error, and amplitude a factor of 2 to 3 larger than the rest of the time series. This point lies right at the quadrature phase for Kepler-138~c, and biases its RV solution if included, but not those of the other planets. Alternatively to the trained GP model, we performed fits to the RV observations where stellar activity was only modeled as a residual jitter, or with a GP model fitted to the RV or S-index time series themselves. However, the sparsity of the datasets compared to the $\sim 20$\,d rotation period of the star hindered satisfactory modeling of the stellar activity component (\figRVpdgm).

Our RV observations provide stringent upper limits on the masses of Kepler-138 b, c, d, and e, in agreement with the mass constraints from the photodynamical and TTV fits (\figRVcorner). In particular, the mass of $>3.5\,M_\oplus$ required to invoke a rocky composition in the extreme case of an iron-free interior for Kepler-138~d is excluded with $>91$\% confidence by the RV fit alone. Continued precise-RV follow-up of the system with instruments that can reach sub-m/s precision (e.g. MAROON-X) holds the potential to provide precise mass estimates independent from TTV measurements.

\subsection{Transmission spectrum.} We construct the transmission spectrum of Kepler-138~d by combining the individual transit-depth measurements, $(R_p/R_\star)^2(\lambda)$ in the \textit{Kepler}, \textit{HST}/WFC3, and \textit{Spitzer}/IRAC bandpasses. \textit{Kepler} and \textit{Spitzer}/IRAC deliver broadband photometric measurement without spectroscopic information and we directly take the inferred transit depths from our light curve analysis discussed above. For \textit{HST}/WFC3, however, we divide the overall bandpass of the G141 grism observations into four wavelength bins of equal width (see \figretrieval). We then determine individual transit depths measurements from the spectrophotometric light curves extracted from each wavelength bin. When fitting these spectrophotometric \textit{HST} light curves, we take advantage of the fact that systematics can be considered wavelength-independent to first order and start by dividing each spectroscopic light curve by the ratio of the white light curve to its best-fitting transit model \citep{deming_infrared_2013,kreidberg_clouds_2014,benneke_sub-neptune_2019}. We then model the residual systematics in each spectroscopic light curve as a linear function of the $x$ position on the detector\citep{benneke_sub-neptune_2019,benneke_water_2019}: 
\begin{equation}
S_{\mathrm{WFC3, spec.}}(t) = a d(t) + m(x-x(t=0)).
\end{equation}
\noindent Here, $d(t)$ is defined similarly to Eq. \ref{eq:sys_wlc} and $m$ is the slope of the linear dependence. The systematics model for spectroscopic fits has 3 free parameters: $a$, $d$ and $m$.
For the final transmission spectrum, we use the weighted average of the results from fits to individual visits in each observed spectroscopic channel and bandpass (\figretrieval, \tabspectrum).

\subsection{Atmospheric retrievals.}
We model the atmosphere and transmission spectrum of Kepler-138~d using line-by-line radiative transfer within the SCARLET framework \citep{benneke_strict_2015,benneke_atmospheric_2012,benneke_how_2013,benneke_sub-neptune_2019}. We use the \texttt{nestle}(\url{https://github.com/kbarbary/nestle}, Refs.\citealp{feroz_multinest_2009,shaw_efficient_2007,mukherjee_nested_2006,skilling_nested_2004}) nested sampling package to perform atmospheric retrievals and determine the range of physically-plausible scenarios that can give rise to the observed spectrum. The nested sampling method additionally enables us to perform Bayesian model comparison based on the Bayesian evidence to assess which parameters are required to explain the observed data \citep{skilling_nested_2004}. We use a total of 30,000 active samples to explore the parameter space. A new sample is drawn at each iteration using the multi-ellipsoid method \citep{feroz_multinest_2009}. Our stopping criterion for the drawing of new samples is a threshold placed on the ratio between the estimated total evidence and the current evidence:
\begin{equation}
    \log (\mathcal{Z}_i + \mathcal{Z}_\mathrm{est}) - \log \mathcal{Z}_i < 0.5,
\end{equation}
\noindent where $\mathcal{Z}_\mathrm{est}$ is the estimated remaining evidence from the highest likelihood reached so far $\mathcal{L}_\mathrm{max}$ and the remaining prior volume at step $i$, $X_i$ ($\mathcal{Z}_\mathrm{est} = \mathcal{L}_\mathrm{max} X_i$), and $\mathcal{Z}_i$ the calculated evidence at step $i$.

None of the main infrared absorbers (H$_2$O, CH$_4$, CO, CO$_2$, NH$_3$, HCN) are significantly detected when their abundances are fitted independently. Therefore, we opt for chemically-consistent retrievals where the atmospheric metallicity and C/O ratio dictate the composition of each layer in chemical equilibrium, given a  temperature structure \citep{benneke_strict_2015}. We parametrize the atmospheric metallicity as the ratio  $\left(n_\mathrm{Z}/n_\mathrm{H}\right)_\mathrm{atm}/\left(n_\mathrm{Z}/n_\mathrm{H}\right)_\odot$ where $n_\mathrm{H}$ is the number density of hydrogen and Z stands for all metals. Transmission spectroscopy only provides weak constraints on temperature gradients in the atmosphere \citep{benneke_atmospheric_2012} and we elect to fit in the retrieval for the temperature of an isothermal profile. This single fitted temperature is physically representative of the atmospheric terminator region probed by our observations. We consider the presence of clouds at the terminator and parametrize them with the pressure $P_{\mathrm{cloud}}$ where the gray cloud deck becomes optically opaque to grazing light beams (\figretrieval). 
All Bayes factors for a more complex cloud model versus the baseline gray cloud model (including hazes, or a full Mie scattering description \citep{benneke_sub-neptune_2019})  are $<1.5$ and thus inconclusive as to whether the data supports the added model complexity (``Jeffreys' scale''; Jeffreys 1961, \citealp{trotta_bayes_2008}). The constraints obtained on the pressure level of a homogeneous cloud are inherently tied to the model prescription and would have to be updated for future analyses of more precise data to account for the possibility of variable particle sizes or a non-uniform cloud coverage \citep{benneke_sub-neptune_2019, line_influence_2016}. Our final retrieval therefore explores a wide range of C/O ratios, metallicities, terminator temperatures and $P_{\mathrm{cloud}}$ in order to determine the range of scenarios consistent with the observed spectrum of Kepler-138~d.

The observed transmission spectrum is consistent with the volatile-rich ``water world'' scenario inferred from the planet's mass, radius, insolation, as well as atmospheric loss considerations. Such a high metallicity of $>100 \times$ solar at 2$\sigma$ (\figretrieval) results in a high mean molecular weight atmosphere that does not show strong features in the transmission spectrum. For any H$_2$/He-dominated atmosphere to match the transmission spectrum, one would need to additionally invoke clouds above the 0.1 bar level at 2$\sigma$ (\figretrieval). These two scenarios are degenerate in terms of the amplitude of the resulting spectral features in transmission \citep{miller-ricci_atmospheric_2009,benneke_how_2013}. Meanwhile, atmospheric compositions that would produce large features, e.g. if the atmosphere was cloud-free and had a near-solar metallicity, are disfavored by the observations at $>2 \sigma$ (\figretrieval).

\subsection{Coupled interior-atmosphere structure modeling: hydrogen-rich atmospheres.}

We compare the measured mass and radius of Kepler-138~d to a grid of self-consistent four-layer (iron, silicates, water, and hydrogen) coupled interior+atmosphere structure models to account for the size of the radiative layer, the link with interior models, and the effect of non-gray opacities resulting from the atmospheric composition. These full-planet models are modular and can be adapted to predict radii for a variety of atmospheric compositions and relative fractions of iron, silicates, and water in the planetary interior, as well as across planet ages if coupled with a thermal evolution model. 

The first step in constructing the full-planet models is building a grid of interior models (methods outlined in \citealp{thorngren_intrinsic_2019}). Our interior models grid spans a wide range of planet masses, H$_2$/He mass fractions $f_\mathrm{H_2/He}$, internal water mass fractions $f_\mathrm{H_2O}$, and specific entropies. The water mass fractions are parametrized in such a way that $f_\mathrm{H_2O}=0$ corresponds to a dry iron+silicates interior while for $f_\mathrm{H_2O}=1$, the interior of the planet is modeled as a pure H$_2$O composition underlying the hydrogen layer. The total water mass fraction is therefore $\left(1-f_\mathrm{H_2/He}\right) \times f_\mathrm{H_2O}$ , while the total rock/iron mass fraction is $\left(1-f_\mathrm{H_2/He}\right) \times \left(1-f_\mathrm{H_2O}\right)$ .
The interior models are in layers of iron, silicates, water, and H$_2$/He.
The rock/iron component is modeled as an Earth-like mixture of 1/3 iron and 2/3 olivine. We use the ANEOS equations of state (EOS) for the iron core and the rock layer. We adopt state-of-the-art EOS for the H$_2$/He (solar composition;  \citep{chabrier_new_2019}) and the water layer \citep{mazevet_ab_2019}. Adiabatic temperature-pressure profiles are computed within the water and H$_2$/He layers, while a uniform temperature is used for the rock/iron interior. For each interior model in the grid, we record the pressure, temperature, and radius as a function of the mass interior to a given mass bin and calculate profiles up to a pressure of 10 bar.

We then compute a grid of self-consistent models using SCARLET \citep{benneke_atmospheric_2012,benneke_how_2013, benneke_strict_2015, benneke_sub-neptune_2019}, from which the appropriate non-gray atmosphere model will be added on top of an interior model to form one full-planet model for each composition. For a fixed total planet mass, internal temperature of 30 K \citep{valencia_bulk_2013,madhusudhan_interior_2020,lopez_understanding_2014}, and target $f_\mathrm{H_2/He}$, we compute self-consistent non-gray atmosphere models from 10 kbar to $10^{-10}$ bar for a range of reference radii $R_\mathrm{ref}$ at 1 kbar. We improve SCARLET models upon previous work by lifting the assumption of a constant mass throughout the atmosphere in the hydrostatic equilibrium calculation. We rather ensure hydrostatic equilibrium self-consistently at each iteration by accounting for the mass contained in each atmosphere layer for the given temperature-pressure profile and chemical composition, and its impact on the gravitational field in the other layers.  

For a given planet mass, $f_\mathrm{H_2/He}$ and $f_\mathrm{H_2O}$, we couple the interior and atmosphere models such that their temperature and radius match at the pressure of the radiative-convective boundary (RCB; see \figstrucmodel). To this end, the location of the RCB is identified within the atmosphere model for each $R_\mathrm{ref}$ and $f_\mathrm{H_2O}$. Therefore, for different planet model parameters, the atmosphere model at a fixed internal temperature of 30 K will be matched with an interior model that has different specific entropies\citep{madhusudhan_interior_2020}. For water-free models, in cases where $f_\mathrm{H_2/He}$ is so low that the total mass in the SCARLET atmosphere down to the identified RCB exceeds the total H$_2$/He mass of the planet, we integrate the atmosphere mass from the top until we identify the pressure $p_\mathrm{bottom}$ above which the expected H$_2$/He mass is contained and compute the extent of the atmosphere using only the layers above. In all cases, the planet's radius as measured by the transit is assumed to be the radius at 20 mbar\citep{hubbard_theory_2001} in the combined model.
Finally, we obtain a grid of full-planet models that maps the planet's photosphere radius as a function of planet mass, water mass fraction and hydrogen mass fraction. The grid is equally spaced in log planet mass and water mass fraction, and equally spaced in log-space for the hydrogen mass fraction.

For low $f_\mathrm{H_2/He}$, the boundary between the hydrogen and water layers (HHB hereafter) can be located within the atmosphere model. For such a scenario, the H$_2$/He mass in the atmosphere layers above the HHB reaches $M_\mathrm{p}\times f_\mathrm{H_2/He}$. In this case, we alter the atmospheric composition such that $1\times$ solar metallicity is used above the HHB, and $1000 \times$ solar metallicity is prescribed below. 
We acknowledge that a sharp transition from a metallicity of $1000 \times$ the solar value to a solar metallicity in the atmosphere is physically unrealistic. Instead, one would expect vertical mixing to increase the metallicity and even result in metallicity gradients throughout the atmosphere (e.g. Refs.\citealp{otegi_impact_2020,lozovsky_threshold_2018}, Piaulet et al. in prep). Changes in the envelope metallicity are expected to impact significantly the radius of the model planet. Therefore, we perform a two-step analysis considering edge cases: we first estimate how much solar metallicity H$_2$/He-dominated gas could be accommodated by the planet properties, and then estimate the range of bulk compositions compatible with a high metallicity, volatile-rich steam atmosphere (see next paragraph and \figstructure a and b).

We adopt a fixed internal temperature instead of a fixed internal specific entropy because the internal temperature and atmospheric composition could be constrained by observations in transmission spectroscopy, and do not depend on the adopted thermal evolution model. Furthermore, we recognize that a fixed specific entropy would not result in mass-radius relations that represent a snapshot in time in terms of planet age, as lower-mass planets cool down much quicker than their more massive counterparts \citep{lopez_understanding_2014}.

\subsection{Coupled interior-atmosphere structure modeling: pure water atmospheres.}

We constrain the water content of Kepler-138~c and d using coupled interior+atmosphere three-layer models \citep{aguichine_mass-radius_2021} with rock+iron cores underlying water layers and steam atmospheres. These models use the Ref. \citep{mazevet_ab_2019} EOS for water and are appropriate for irradiated water worlds. In particular, similarly to our coupled four-layer models, they take into account the presence of a supercritical water layer, which puffs up the radii of close-in planets, even with low amounts of water. The grid covers masses from 0.2 to 20\,$M_\oplus$, irradiance temperatures $T_\mathrm{irr}$ from 400\,K to 1300\,K and water mass fractions $f_\mathrm{H_2O}$ of 10 to 100\% on top of a core+mantle composed of any relative fractions of rock and iron. The parameter $f'_\mathrm{core}$ describes the fraction of the rock+iron portion of the planet composed of iron, by mass. For example, $f'_\mathrm{core}=0$ in the absence of an iron core and $f'_\mathrm{core}=0.325$ for an Earth-like composition. The radius at a pressure of 0.1 Pa is taken as the observable transit radius following previous work, and corresponds to the top of the moist convective layer for a pure water atmosphere \citep{aguichine_mass-radius_2021,turbet_runaway_2019,turbet_revised_2020}. We augment this grid using rocky planet models with various relative amounts of rock and iron \citep{zeng_mass-radius_2016} to obtain planet radii down to $f_\mathrm{H_2O}=0$.

\subsection{Constraints on planetary composition.}

We compute the posterior probability distributions of $f_\mathrm{H_2O}$ and $f_\mathrm{H_2/He}$ within a Bayesian framework. We build a fine grid of models with various water and hydrogen mass fractions across a range of planet masses and compute for each of them the corresponding planet radius.
We evaluate the match of a specific model to the measured mass using the combined constraint from the photodynamical and RV analyses, motivated by the strict upper limit on the planet mass obtained from the RV analysis alone, and leveraging the inclination constraints from the photodynamical analysis. More specifically, we divide the posterior on $M_d \sin i_d$ by sample inclinations $\sin i_d$ drawn from the inclination posterior from the photodynamical fit to obtain a distribution of $M_d$ from the RV fit. We emphasize that this approach leads to higher derived absolute masses compared to the assumption of $\sin i_d=1$, as $i_d = 89.04 \pm 0.04$ from the photodynamical fit. The resulting constraints on the significance with which rocky scenarios are excluded from this distribution are therefore conservative. We then compute kernel density estimates (KDEs) for both the distribution of $M_d$ from the RV and the photodynamical fit, and multiply these KDEs together. Finally, we normalize the resulting distribution and use this as the observed mass distribution to obtain our final constraints on planetary composition.
The match to the planet radius is evaluated using a Gaussian likelihood. 
This Bayesian analysis provides the two-dimensional posterior distribution of $f_\mathrm{H_2/He}$ and $f_\mathrm{H_2O}$ (\figstructure a). Our constraints on the H$_2$/He mass fractions were obtained using planet models with an internal temperature of only 30~K and therefore serve as upper limits on the amount of hydrogen that can be accounted for by Kepler-138~d's mass and radius. 
If both a hydrogen envelope and a water layer are to be invoked for Kepler-138~d, such a ``Hycean'' world\citep{madhusudhan_habitability_2021} could allow for a range of states in the water layer, ranging from vapor form to the surface of a supercritical or even a liquid water ocean (\fighhb), depending on the water mass fraction, the planetary albedo, and the details of the atmosphere's composition. 

Meanwhile, in the case of a volatile-rich atmospheric composition (in the absence of hydrogen), we use the grid of pure-water atmosphere models described above to constrain the range of water fractions consistent with the observations for various internal compositions. We compute the posterior distribution of $f_\mathrm{H_2O}$, $f'_\mathrm{core}$ and $T_\mathrm{irr}$ for Kepler-138~c and d using MCMC sampling of the parameter space with \texttt{emcee} within the open-source \texttt{smint} package (see e.g. \citealp{piaulet_wasp-107bs_2021,kosiarek_physical_2020}). 
For each combination of parameters, we interpolate within the grid (linear interpolation in the dimensions of $f_\mathrm{H_2O}$, $f'_\mathrm{core}$ and $T_\mathrm{irr}$, log interpolation for planet mass) to obtain the theoretical planet radius, which is compared with the observed transiting radius via a Gaussian likelihood. We adopt a Gaussian prior on $T_\mathrm{irr}$ informed by the system properties and use uniform priors on $f_\mathrm{H_2O}$ and $f'_\mathrm{core}$ from 0 to 1. We adopt as a prior on the planet mass the combined RV+photodynamical posterior distribution. We rule out as having zero probability unphysical models flagged as such in the model grid and extend the prior on the irradiance temperature to allow for temperatures down to 285 K, i.e. still within the regime where only one planetary structure corresponds to one irradiance temperature \citep{aguichine_mass-radius_2021}. This allows us to fully encompass the prior on the irradiance temperature of Kepler-138~d ($T_\mathrm{irr, d} = 377 \pm 7$\,K). For models with $T_\mathrm{irr}<400$\,K, we compute radii assuming $T_\mathrm{irr}=400$\,K. This has no significant impact on our conclusions given the slow dependence of water mass fraction on $T_\mathrm{irr}$ and the proximity of Kepler-138~d's temperature to the grid computation range (see \figcompc and \figcompd). 
We use 100 walkers and run the chains for 10,000 steps, 60\% of which are discarded as burnin. We ensure convergence was attained by calculating the autocorrelation timescales for the chains, which are all at least 60 times shorter than the post-burnin chain length. We obtain the posterior probability distribution on the water mass fraction by marginalizing over planet mass and temperature. Finally, we extract the 1D distribution of allowed water mass fractions marginalized over the full range of interior iron fractions explored by our Bayesian analysis (\figstructure b).

\subsection{Stellar age.}

We revisit the age of the M dwarf Kepler-138 using open cluster ages and find that this model-independent approach robustly constrains its age between 1 and 2.7 Gyr.  We compare Kepler-138 to the stellar population of known open clusters\citep{curtis_when_2020} in the $T_\mathrm{eff}$-$P_\mathrm{rot}$ space (\figstellarage). We use for the equatorial rotation period $P_\mathrm{rot, eq}$ the value inferred from the detailed modeling of the stellar surface to reproduce the rotational modulations observed in the \textit{Kepler} light curve\citep{almenara_absolute_2018}, and for the  effective temperature $T_\mathrm{eff}$ a value inferred from stellar spectroscopy (Ref. \citep{muirhead_characterizing_2012}, Ext. Data Table \ref{\tabstarparams}). Kepler-138 falls above the precise 1 Gyr NGC 6819 sequence and below the 2.7 Gyr Ruprecht 147 sequence, from which we infer a model-independent age in the range between 1 and 2.7 Gyr.

% \newpage
\subsection{Atmospheric escape.}\label{sec:int_comp}

We investigate the longevity of a 0.01 wt\% H$_2$/He envelope atop Kepler-138~d using the formula for energy-limited escape as well as a detailed self-consistent 1D hydrodynamic upper atmosphere model. In both cases, we find that the atmosphere is swiftly lost to space on timescales of 10-100 Myr, indicating that a H$_2$/He atmosphere is not stable on Kepler-138~d.

For the case of hydrodynamic escape, we estimate the escape flux using the energy-limited formula \citep{watson_dynamics_1981}:
\begin{equation}
    \dot{M} = \eta \frac{\pi\,R_\mathrm{p}\,R_\mathrm{eff}^2 L_\mathrm{HE}}{4 \pi a^2 G\,M_\mathrm{p}} f(A)
\end{equation}
\noindent where $\eta$ is the mass loss efficiency that accounts for any energy losses (e.g. radiative, hydrodynamic or due to ionization), $R_\mathrm{eff}$ is the radius of the effective XUV photosphere, $R_\mathrm{p}$ and $M_\mathrm{p}$ are the radius and mass of the planet, $a$ is its semi-major axis and $f(A)$ is a factor that depends on the amplitude of flares \citep{owen_evaporation_2017,feinstein_flare_2020}. For the high-energy luminosity $L_\mathrm{HE}$, we adopt a prescription with a constant value for the first 100 Myr, followed by a decay with $t^{-1.5}$ (Refs. \citep{owen_evaporation_2017,ribas_evolution_2005,jackson_coronal_2012,tu_extreme_2015,gudel_x-ray_1997}). Although the mass loss efficiency is not a constant \citep{murray-clay_atmospheric_2009, owen_planetary_2012,owen_uv_2016}, we adopt an approximate value of $\eta=10$\% for the present calculation, appropriate for super-Earths and sub-Neptunes \citep{owen_planetary_2012}. 
At the present age of Kepler-138, we find short envelope loss timescales of 70 to 300 Myr depending on where exactly the XUV photosphere lies (considering a range of $R_\mathrm{eff}$ from 1 to 2$R_\mathrm{p}$), and whether or not we fold in the 4--7\% mass-loss rate increase from stellar flares \citep{feinstein_flare_2020}.

For a more detailed analysis, we additionally simulate the escape of a hydrogen-dominated atmosphere using a 1D hydrodynamic upper atmosphere model \citep{kubyshkina_grid_2018,erkaev_euv-driven_2016}.  The upper atmosphere model accurately accounts for transitions from hydrodynamic boil-off to blow-off and hydrostatic Jeans escape regimes, and includes hydrogen dissociation, recombination and ionisation as well as stellar X-ray and EUV heating, and H$_3^+$ and Ly $\alpha$ cooling. The gravitational potential includes Roche lobe effects \citep{erkaev_roche_2007}. The stellar EUV radiation is assumed to be emitted at 60 nm \citep{murray-clay_atmospheric_2009}, while X-ray is modeled as emission from a single wavelength at 5 nm. The EUV and X-ray stellar luminosities are computed using evolutionary tracks calibrated with X-ray and UV measurements for stars with similar masses as Kepler-138 \citep{johnstone_active_2021}. We obtain upper self-consistent atmosphere profiles up to the Roche lobe at $\sim 32\,R_p$, which lies below the exobase in this model (\figescape). We find a mass-loss rate of $2.4 \times 10^9$ g/s. For a H$_2$/He mass fraction of 0.01\%, the envelope could thus not be sustained in its blow-off state for more than about 20 Myr.

\subsection{Outgassed secondary atmosphere.}
Besides a primary H$_2$/He-dominated atmosphere accreted from the protoplanetary nebula, which would be quickly lost to space for a planet such as Kepler-138 c and d, rocky planets can replenish their atmospheres from the inside-out. This secondary origin would also result in volatile-rich atmosphere compositions for Kepler-138~c and d.

Secondary atmospheres can be outgassed from solid material or after lid formation\citep{bower_retention_2021}, but form most efficiently during the early phases of the planet's lifetime when the interior is warm enough to maintain a molten, or at least partially molten mantle\citep{elkins-tanton_ranges_2008,schaefer_chemistry_2010,lichtenberg_vertically_2021}.
The crystallization of the molten magma can be delayed or stalled by the greenhouse effect from the atmosphere \citep{kite_exoplanet_2020}, strong stellar irradiation \citep{dorn_hidden_2021}, or tidal heating. The longer the timescale over which the magma ocean is maintained, the larger the potential for replenishment of even an escaping hydrogen atmosphere from interior outgassing.  At present, both the outgassing rates and timescales of hydrogen and carbon-bearing molecules and the depth and composition of resulting secondary atmospheres remain largely unknown. They depend not only on the volatiles' solubility in the magma, but on other factors that remain observationally unconstrained such as the redox state of the interior\citep{sossi_atmospheres_2021}, the initial volatile budget, or the timescale for surface lid formation\citep{bower_retention_2021} which is linked to the efficiency of the melt-solid separation\citep{andrault_deep_2016}.

If the low densities of Kepler-138~c and d were due to a stable secondary atmosphere, outgassing rates would need to be large enough to balance out the atmospheric mass loss and to sustain thick gas envelopes. 
This would point to a mantle composition drastically different from other rocky super-Earth size planets which are not found to harbor such thick outgassed envelopes, and point to large initial volatile budgets\citep{kite_exoplanet_2020}. The resulting atmospheres would still be rich in volatiles such as H$_2$O, CH$_4$, CO and CO$_2$. A large atmosphere buildup of molecular H$_2$ can be expected especially if the melt-solid separation is fast, for reducing mantle compositions or low C/H ratios\citep{bower_retention_2021}. Other volatiles also accumulate in parallel (H$_2$O for low C/H ratios), resulting in volatile-dominated compositions. Even after the formation of a surface lid, large amounts of ``trapped'' water dissolved in the mantle could be outgassed over geological timescales.

\subsection{Impact of stellar contamination on the radius of Kepler-138~d.}

Unocculted stellar spots\citep{rackham_transit_2018} can result in an overestimate of the radius of a transiting planet due to the fact that the transit light source is not accurately represented by the out-of-transit spectrum. For Kepler-138~d, the possible levels of stellar contamination are small compared to the uncertainty on the \textit{Kepler} radius measurement.

We model the transmission spectrum of Kepler-138~d assuming that any transit depth variations are due to unocculted stellar spots\citep{rackham_transit_2018}. The transmission spectrum is therefore computed as:
\begin{equation}
    D_\mathrm{\lambda, obs}= \frac{D}{1 - f_
    \mathrm{spot}\left(1 - \frac{F_{\lambda, \mathrm{spot}}}{F_{\lambda, \mathrm{phot}}}\right)}
\end{equation}
\noindent where $f_\mathrm{spot}$ is the spot covering fraction, $F_{\lambda, \mathrm{phot}}$ and $F_{\lambda, \mathrm{spot}}$ are respectively the spectrum of the star at the effective temperature and the spot temperature and $D$ is a scaling factor, here fitted to obtain the best match to the observed transmission spectrum. The stellar spectra are taken from the PHOENIX\citep{husser_new_2013} spectral library and correspond to the properties ($\log~g$, [Fe/H], $T_\mathrm{eff}$) of Kepler-138 (\tabstarparams).

We do not include faculae in these models, as the impact of unocculted hotter photospheric regions on the inferred radius would be an underestimate, rather than an overestimate. Previous modeling of the spots of Kepler-138 based on the \textit{Kepler} light curves constrained a spot-to-photosphere temperature difference of about 240~K, and a spot covering fraction in the range 0.1\% to 3\%\citep{almenara_absolute_2018}. We therefore calculate three models with spots 240~K cooler than the effective temperature of Kepler-138 (\tabstarparams), and $f_\mathrm{spot}=0.1$, 3 and 10\%. We compute the 10\% case in order to account for the fact that spots that do not cause rotational modulation of the light curve (e.g. polar spots) are not accounted for in the estimated range of 0.1\% to 3\% spot covering fractions. We find that the effect of stellar contamination on the radius estimate is negligible: even for 10\% spot covering fraction, the bandpass-integrated stellar contamination signal is smaller than the 1$\sigma$ uncertainty on the \textit{Kepler} transit depth (\figspots). Additionally, our \textit{HST} and \textit{Spitzer} infrared transit depths measurements do not indicate the presence of any strong upwards slope towards the optical, while this is one of the telltale signs of contamination by unocculted stellar spots\citep{rackham_transit_2018}.

\subsection{Potential impact of photochemical hazes on the pressure level probed by Kepler.}

Photochemical hazes have the potential to significantly increase the apparent radius of a transiting exoplanet \citep{gao_deflating_2020} in the \textit{Kepler} wavelength range, with $\sim$nbar, instead of mbar, pressures being probed in transmission. The impact of such a bias on planet radius is an overestimate of the mass fractions of H$_2$/He or water compared to a planet's true volatile content.

For Kepler-138~d, however, this would result in an even smaller hypothetical H$_2$/He atmosphere mass fraction than what we infer above, which would be even more susceptible to escape and therefore physically implausible. In the ``water world'' case, not only is the impact of hazes on the near-infrared spectrum less pronounced due to shorter mixing timescales in high-metallicity atmospheres \citep{lavvas_photochemical_2019}, but the planet parameters would remain inconsistent with a bare rock scenario, given that the presence of hazes presupposes that of a gas layer as a source of haze precursors. Furthermore, small amounts of water would be unstable against early loss in the early active stages of the star's evolution \citep{luger_extreme_2015}, which suggests that if the planet retained any water to this date, it must have formed water-rich.

\subsection{Planetary rings.}

Circumplanetary rings are another explanation for anomalously large inferred planet radii \citep{piro_can_2018,piro_exploring_2020}. We find that for Kepler-138~d, rings cannot explain its low density.

The tidal synchronization timescale is very short for Kepler-138~d, and it is therefore expected to be tidally-locked. In particular, for $Q_p=10^{6.5}$ typical of a gas-enveloped planet (or $Q_p=10$ to $100$ characteristic or rocky planets\citealp{clausen_dissipation_2015}), Kepler-138~d becomes tidally-locked within 440 Myr (or mere \textit{thousands} of years). This tidally-locked state corresponds to a quadrupole gravitational harmonic $J_2$ of $J_{2, \mathrm{tidal~locking}}=3.9_{-0.9}^{+1.6} \times 10^{-7}$ (Refs.  \citealp{chandrasekhar_ellipsoidal_1969,piro_exploring_2020}). Therefore, Kepler-138~d does not fulfill the criterion $J_2>J_{2, \mathrm{min}}=1.8_{-0.4}^{+0.7} \times 10^{-5}$ \citep{tremaine_satellite_2009,schlichting_warm_2011,piro_exploring_2020} required to prevent tidal warping of the rings from the parent star, and could not maintain rings.

\end{methods}

%################################################################################################################

\clearpage
\begin{addendum}

 \item[Data Availability] The data used in this paper are deposited on publicly-available servers. The data from the Hubble and Spitzer space telescope used in this work can be downloaded from the Mikulski Archive for Space Telescopes (MAST). The Keck/HIRES radial velocities are available online as a Supplementary Dataset. The planet population plots used data from the public NASA Exoplanet Archive, which also hosts an interface where the Kepler photometry can be downloaded.
 \item[Code Availability] The \texttt{smint} code is publicly-available on GitHub at \url{https://github.com/cpiaulet/smint}. The radial velocity analysis is based on the publicly-available package \texttt{george} as well as \texttt{RadVel} and \texttt{emcee}. Further scripts can be provided by the corresponding author upon reasonable request. 
 
 \item[Acknowledgements]
We thank the three reviewers for valuable comments that improved this manuscript. We gratefully acknowledge the open-source software which
made this work possible: LDTK \citep{parviainen_ldtk_2015}, batman \citep{batman}, emcee \citep{foreman-mackey_emcee_2013}, TTVFast \citep{deck_ttvfast_2014}, \texttt{REBOUND} \citep{rein_rebound_2012}, \texttt{WHFast} \citep{rein_whfast_2015}, \texttt{nestle} (\url{https://github.com/kbarbary/nestle}, \citealp{feroz_multinest_2009,shaw_efficient_2007,mukherjee_nested_2006,skilling_nested_2004}), astropy \citep{astropy:2013, astropy:2018}, numpy \citep{harris2020array}, ipython \citep{ipython}, matplotlib \citep{matplotlib}, \texttt{RadVel} \citep{fulton_radvel:_2018}, \texttt{george} \citep{ambikasaran_fast_2015}, \texttt{smint}\citep{piaulet_wasp-107bs_2021}, GNU parallel \citep{tange_2020_3956817}. 
This work is based on observations with the NASA/ ESA HST, obtained at the Space Telescope Science Institute (STScI) operated by AURA, Inc. We received support for the analysis by NASA through grants under the HST-GO-13665 program (PI Benneke). 
This work relies on observations made with the Spitzer Space Telescope, which was operated by the Jet Propulsion Laboratory, California Institute of Technology, under a contract with NASA.
This paper includes data collected by the Kepler mission. Funding for the Kepler mission is provided by the NASA Science Mission directorate.
This study has made use of data from the European Space
Agency (ESA) mission Gaia (\url{https://www.cosmos.esa.int/
gaia}), processed by the Gaia Data Processing and Analysis
Consortium (DPAC, \url{https://www.cosmos.esa.int/web/gaia/
dpac/consortium}). Funding for the DPAC has been provided
by national institutions, in particular, the institutions participating in the Gaia Multilateral Agreement. 
Data presented in this paper were obtained from the Mikulski Archive for Space Telescopes (MAST).
This research has made use of NASA’s Astrophysics Data System and the NASA Exoplanet Archive, which is operated by the California Institute of Technology, under contract with NASA within the Exoplanet Exploration Program. Parts of this analysis have been run on the {\it Lesta} cluster kindly provided by the Observatoire de Gen\`{e}ve.
C.P. acknowledges financial support by the Fonds de Recherche Qu\'{e}b\'{e}cois—Nature et Technologie (FRQNT; Qu\'{e}bec), the Technologies for Exo-Planetary Science (TEPS) Trainee Program and the Natural Sciences and Engineering Research
Council (NSERC) Vanier Scholarship. 
D. D. acknowledges support from the TESS Guest Investigator Program grant 80NSSC19K1727 and NASA Exoplanet Research Program grant 18-2XRP18\_2-0136.
B.B. acknowledges financial
support by the NSERC of Canada, and the FRQNT. I.W. is supported by an appointment to the NASA Postdoctoral Program at the NASA Goddard Space Flight Center, administered by Oak Ridge Associated Universities under contract with NASA. C.V.M. acknowledges HST funding through grant HST-AR-15805.001-A from the Space Telescope Science Institute. 
 
\item[Author Contributions] C.P. and B.B. conceived the project. C.P. wrote the manuscript and carried out the reduction of the \textit{HST} and \textit{Spitzer} data as well as the TTV, radial-velocity, atmospheric escape, atmospheric retrieval and planetary structure analyses, under the supervision of B.B and with the help of M.P. for the TTV analysis and the contribution of D.K. for the upper atmosphere modeling. J.M.A. realized the photodynamical analysis and the transit search for Kepler-138~e. D.D. provided the \textit{Spitzer} observations. H.A.K., A.W.H., H.I., L.M.W. and C.B. conducted the observations and reduction of the HIRES RVs. D.T. provided the grid of interior models. R.A. constrained the stellar age. All co-authors provided comments and suggestions about the manuscript.

 \item[Competing Interests] The authors declare no competing interests.
 \item[Correspondence] Correspondence and requests for any materials presented in this work should be addressed to Caroline Piaulet.~(email: caroline.piaulet@umontreal.ca).

 \clearpage

\begin{table*}
\begin{tabular}{cccc}
\hline
\hline
Start Date               &  Epoch            &  Instrument  & Transit time\\
UT &  & & BJD$_\mathrm{TDB}$- 2450000\\
\hline
2014-12-21  &  89  &  \textit{HST}/WFC3 G141 & $7012.74875_{-0.00054}^{+0.00054}$\\
2015-04-15  &  94  &  \textit{HST}/WFC3 G141 & $7128.19741_{-0.00043}^{+0.00045}$\\
2015-08-08  &  99  &  \textit{Spitzer}/IRAC Ch1 & $7243.6457_{-0.0056}^{+0.0042}$
\\ %53911296ch1
2015-09-01  &  100  &  \textit{Spitzer}/IRAC Ch1 & $7266.7348_{-0.0025}^{+0.0026}$\\ %53911040ch1
2015-09-24  &  101  &  \textit{Spitzer}/IRAC Ch2 & $7289.8260_{-0.0026}^{+0.0026}$\\ %53902080ch2
2015-10-17  &  102  &  \textit{HST}/WFC3 G141 & $7312.9152_{-0.0175}^{+0.0018}$\\
2015-10-17  &  102  &  \textit{Spitzer}/IRAC Ch2 & $7312.9147_{-0.0046}^{+0.0033}$\\ %53901824ch2
2015-11-09  &  103  &  \textit{Spitzer}/IRAC Ch1 & $7336.0079_{-0.0021}^{+0.0023}$\\ %53910784ch1
2015-12-25  &  105  &  \textit{Spitzer}/IRAC Ch1 & $7382.1869_{-0.0052}^{+0.0051}$\\ %53910528ch1
2016-01-17  &  106  &  \textit{Spitzer}/IRAC Ch2 & $7405.2780_{-0.0026}^{+0.0022}$\\ %53901056ch2
2016-02-09  &  107  &  \textit{Spitzer}/IRAC Ch2 & $7428.3685_{-0.0029}^{+0.0047}$\\ %53901568ch2

2016-08-12  &  115  &  \textit{Spitzer}/IRAC Ch1 & $7613.0978_{-0.0070}^{+0.0092}$\\ %53910272ch1
2016-09-04  &  116  &  \textit{Spitzer}/IRAC Ch2 & $7636.1772_{-0.0093}^{+0.0091}$\\ %53900800ch2
\hline
\hline
\end{tabular}
\caption{\textbf{Measured transit times of Kepler-138~d.} Quoted errors encompass the 68\% (1-$\sigma$) confidence region. The transit epoch is relative: epoch 0 corresponds to the first transit of Kepler-138~d observed by \textit{Kepler} on 2009-05-06. On 2015-10-17, both \textit{HST} and \textit{Spitzer} observed the same transit of Kepler-138~d and we retrieve consistent transit times for this visit.\label{tab:tr_times}}
\end{table*}

% % \begin{sidewaystable}
% \begin{table}
% % \begin{tabular}{p{3.5cm}p{8cm}p{5cm}p{3.5cm}p{3.5cm}}
% \begin{tabular}{lcccc}
% \hline
% \hline
% Parameter &  Planet b  & Planet c & Planet d & Planet e\\
% \hline
% \textbf{\textit{Fitted parameters}}  & & & & \\
% Scaled semimajor axis, $a/R_\star$  & $30.3 \pm 0.8$ & $37 \pm 1$ & $51.8 \pm 1.4$ & $73 \pm 2$ \\
\begin{table}
\resizebox{\textwidth}{!}{
\begin{tabular}{lcccc}
\hline
\hline
Parameter &  Planet b  & Planet c & Planet d & Planet e\\
\hline
\textbf{\textit{Fitted parameters}}  & & & & \\
$a/R_\star$  & $30.3 \pm 0.8$ & $37 \pm 1$ & $51.8 \pm 1.4$ & $73 \pm 2$ \\
$i$ ($\degree$) & $88.67 \pm 0.08$ & $89.02 \pm 0.07$ & $89.04 \pm 0.04$ & $88.53 \pm 1.0$ \\
$b$ & $0.7 \pm 0.03$ & $0.6 \pm 0.03$ & $0.87_{-0.009}^{+0.008}$ & $1.8_{-1.2}^{+1.9}$ \\
$\Omega$ ($\degree$) & $181.3 ^{+ 1.1}_{ - 2.6}$ & $\equiv 180$ & $180.5 \pm 0.5$ & $178.5 \pm 1.2$\\
$M_0$ ($\degree$) & $40 \pm 30$ & $40 \pm 20$ & $160^{+20}_{-30}$ & $300^{+90}_{-60}$ \\
$\sqrt{e} \cos \omega$   & $0.10 \pm 0.04$ &  $0.11 \pm 0.03$ & $-0.04^{+0.05}_{-0.03}$ & $0.08^{+0.07}_{-0.06}$ \\
$\sqrt{e} \sin \omega$   & $0.09^{+0.03}_{-0.04}$& $0.06 \pm 0.04$ & $-0.09^{+0.04}_{-0.03}$ & $0.32^{+0.04}_{-0.05}$ \\
$\left(R_\mathrm{p}/R_\star\right)_\mathrm{Kepler}$  & $0.0109 \pm 0.0003$ & $0.0258 \pm 0.0003$ & $0.0259 \pm 0.0003$ & --\\
$\left(R_\mathrm{p}/R_\star \right)_\mathrm{HST}$  & -- & -- & $0.0255^{+0.0013}_{-0.0009}$ & --\\
$\left(R_\mathrm{p}/R_\star\right)_{\mathrm{Spitzer} 3.6 \mu m}$  & -- & -- & $0.0226 \pm 0.0012$ & --\\
$\left(R_\mathrm{p}/R_\star\right)_{\mathrm{Spitzer} 4.5 \mu m}$  & -- & -- & $0.0238 \pm 0.0013$ & --\\
$M_\mathrm{p}/M_\star$  $\left(\times 10^{-5}\right)$& $0.044 \pm 0.013$ & $1.3 \pm 0.3$ &  $1.3 ^{+0.3}_{-0.4} $ & $0.23^{+0.10}_{-0.05} $\\
\textbf{\textit{Derived parameters}}  & & & & \\
$P'$  (d) & $10.3134 \pm 0.0003$ & $13.78150^{+0.00007}_{-0.00009}$ & $23.0923 \pm 0.0006$ & $38.230 \pm 0.006$\\
$T_0'$  (BJD-2454000) & $956.236 \pm 0.003$ & $955.7288 \pm 0.0006$ & $957.8160 \pm 0.0009$ & $924.1^{+0.3}_{-0.2}$\\
$a$  (au) & $0.0753 \pm 0.0006$ & $0.0913 \pm  0.0007$ & $0.1288 \pm 0.0010$ & $0.1803 \pm 0.0014$\\
$e$   & $0.020 \pm 0.009$ & $0.017^{+0.008}_{-0.007}$ & $0.010 \pm 0.005$ & $0.112^{+0.018}_{-0.024}$\\
$\omega$ ($\degree$) & $40 \pm 20$ & $34 \pm 19$ & $250^{+30}_{-20}$ & $76^{+10}_{-14}$\\
$R_\mathrm{p}$ $\left(R_\oplus\right)$ & $0.64 \pm 0.02$ & $1.51 \pm 0.04$ & $1.51 \pm 0.04$ & --\\
$M_\mathrm{p,phot.}$ $\left(M_\oplus\right)$ & $0.08 \pm 0.02$ & $2.2^{+0.6}_{-0.5}$ & $2.3^{+0.6}_{-0.7}$ & $0.42^{+0.18}_{-0.10}$\\
$M_\mathrm{p, phot.+RV}$ $\left(M_\oplus\right)$ & $0.07 \pm 0.02$ & $2.3^{+0.6}_{-0.5}$ & $2.1^{+0.6}_{-0.7}$ & $0.43^{+0.21}_{-0.10}$\\
$\rho_\mathrm{p}$ (g~cm$^{-3}$) & $1.7 \pm 0.5$ & $3.6^{+1.1}_{-0.9}$ &$3.6 \pm 1.1$ & -- \\
$\log g_\mathrm{p}$ (cgs) & $2.28^{+0.11}_{-0.14}$ & $2.98^{+0.11}_{-0.12}$ & $2.99^{+0.11}_{-0.15}$ & -- \\
$S_\mathrm{inc}$ $\left(\mathrm{kW/m}^2\right)$ & $13.5 \pm 1.0$ & $9.2 \pm 0.7$ & $4.6 \pm 0.3$ & $2.36 \pm 0.17$ \\
$S_\mathrm{inc}$ $\left(S_\oplus\right)$ & $9.9 \pm 0.7$ & $6.8 \pm 0.5$ & $3.4 \pm 0.2$ & $1.73 \pm 0.12$ \\
$T_\mathrm{eq, A_B=0.3}$ (K) & $452^{+8}_{-9}$ & $410 \pm 8$ & $345 \pm 7$ & $292^{+5}_{-6}$ \\
\hline
\hline
\end{tabular}}
\caption{\textbf{Kepler-138 planetary parameters from the 4-planet photodynamical fit.} The quoted errors encompass the 68\% confidence region. The astrocentric orbital elements are given for the reference time $t_\mathrm{ref}~[\mathrm{BJD}_\mathrm{TDB}]=2454955$. For the planet masses, we include both the constraints from the photodynamical fit alone (``phot'') and the combined constraints from the photodynamical fit and the RV analysis (``phot+RV''). The constraints on the inclination are reported after folding the posteriors around 90$^\circ$ in order to account for degenerate inclination configurations.\label{tab:photodyn_results}}
\end{table}

\end{addendum}

\begin{figure}[t!]
\begin{center}
\includegraphics[width=0.99\linewidth]{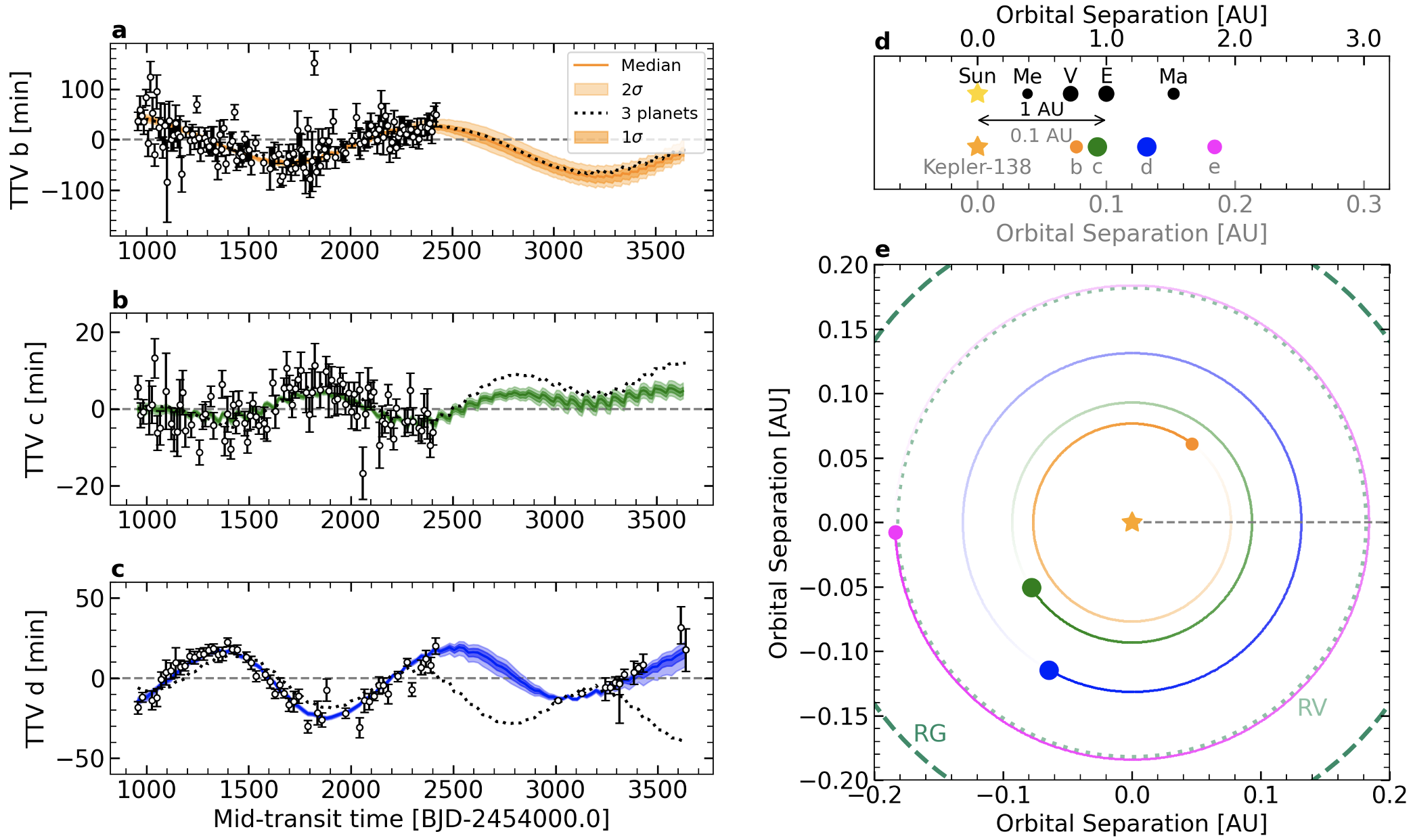}
\end{center}
\caption{\textbf{Results from the 4-planet photodynamical analysis of the \textit{HST}, \textit{Spitzer}, and \textit{Kepler} light curves of Kepler-138.} \textbf{a,b,c}, TTVs are shown as the residuals from a linear ephemeris fit for Kepler-138~b, c, and d, with error bars encompassing the 68\% confidence region. The median model is shown (solid, colors), along with the 1 and 2$\sigma$ contours (color shading) obtained from the MCMC samples, and the best fit 3-planet TTV model is overlaid (dotted, black). The \textit{HST} and \textit{Spitzer} transit times of Kepler-138~d cannot be reproduced with the 3-planet model but can be matched in the presence of a fourth planet with a mass of $\sim 0.4\,M_\oplus$ on a 38-day orbit. \textbf{d}, Comparison to the inner solar system. Planet relative sizes and relative distances are to scale, with a 10:1 ratio for solar system distances compared to Kepler-138. For Kepler-138~e, a size corresponding to an Earth-like composition was used. With a small inner planet, two ``twin'' larger planets, and a lighter outer planet, the sizes of the Kepler-138 planets resemble a scaled version of the inner solar system around a colder star. \textbf{e}, Top-down view of the Kepler-138 system at BJD$_\mathrm{TDB}=2455057.83$. The direction of the observer, corresponding to the phase where the planets transit in our line of sight, is shown as the grey dashed line. The star is at the center, and circles show planets b (orange), c (green), d (blue), and e (magenta), with their relative sizes to scale. The inner contours of the conservative (runaway greenhouse, ``RG'') and optimistic (recent Venus, ``RV'') habitable zones are highlighted in green \citep{kopparapu_habitable_2013}.}
\label{fig:photodyn_results}
\end{figure}

\begin{figure}
\centering
   \includegraphics[width=0.6\linewidth]{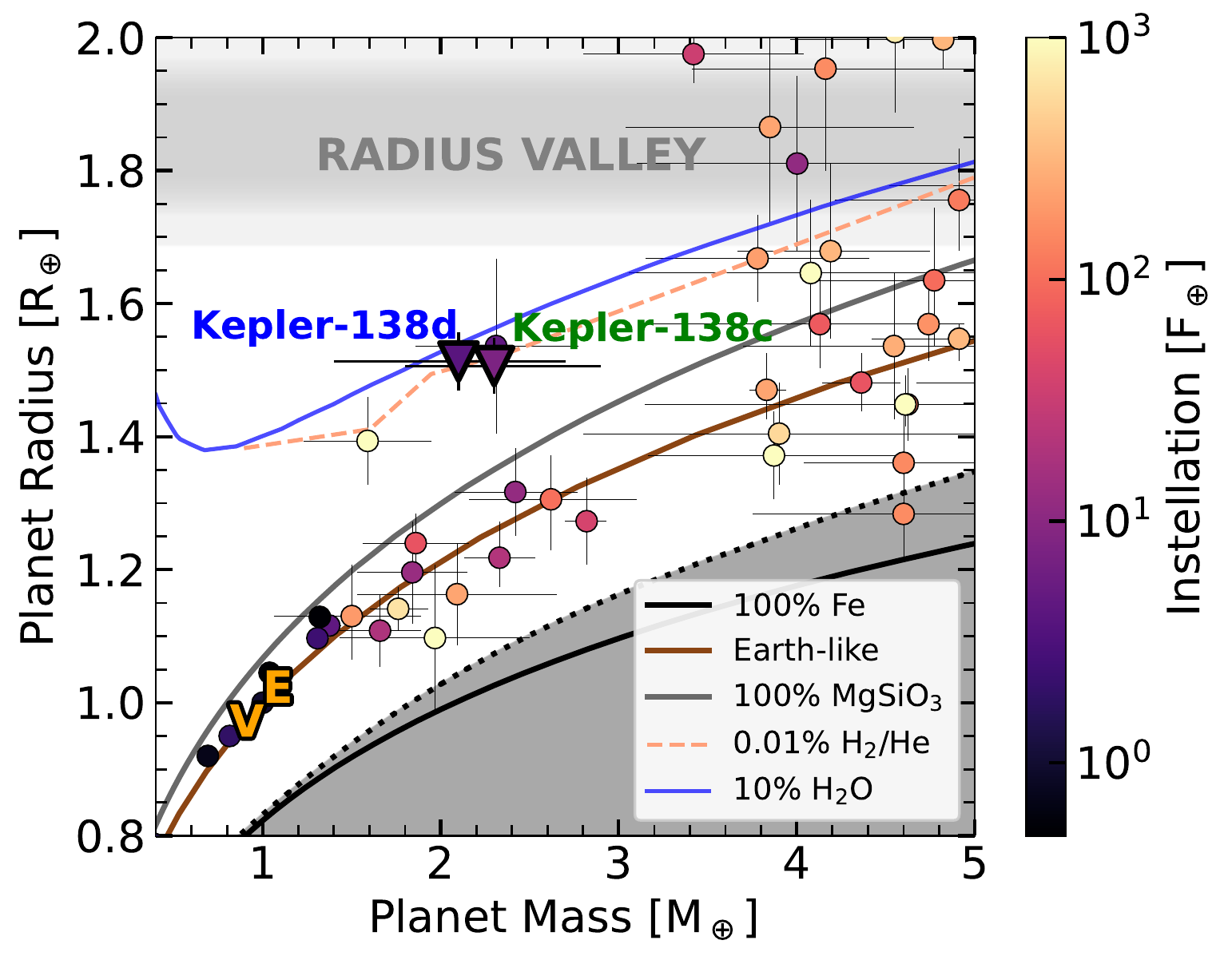}

\caption{\textbf{Comparison of Kepler-138~c and d to the population of super-Earth size planets.}
Mass-radius plot of planets with super-Earth sizes and masses below 5 $M_\oplus$. Kepler-138~c and d (bold triangles) stand out as having low densities compared to the population of small transiting exoplanets (circles colored according to instellation, planets with masses measured at better than 3$\sigma$) and solar system planets (orange letters). Error bars correspond to the 68\% confidence region for the mass and radius of each planet. Modeled mass-radius curves are displayed for rocky planets and gas-enveloped planets with an Earth-like composition core (models described in the Methods and Refs. \citep{zeng_mass-radius_2016,zeng_detailed_2013, aguichine_mass-radius_2021}). The transparent grey region corresponds to the ``radius valley'' while the solid grey region in the bottom right corner is forbidden according to models of maximum mantle stripping via giant impacts \citep{marcus_minimum_2010}. The best match to the mass and radius of Kepler-138~c and d is obtained for a volatile-rich composition with approximately $10$\% water. Alternatively, Kepler-138~d's low density could be explained by a light 0.01 wt\% (or about $0.0003 M_\oplus$) H$_2$/He atmosphere, but such an envelope would be rapidly lost to space (see text).}
\label{fig:mass_radius}
\end{figure}

\begin{figure}[t!]
\begin{center}
\includegraphics[width=0.6\linewidth]{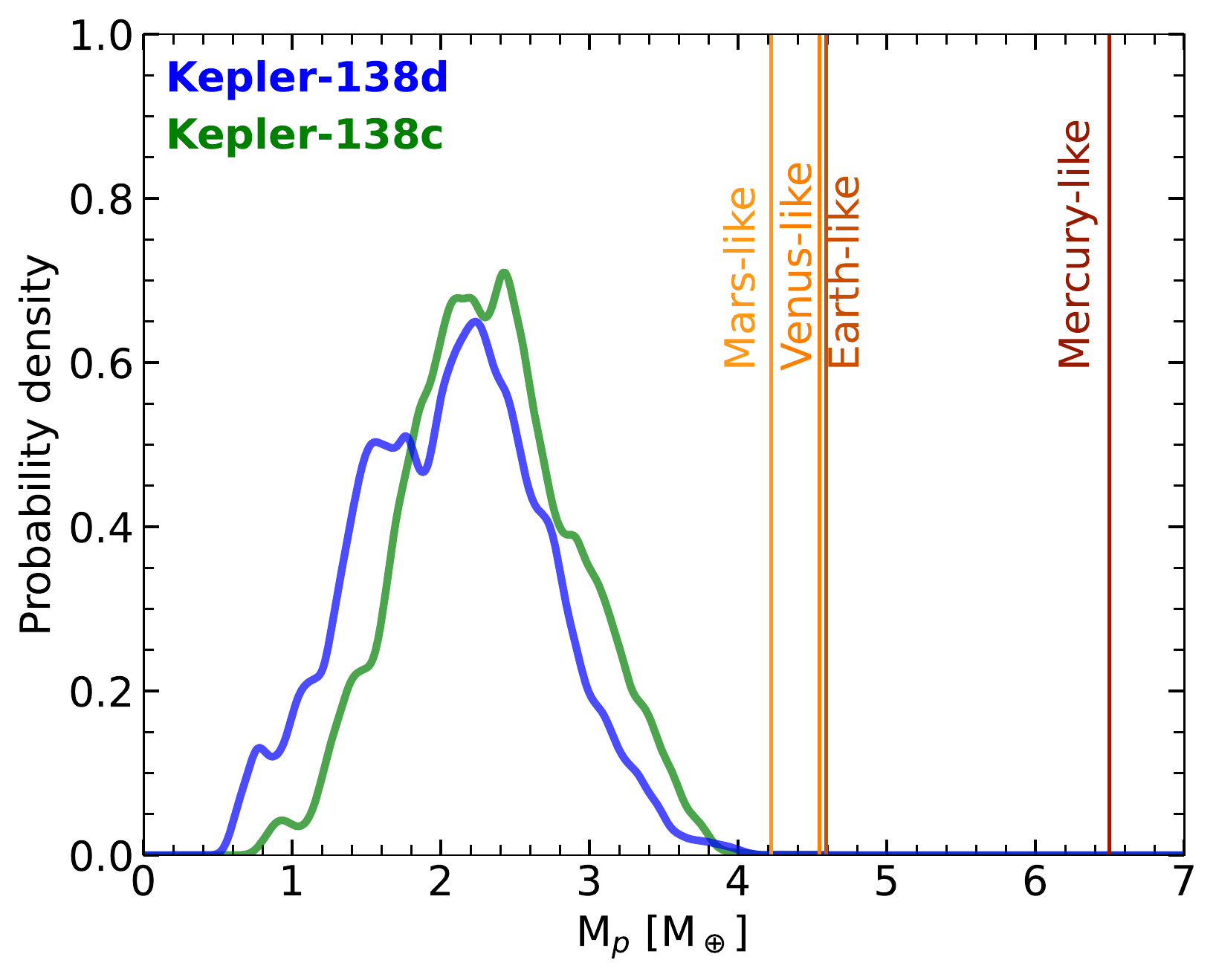}
\end{center}
\caption{\textbf{Low density of Kepler-138~c and d compared to rocky compositions.} Combined posterior distributions on the masses of Kepler-138~c (green) and d (blue) from the RV and photodynamical fits compared to the expected masses (vertical lines) for planets with the same size, but bulk compositions similar to the rocky planets in the inner solar system: Mars-like (24\% iron), Venus-like (32\% iron), Earth-like (33\% iron) and Mercury-like (65\% iron).}
\label{fig:mass_distri_comparison}
\end{figure}

\begin{figure}
\centering

\includegraphics[width=1\linewidth]{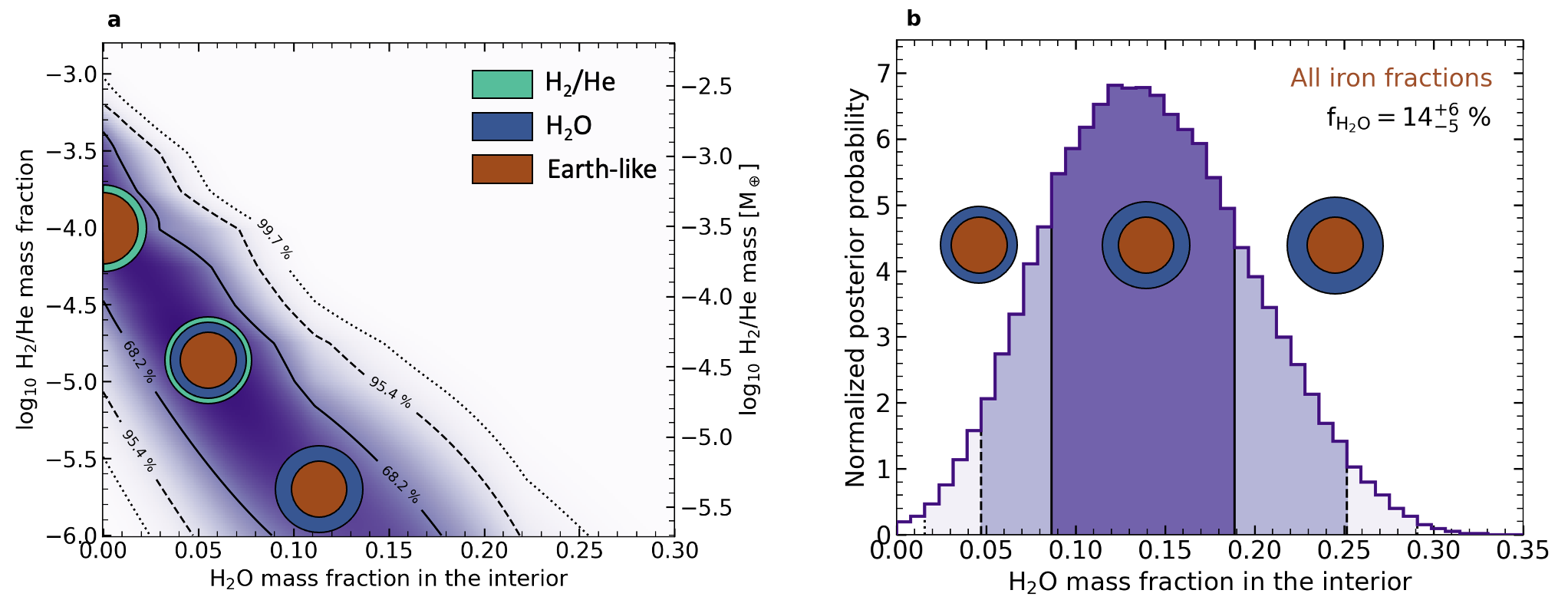}

\caption{\textbf{Planet structure modeling results for Kepler-138~d.}
\textbf{a,} Posterior probability density (purple shading) as a function of the H$_2$O mass fraction in Kepler-138~d's interior, and the mass fraction of a hypothetical H$_2$/He layer atop Kepler-138~d for an interior composed of a mixture of iron and silicates in Earth-like ratios. The contours of 1, 2, and 3$\sigma$ confidence are outlined. \textbf{b,} 1-D posterior probability density of water mass fractions for the hydrogen-free composition scenario, with a rocky core underlying a water layer and a steam atmosphere. The distribution is marginalized over the full range of iron/silicate ratios in the interior. The different purple shadings correspond to the 1, 2, and 3$\sigma$ confidence regions. Concentric circles schematically illustrate the planetary composition, where brown represents an Earth-like interior, blue represents the water layer, and green indicates hydrogen. Best fits to the observed mass and radius are obtained with either a water mass fraction of $14^{+6}_{-5}$\% -- or $11^{+3}_{-4}$\% for an Earth-like composition core or by adding a hydrogen layer of no more than 0.1 wt\% at 3$\sigma$ (or about $0.003\,M_\oplus$) atop Kepler-138~d. The latter would be rapidly lost to space (see text).}
\label{fig:interior_results}
\end{figure}

%---\figurenames-----------------------------------------------------------
\clearpage

\begin{supplementary}
\setcounter{figure}{0}
\renewcommand{\figurename}{Extended Data Figure}

\clearpage

\begin{figure*}[t!]
\begin{center}
\includegraphics[width=0.7\linewidth]{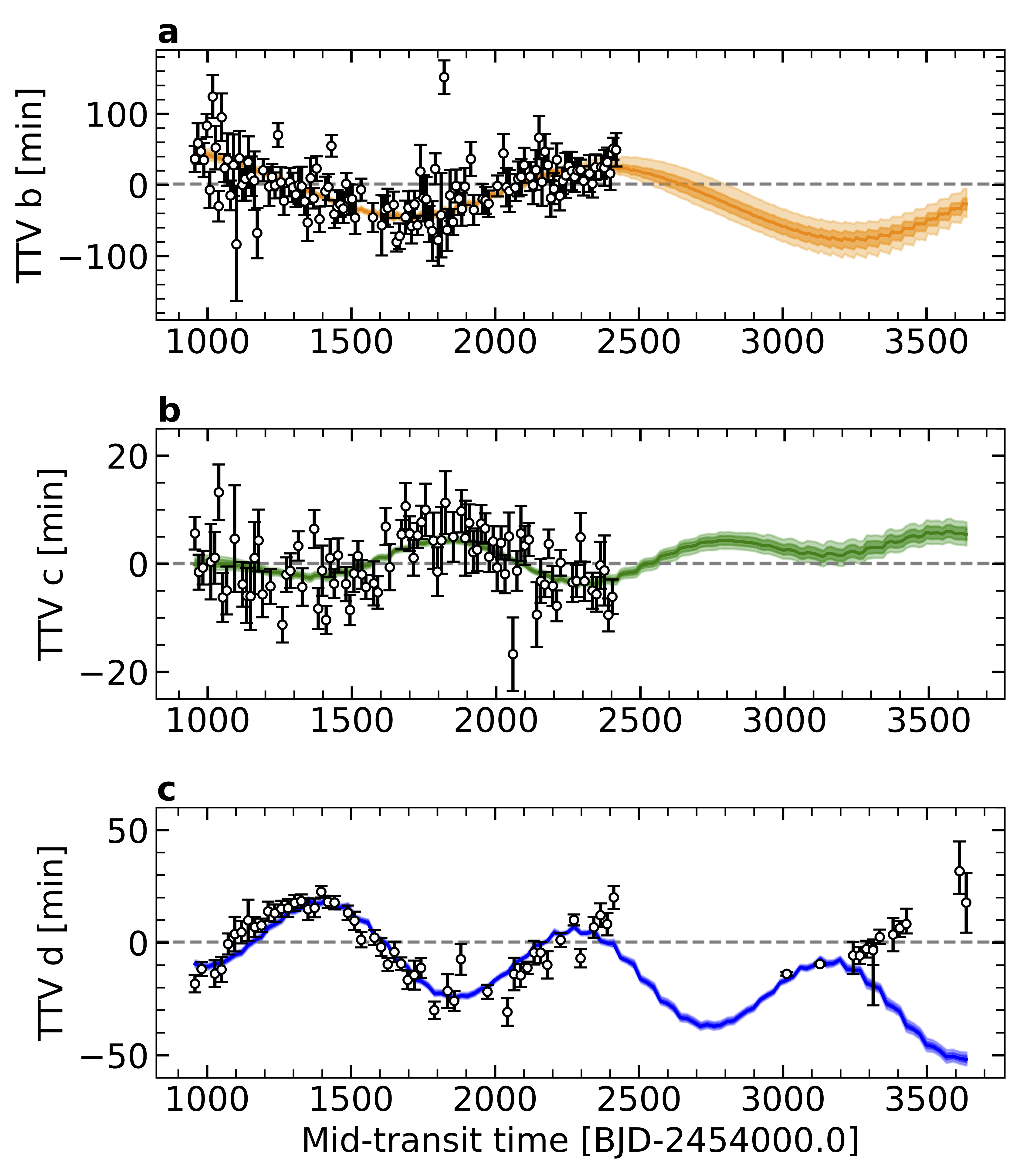}
\end{center}
\caption{\textbf{Three-planet photodynamical fit results.} \textbf{a,b,c} Same as Figure \ref{fig:photodyn_results}\textbf{a,b,c}, for a photodynamical fit including only the three previously-known planets Kepler-138 b,c, and d. This illustrates the extent to which the timescale over which the predicted transit times of Kepler-138~d are modulated (i.e. the super-period) is underestimated by the three-planet solution. This discrepancy was already hinted at by the mismatch with the \textit{Kepler} transit times but revealed at high significance by the now longer baseline over which we obtained transits with \textit{HST} and \textit{Spitzer}, at times beyond BJD=2457000.}
\label{fig:photodyn_3pla}
\end{figure*}

\begin{figure}[t!]
\begin{center}
\includegraphics[width=\textwidth]{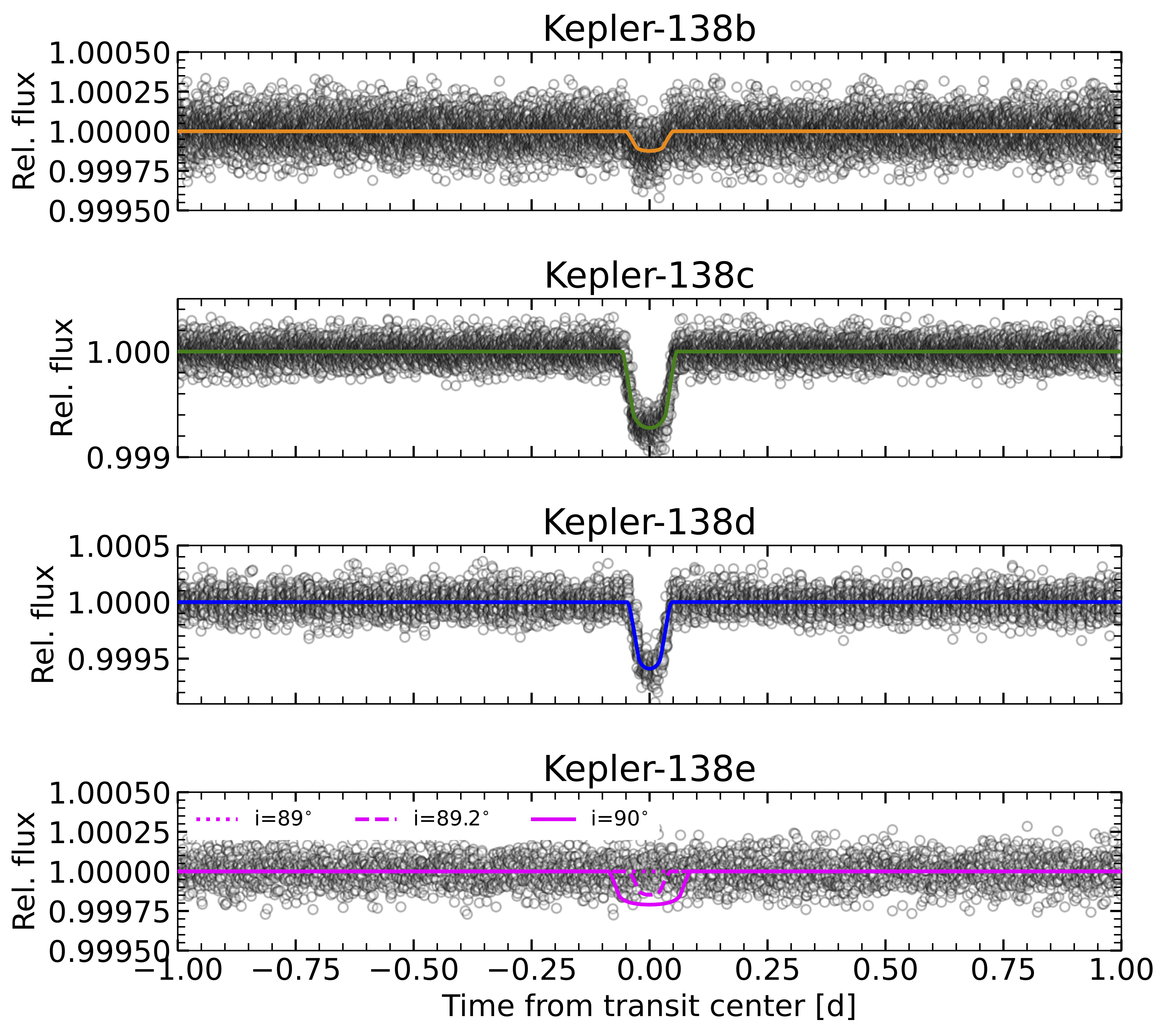}
\end{center}
\caption{\textbf{Folded \textit{Kepler} transits of Kepler-138~b, c, and d, and search for the transit of Kepler-138e.} The four panels show the corrected light curve of Kepler-138 (open circles) folded in a 2 day window around the expected transit epochs of Kepler-138~b, c, d, and e from the photodynamical fit (see Methods). Transit models corresponding to the median retrieved planet parameters are superimposed to the data (solid colored lines), conservatively assuming an Earth-like composition to estimate the radius of Kepler-138e. The transits of Kepler-138~b, c, and d are detected in the \textit{Kepler} light curve, but while Kepler-138e should be larger than Kepler-138~b, its transit is not detected. We interpret this as originating from a likely non-transiting configuration of Kepler-138e's orbit, with an inclination of $\lesssim 89^\circ$ consistent with the photodynamical solution, too low to occult the stellar disk from our perspective.}
\label{fig:planete_transitsearch}
\end{figure}

\begin{figure}[t!]
\begin{center}
\includegraphics[width=0.8\linewidth]{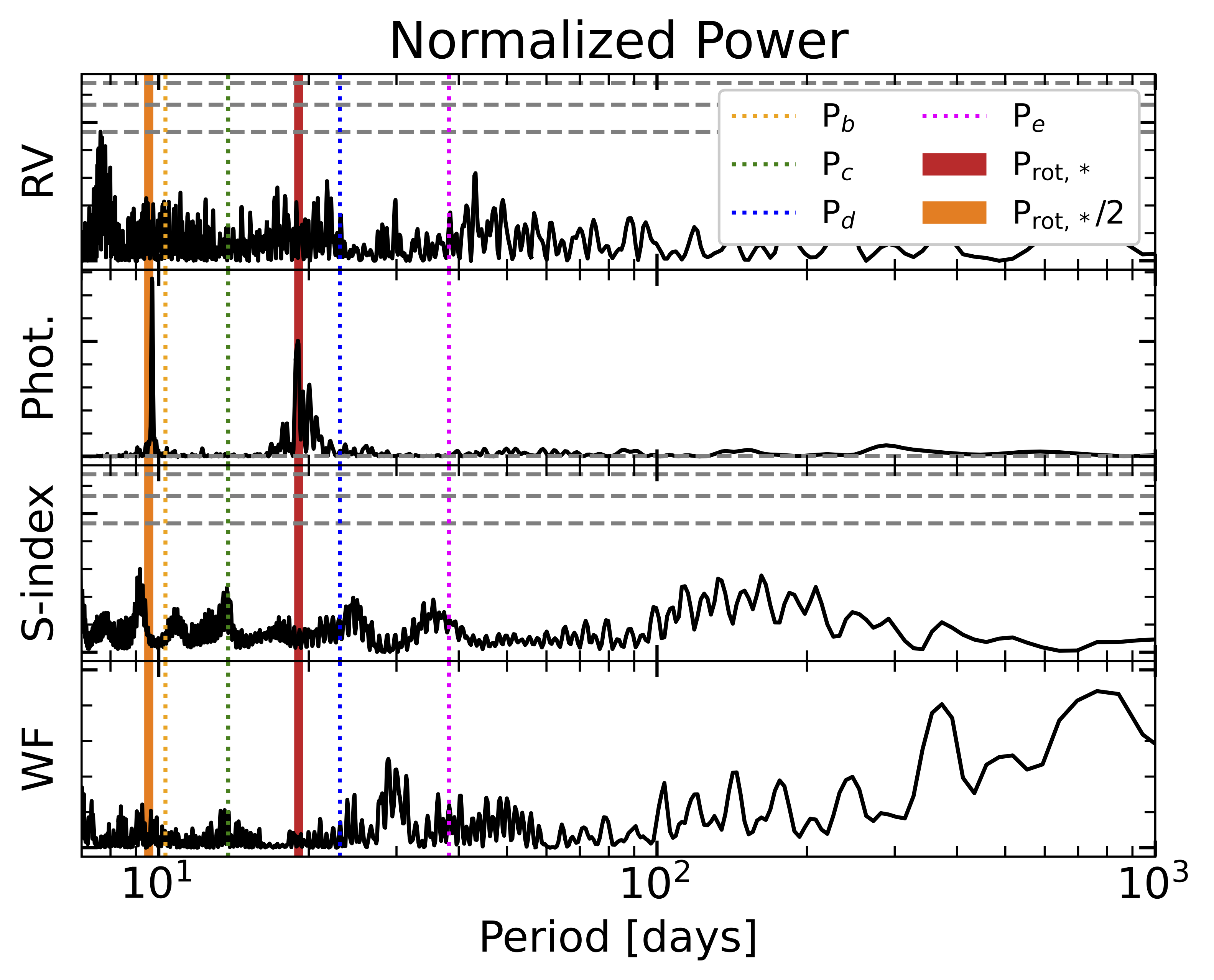}
\end{center}
\caption{\textbf{Search for prominent periodicities in the RV and photometric dataset.} From top to bottom, Lomb-Scargle periodogram of the RV dataset , the \textit{Kepler} light curve , the activity indicator (S-index) and the window function of the RVs. The orbital periods of the four planets, the rotational period of the star and its first harmonic are shown. False-alarm probability levels of 0.1, 1 and 10\% are indicated by dashed gray lines in the top two panels. Significant signals are detected at the stellar period and its first harmonic in the light curve. No significant periodicity was detected in the RV and S-index time series.}
\label{fig:RV_pdgm}
\end{figure}

\begin{figure}[t!]
\begin{center}
\includegraphics[width=0.9\linewidth]{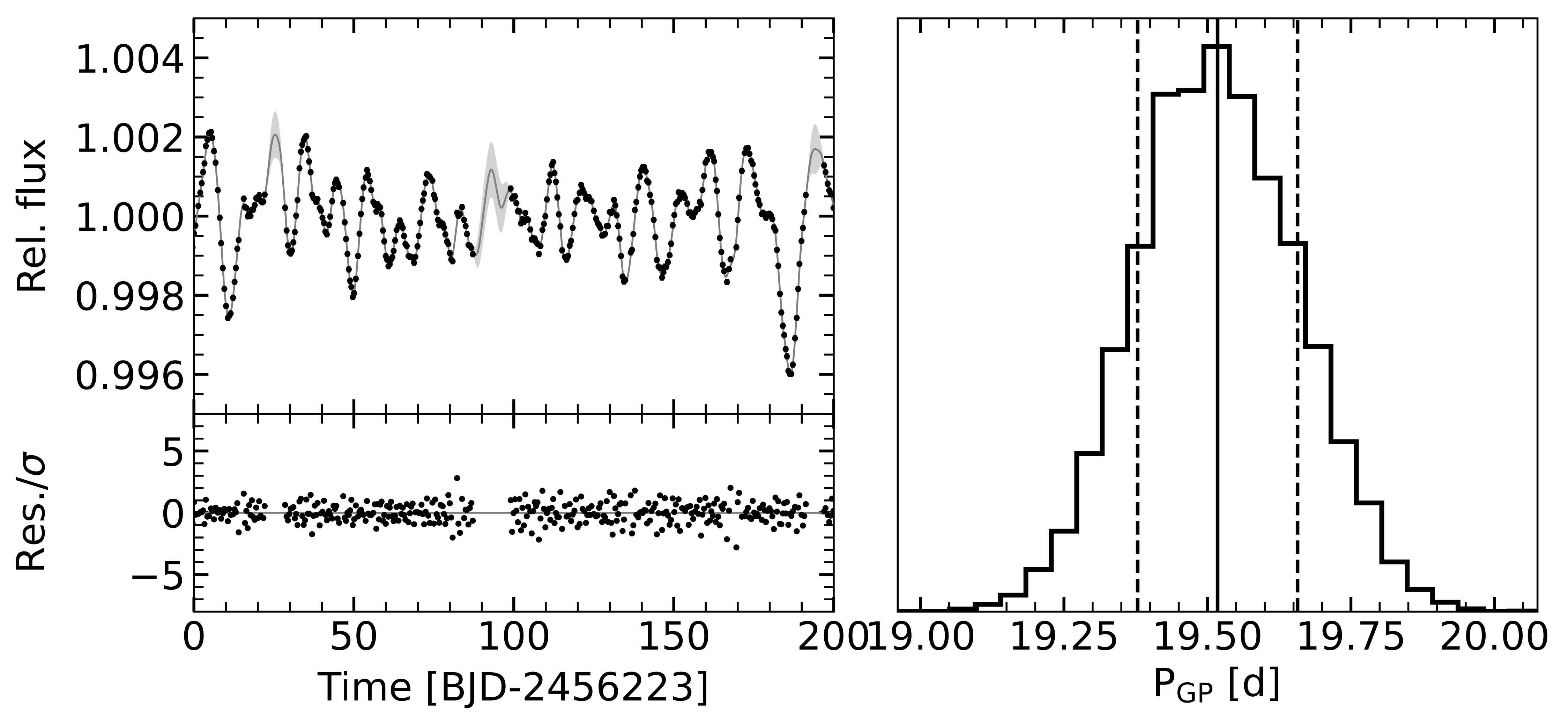}
\end{center}
\caption{\textbf{Gaussian Process fit to the \textit{Kepler} photometry.} Zoom on the last 200 days of the \textit{Kepler} photometric observations (black points) and the best-fitting stellar activity model using a GP (gray shading). The mean is the solid line and the variance is shown as the shaded region. The lower panel shows residuals around the best-fit model divided by the single-point scatter. Posterior constraints on the stellar rotation period from rotational brightness modulations are shown on the right. The GP model reproduces the photometric variability and provides tight constraints on the covariance structure of the stellar signal.}
\label{fig:GP_training}
\end{figure}

\begin{figure}[t!]
\begin{center}
\includegraphics[width=0.8\linewidth]{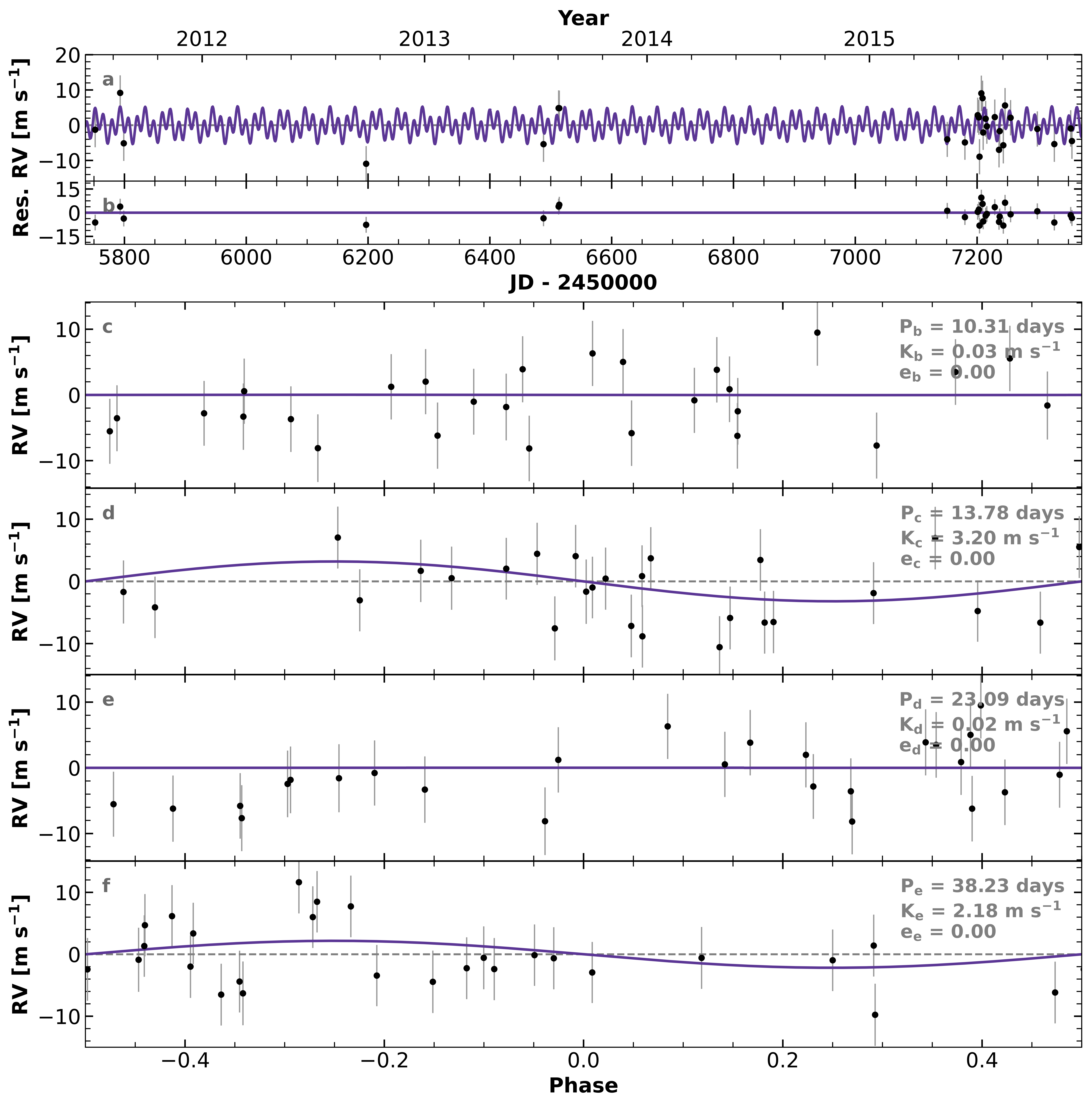}
\end{center}
\caption{\textbf{Median four-planet Keplerian orbital model for Kepler-138.} A trained GP model was used to account for stellar activity in the RV fit. The model corresponding to the median retrieved parameters is plotted in purple while the corresponding parameters are annotated in each panel. We add in quadrature the RV jitter term (Supplementary Table 2) with the measurement uncertainties for all RVs. \textbf{a,} Full HIRES time series. \textbf{b,} Residuals to the best fit model. \textbf{c,} RVs phase-folded to the ephemeris of planet b. The phase-folded model for planet b is shown (purple line), while Keplerian orbital models for all other planets have been subtracted. \textbf{d,e,f,} Same as \textbf{c} for Kepler-138~c, d, and e.}
\label{fig:RV_bestfit}
\end{figure}

\begin{figure}[t!]
\begin{center}
\includegraphics[width=0.99\linewidth]{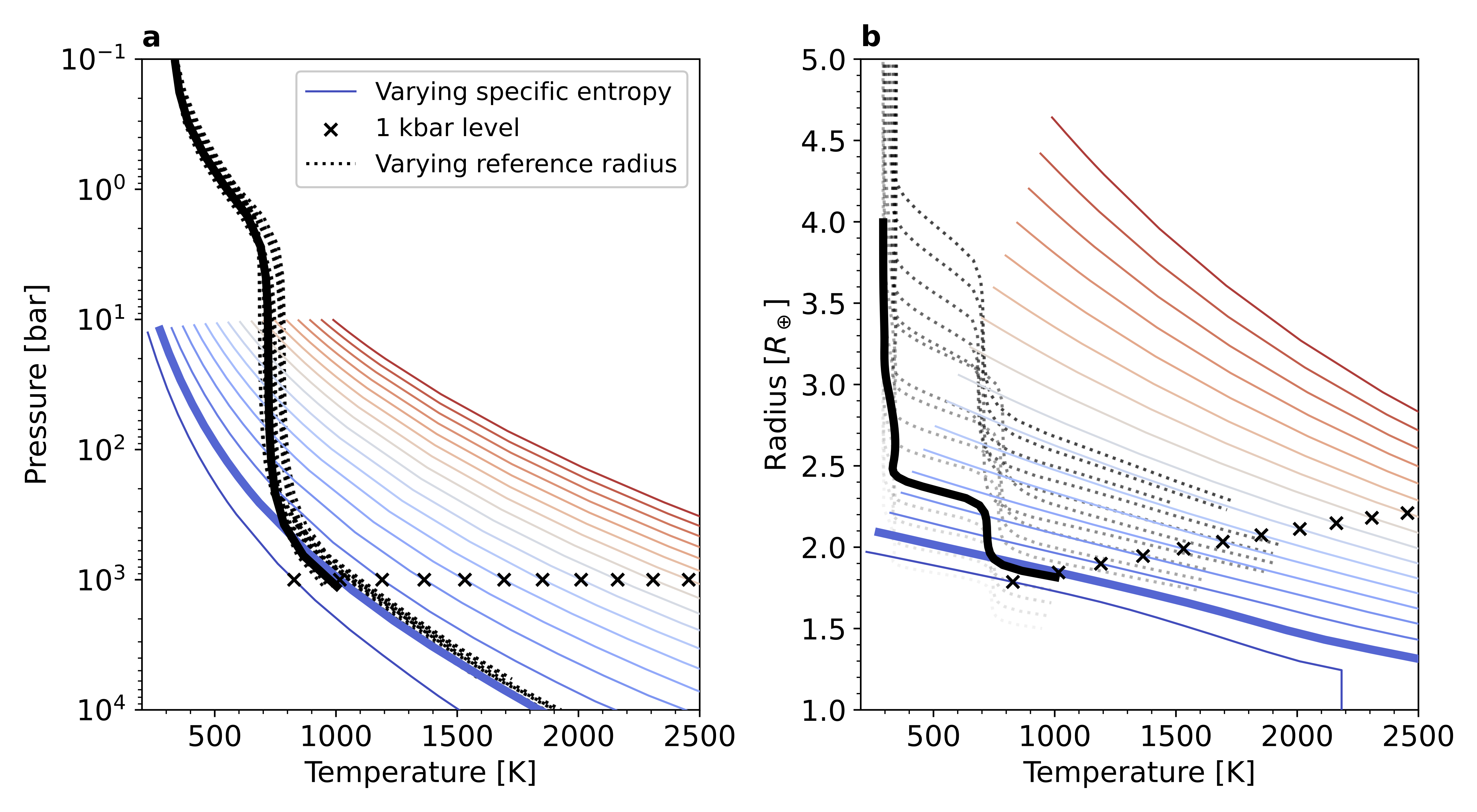}
\end{center}
\caption{\textbf{Illustration of the coupling of interior and atmosphere models.} \textbf{a,} Temperature-pressure and \textbf{b,} temperature-radius profiles computed to generate a complete planet model for a mass of 2.36 M$_\oplus$, a H$_2$/He mass fraction of 3\%, and no water layer. Self-consistent atmosphere models are shown down to the radiative-convective boundary (dotted, black), for the irradiation of Kepler-138~d, but varying the reference radius at a pressure of 1 kbar. Interior models are displayed for the same composition but different specific entropies (solid, colors). For consistency, full-planet models with a given planet mass and composition are obtained from the combination of interior and atmosphere model that have both matching temperatures and radii at the radiative-convective boundary (bold profiles show the closest match in this example).}
\label{fig:structure_modeling}
\end{figure}

\begin{figure}[t!]
\begin{center}
\includegraphics[width=0.7\linewidth]{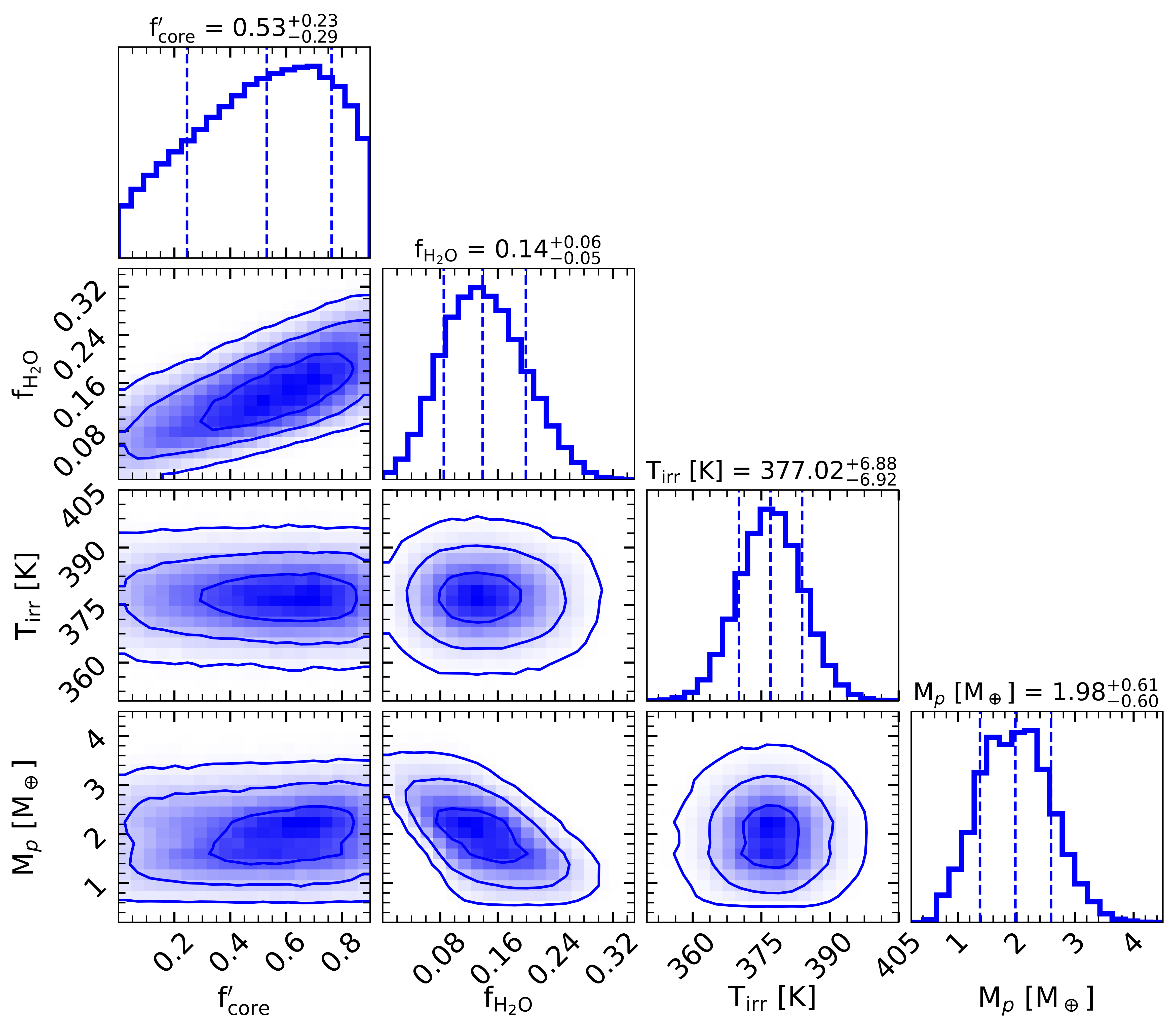}
\end{center}
\caption{\textbf{Composition of Kepler-138~d for the hydrogen-free scenario.} We show the joint and marginalized posterior distributions of the planet structure fit for Kepler-138~d in the case of a hydrosphere lying on top of a rock/iron core. The 1, 2 and 3$\sigma$ probability contours are shown. As expected, the water mass fraction is strongly correlated to the relative amount of rock and iron. The correlation between irradiance temperature and water mass fraction is weak across the considered temperature range.}
\label{fig:corner_interior_d}
\end{figure}

\begin{figure}[t!]
\begin{center}
\includegraphics[width=0.8\linewidth]{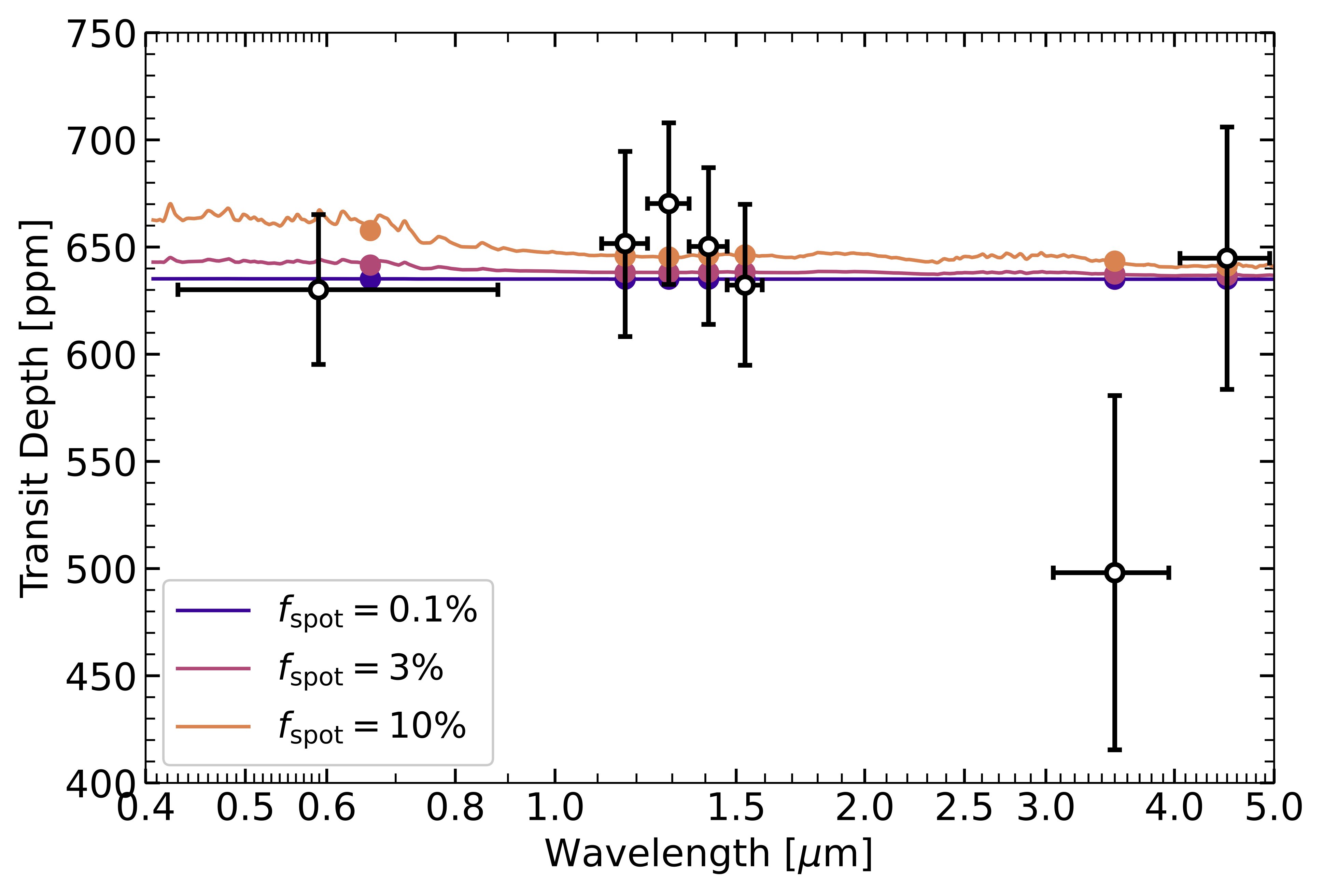}
\end{center}
\caption{\textbf{Impact of unocculted stellar spots on the Kepler transit depth measurement.} Transmission spectrum of Kepler-138~d (black points) superimposed with three scenarios for the level of stellar contamination: spot covering fractions of 0.1, 3 or 10\% (colored lines, colored filled circles show  bandpass-integrated values). The potential impact of unocculted stellar spots on the  radius in the \textit{Kepler} bandpass is small compared to its measurement uncertainty.}
\label{fig:unocculted_spots}
\end{figure}

\begin{figure}[t!]
\begin{center}
\includegraphics[width=0.7\linewidth]{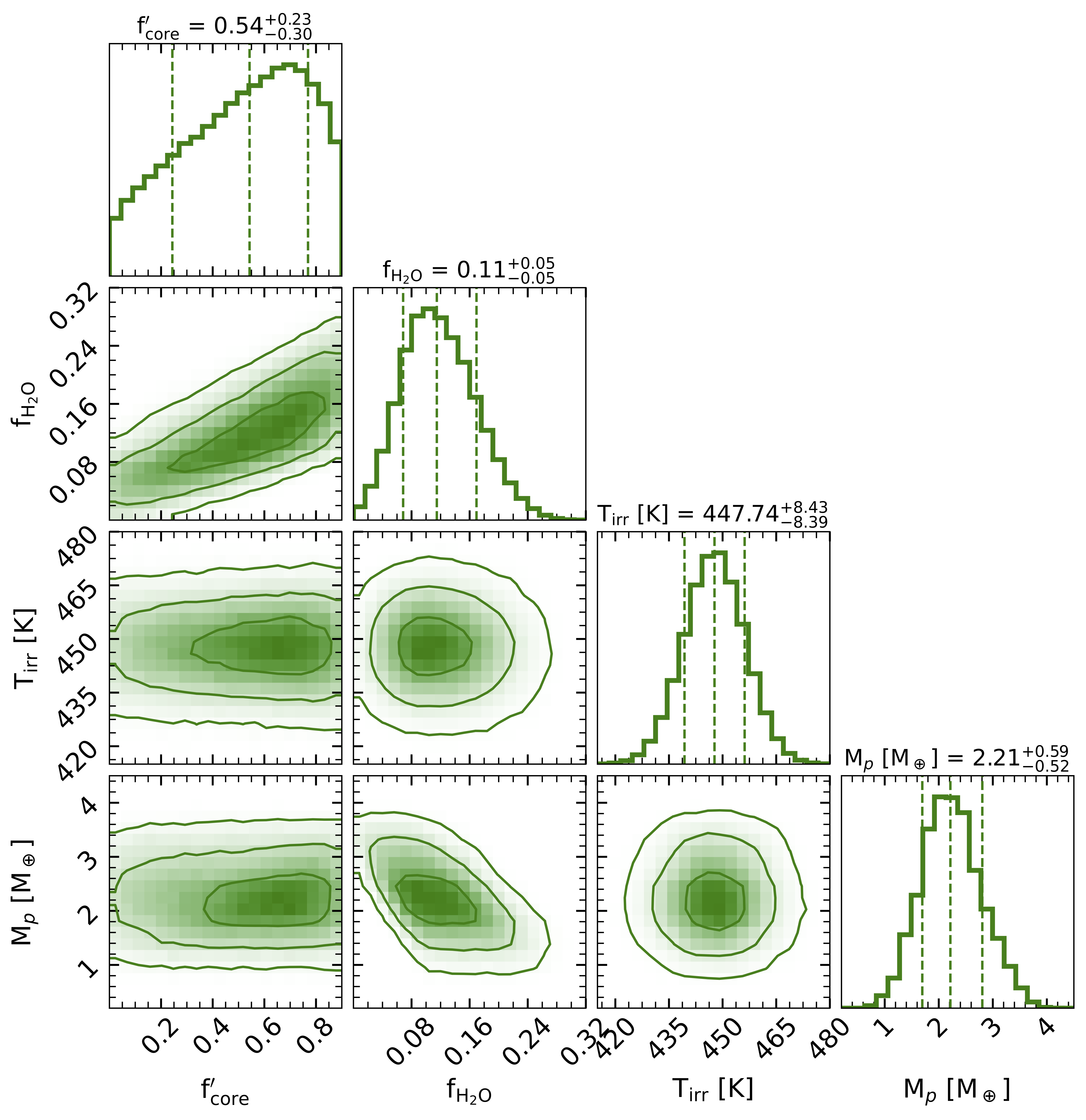}
\end{center}
\caption{\textbf{Composition of Kepler-138~c for the hydrogen-free scenario.} Same as Extended Data Fig. \ref{fig:corner_interior_d}, for Kepler-138~c.}
\label{fig:corner_interior_c}
\end{figure}

\begin{figure}[t!]
\begin{center}
\includegraphics[width=0.999\linewidth]{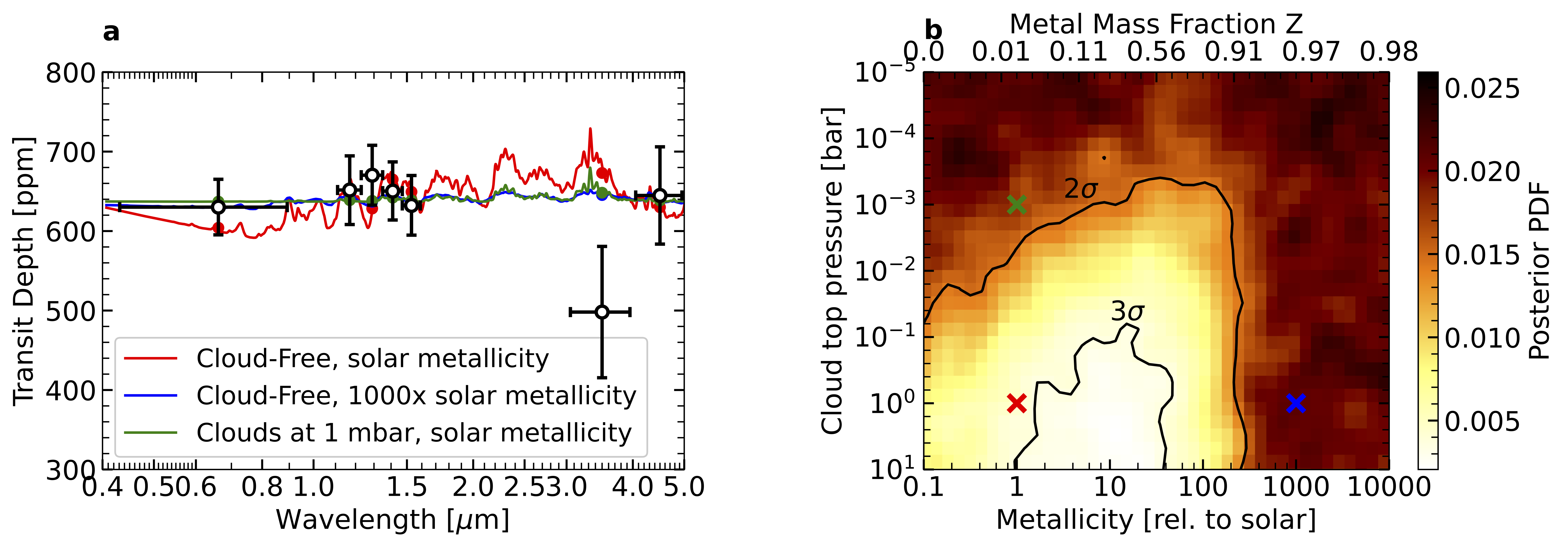}
\end{center}
\caption{\textbf{Constraints on the atmospheric composition from transmission spectroscopy.} \textbf{a,} Optical-to-IR transmission spectrum of Kepler-138~d, compared with three representative forward models: a H$_2$/He atmosphere with a solar composition, a high-metallicity cloud-free atmosphere and a cloudy hydrogen-dominated atmosphere.  \textbf{b,} Joint posterior probability density of the cloud top pressure $P_\mathrm{cloud}$ and atmospheric metallicity, along with the corresponding mass fraction of metals Z assuming a solar C/O ratio. The color encodes the density of posterior samples in each bin and the contours indicate the 2 and 3$\sigma$ constraints. The location in the parameter space of the three  models from panel \textbf{a} is shown with `x' markers. The constraints reflect the well-documented degeneracy between increasing mean molecular weight of the atmosphere and cloud top pressure in terms of the strength of absorption features \citep{benneke_how_2013}. The cloud-free, solar-metallicity scenario is excluded at $2.5 \sigma$. The new planet mass leads to an increased surface gravity which motivates further spectroscopic follow-up to obtain more precise constraints on the atmospheric composition. }      
\label{fig:atmosphere_results}
\end{figure}

\clearpage

\setcounter{figure}{0}
\setcounter{table}{0}
\renewcommand{\figurename}{Supplementary Figure}
\renewcommand{\tablename}{Supplementary Table}

\begin{figure*}[t!]
\begin{center}
\includegraphics[width=0.999\linewidth]{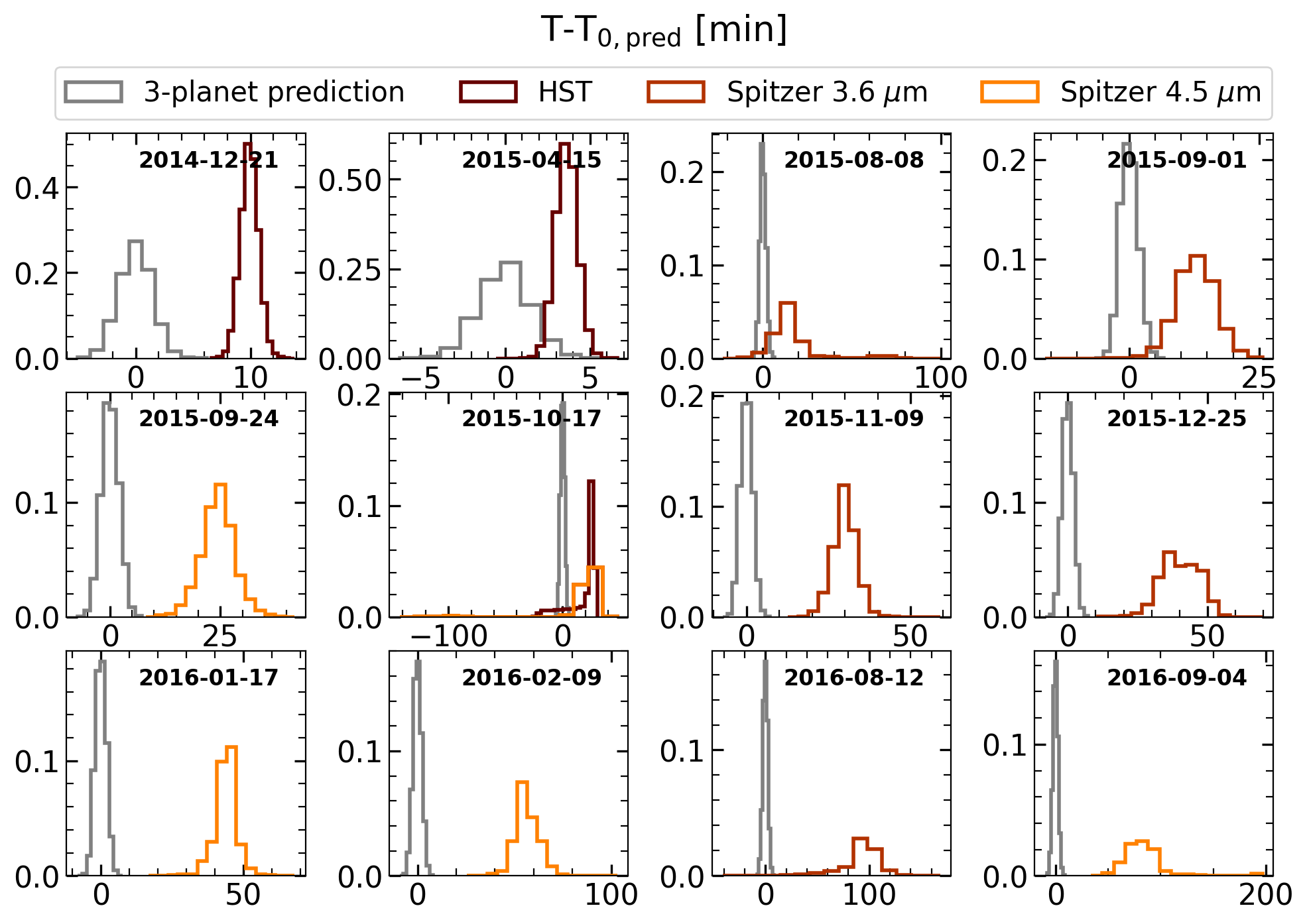}
\end{center}
\caption{\textbf{Inconsistency of the observed \textit{HST} and \textit{Spitzer} transit times with a 3-planet solution.} Comparison of the posterior probability densities on the transit times of Kepler-138~d from our broadband light curve fits to the \textit{HST} and \textit{Spitzer} transits (in color), with the forward prediction from the photodynamical 3-planet model fitted to the \textit{Kepler} transits of Kepler-138~b, c, and d (gray, Ref. \citep{almenara_absolute_2018}). Our measured transit times do not agree with the 3-planet orbital solution. The reference time used for each panel is the median predicted transit time from the fit to the \textit{Kepler} transits. On 2015-10-17, where both \textit{HST} and \textit{Spitzer} observed the same transit, we obtain consistent constraints on the transit time.}
\label{fig:compare_tts}
\end{figure*}

\begin{figure*}[t!]
\begin{center}
\includegraphics[width=0.999\linewidth]{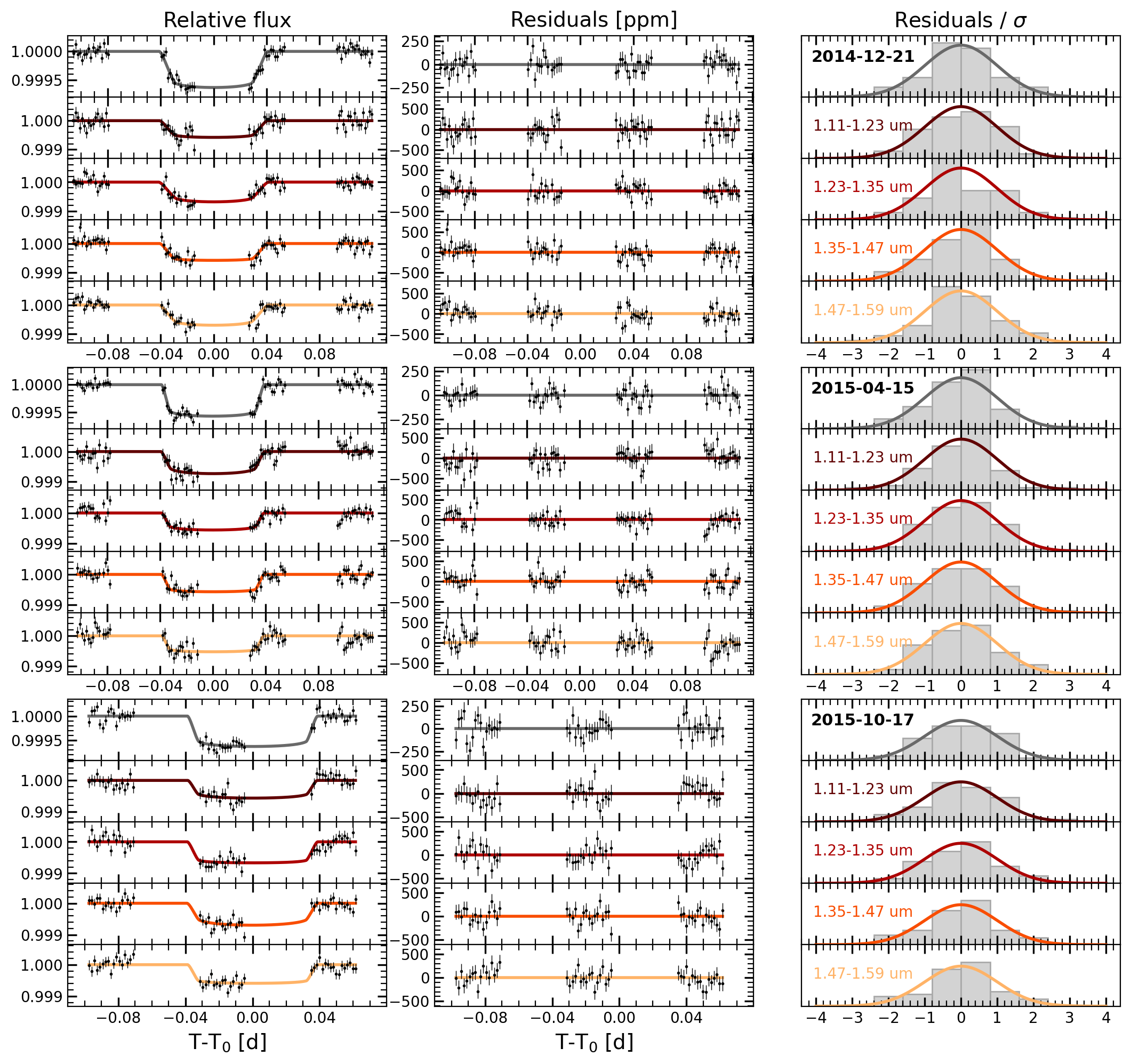}
\end{center}
\caption{\textbf{\textit{HST}/WFC3 light curve fits.} White and wavelength-dependent systematics-corrected light curves, residuals, and their distributions scaled by the white noise photometric uncertainty for the three \textit{HST} visits (top to bottom). We show the best-fitting models as solid curves (grey for white light curves, colored for wavelength-dependent fits). Error bars on individual points in the light curves correspond to their fitted single-point scatter. The residuals generally follow the expected Gaussian distribution for photon-limited precision.}
\label{fig:wlcs_wavelcs}
\end{figure*}

\begin{figure}[t!]
\begin{center}
\includegraphics[width=\linewidth]{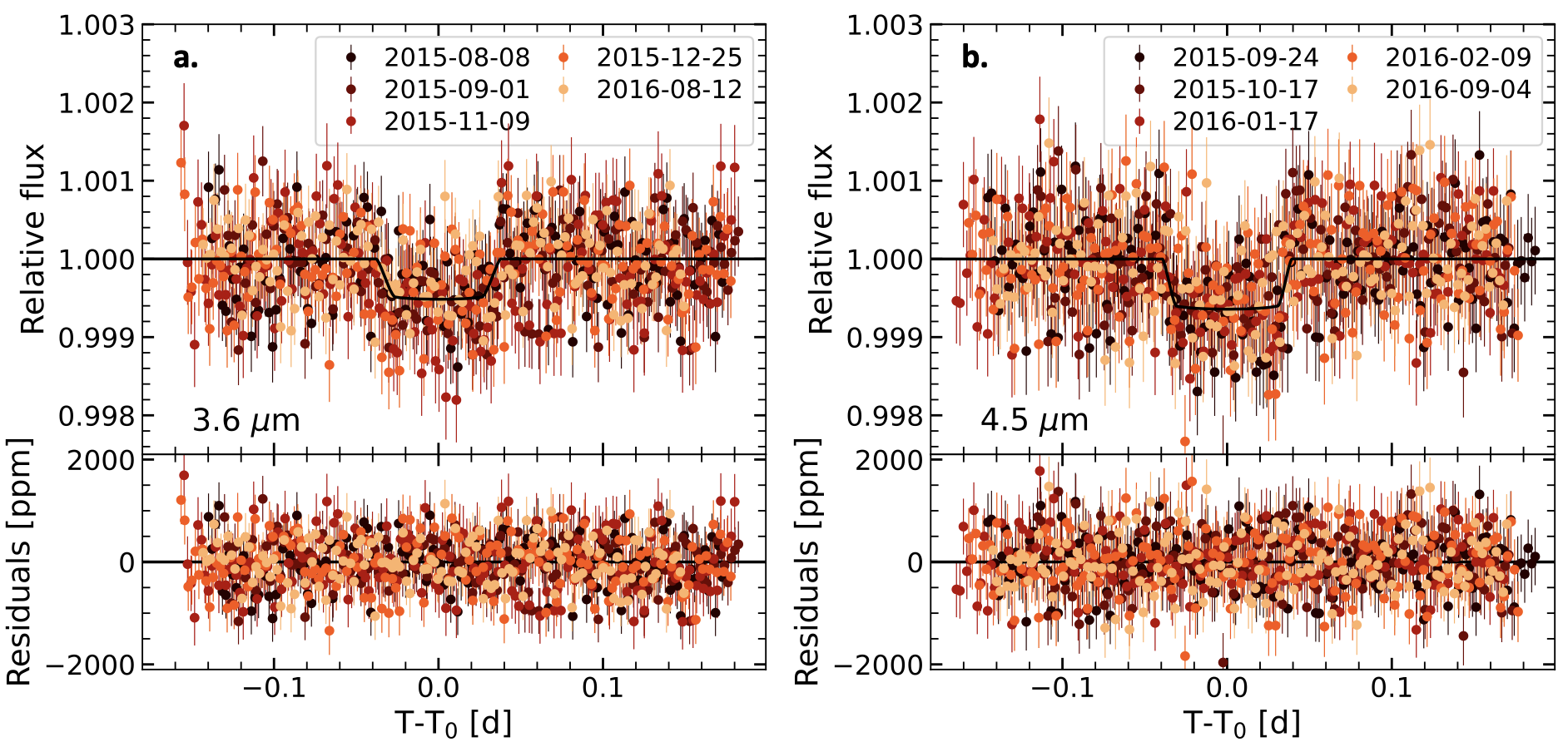}
\end{center}
\caption{\textbf{\textit{Spitzer}/IRAC light curve fits.} The systematics-corrected 3.6 $\mu$m (left) and 4.5 $\mu$m (right) broadband \textit{Spitzer} light curves are shown for each visit, along with the residuals. The light curves are shifted to their best-fit transit time and binned in 4-min increments. The black curve is a transit model with a depth matching the weighted average of the results from individual light curve fits. Error bars on individual points in the light curves correspond to their fitted single-point scatter.}
\label{fig:wlc_spitzer}
\end{figure}

\begin{figure}[t!]
\begin{center}
\includegraphics[width=0.999\linewidth]{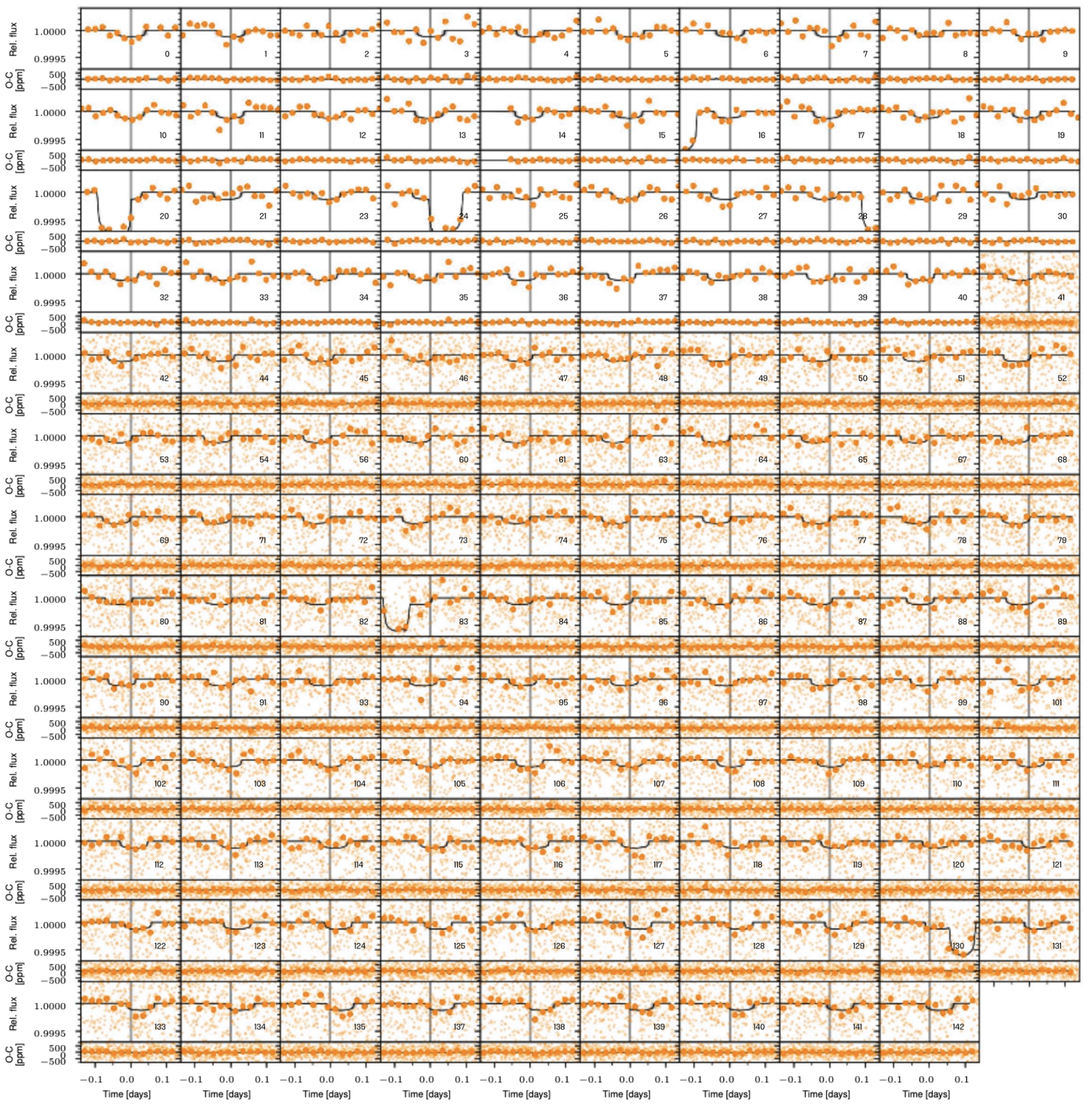}
\end{center}
 \caption{
\textbf{Photodynamical fit to the \textit{Kepler} transits of Kepler-138~b.} The short-cadence \textit{Kepler} observations are shown (dots) with their 30-min averages (circles), as well as the long-cadence light curves (circles). Each panel is labeled with the transit epoch and centered at the predicted transit time for a linear ephemeris (gray vertical lines). We superimpose model predictions from 1000 random MCMC steps. Our transit model accounts simultaneously for all the known transiting planets in the system, as witnessed at the epochs where overlapping transits occur. The median model (black line) and 1, 2, and 3$\sigma$ confidence intervals are shown (three different grey scales). The residuals obtained after subtracting the maximum a posteriori model are shown below each panel.}
\label{fig:photodyn_lcs_b}
\end{figure}

\begin{figure}[t!]
\begin{center}
\includegraphics[width=0.999\linewidth]{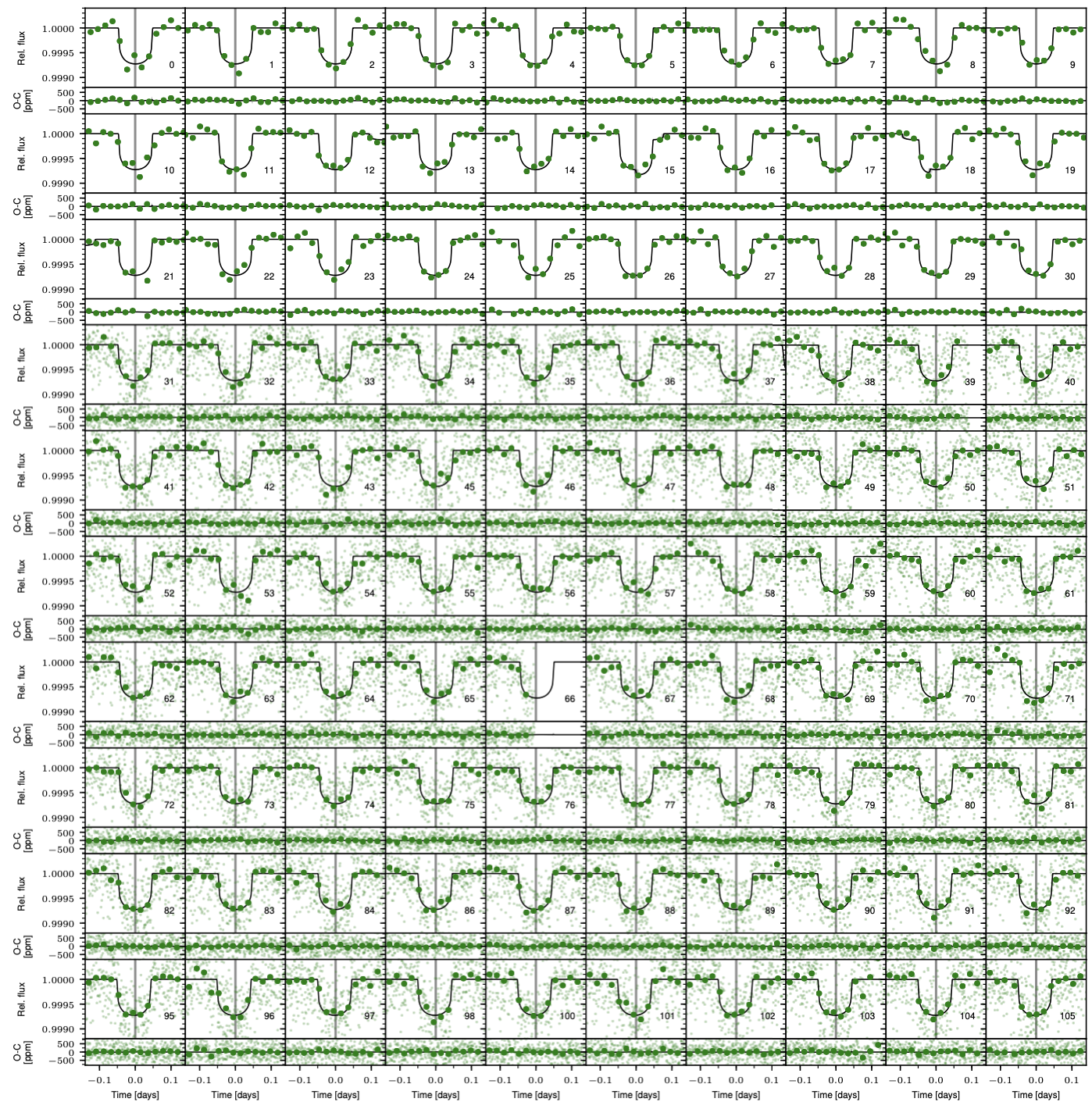}
\end{center}
 \caption{ \textbf{Photodynamical fit to the \textit{Kepler} transits of Kepler-138~c.}
Same as \figurename  \ref{fig:photodyn_lcs_b}, for Kepler-138~c.}
\label{fig:photodyn_lcs_c}
\end{figure}

\begin{figure}[t!]
\begin{center}
\includegraphics[width=0.999\linewidth]{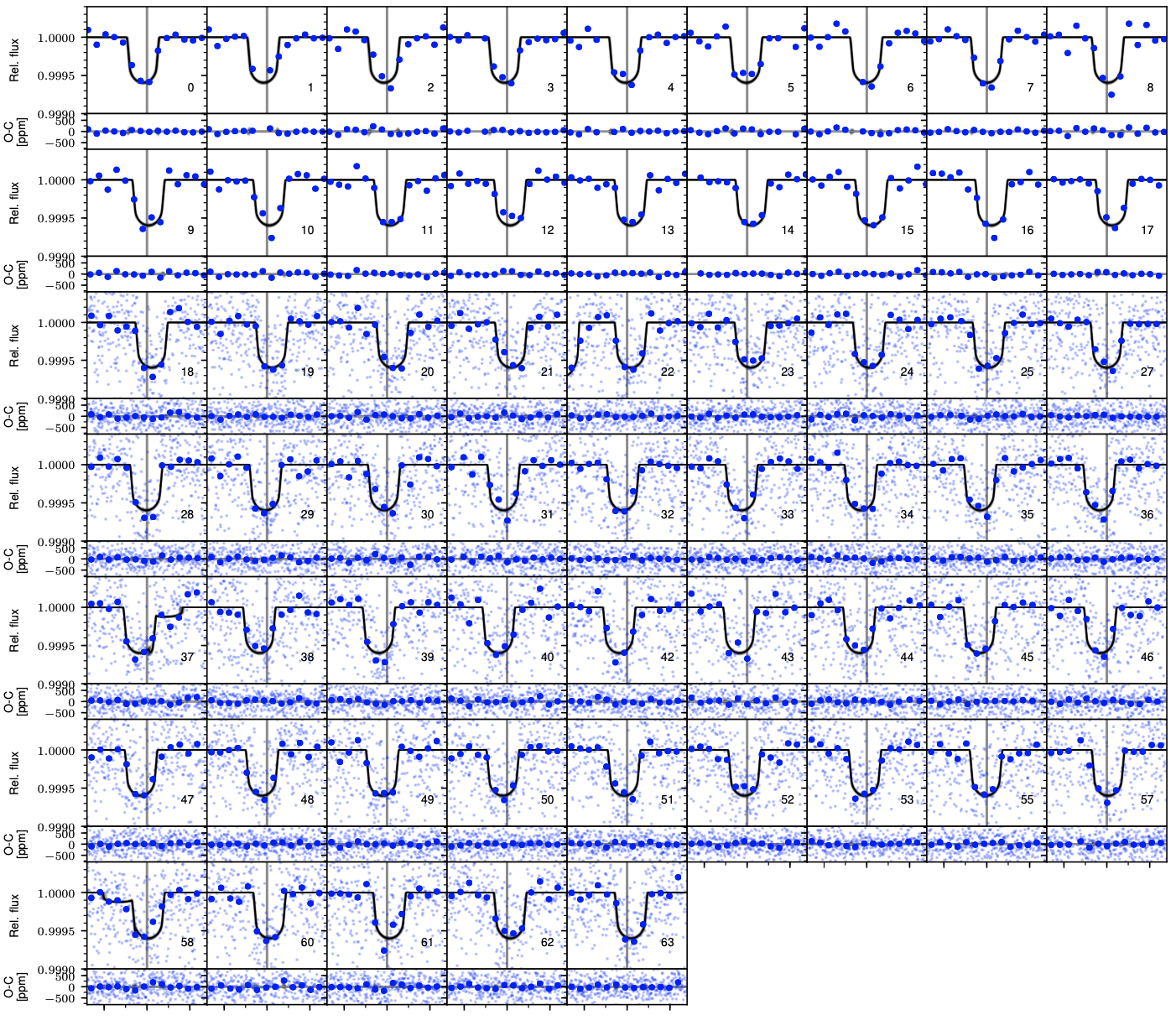}
\end{center}
 \caption{
\textbf{Photodynamical fit to the \textit{Kepler} transits of Kepler-138~d.} Same as Extended Data Fig. \ref{fig:photodyn_lcs_b}, for Kepler-138~d.}
\label{fig:photodyn_lcs_d}
\end{figure}

\begin{figure}
\centering
\begin{center}
\includegraphics[width=0.999\linewidth]{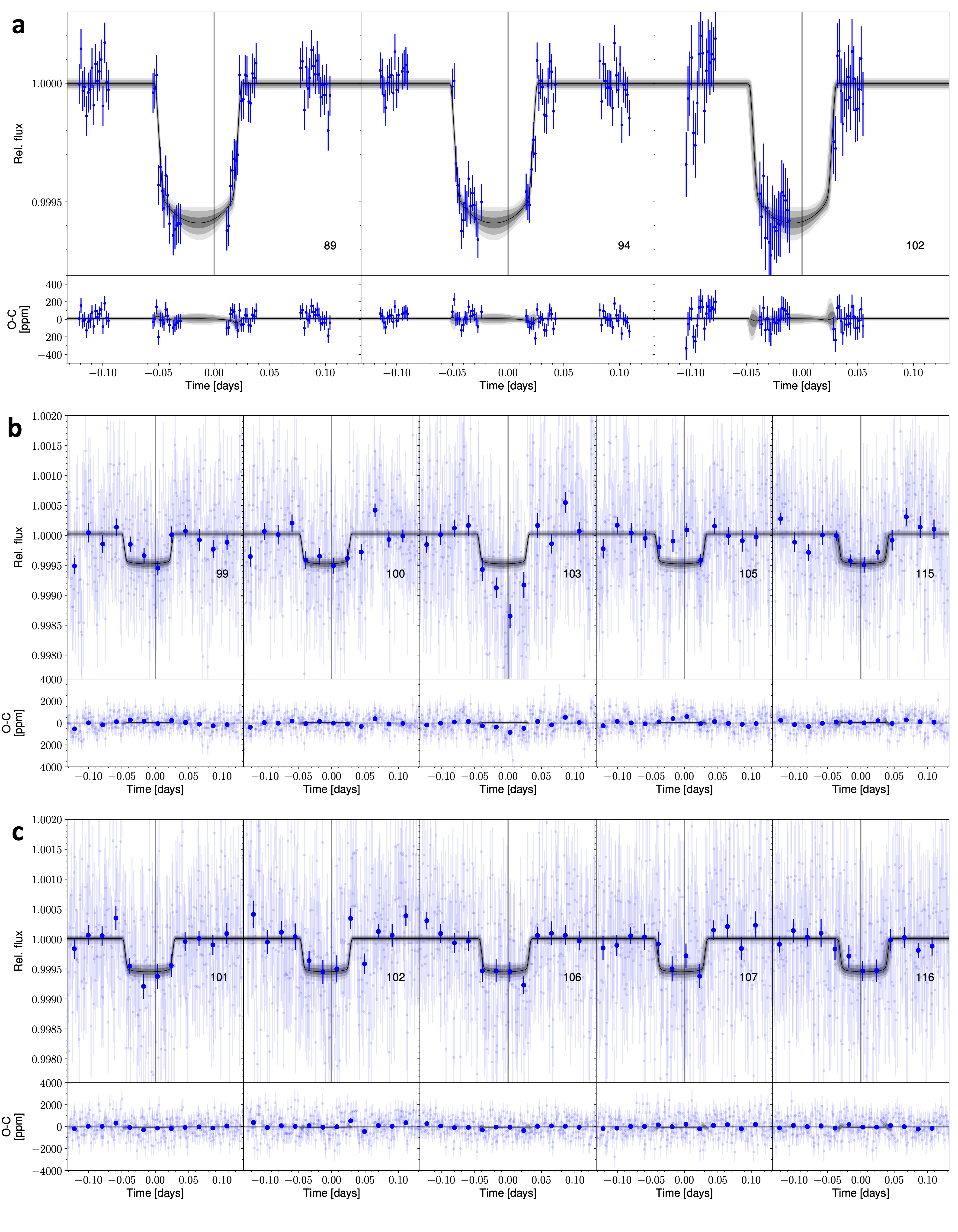}
\end{center}

\caption{\textbf{Photodynamical fit to the \textit{HST} and \textit{Spitzer} transits of Kepler-138~d.} \textbf{a,b,c,} Same as Extended Data Fig. \ref{fig:photodyn_lcs_b}, for the broadband \textit{HST} (\textbf{a}) and \textit{Spitzer} channel 1 (\textbf{b}) and 2 (\textbf{c}) transits of Kepler-138~d. 
}
\label{fig:photodyn_lcs_d_hst_spitzer}
\end{figure}

\begin{figure}[t!]
\begin{center}
\includegraphics[width=\textwidth]{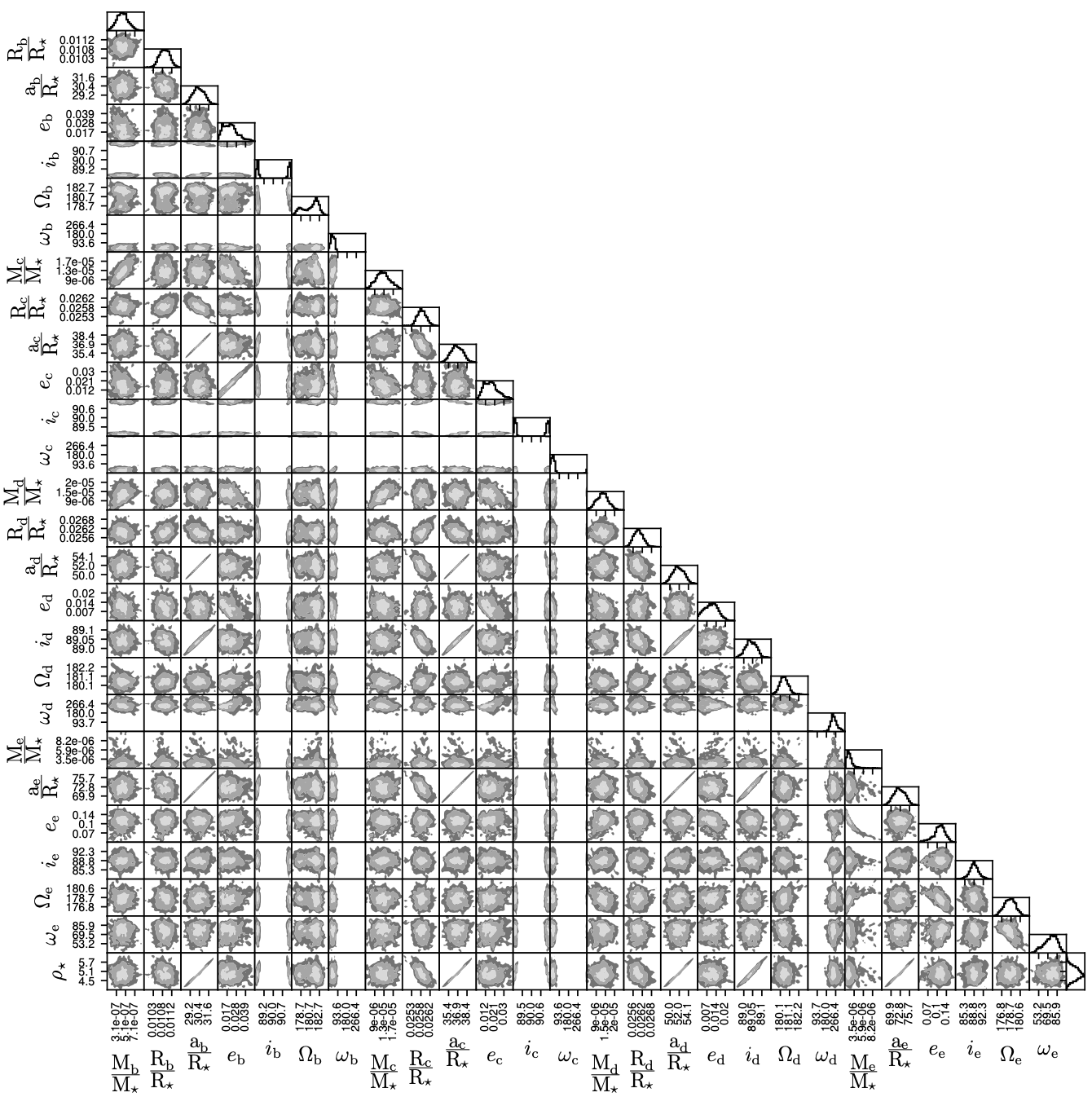}
\end{center}
\caption{\textbf{Joint and marginalized posterior distributions on the system parameters from the 4-planet photodynamical fit.} The 1, 2 and 3$\sigma$ contours are highlighted on the joint distributions (three grey shadings).}
\label{fig:photodyn_corner}
\end{figure}

\begin{figure}[t!]
\begin{center}
\includegraphics[width=0.999\linewidth]{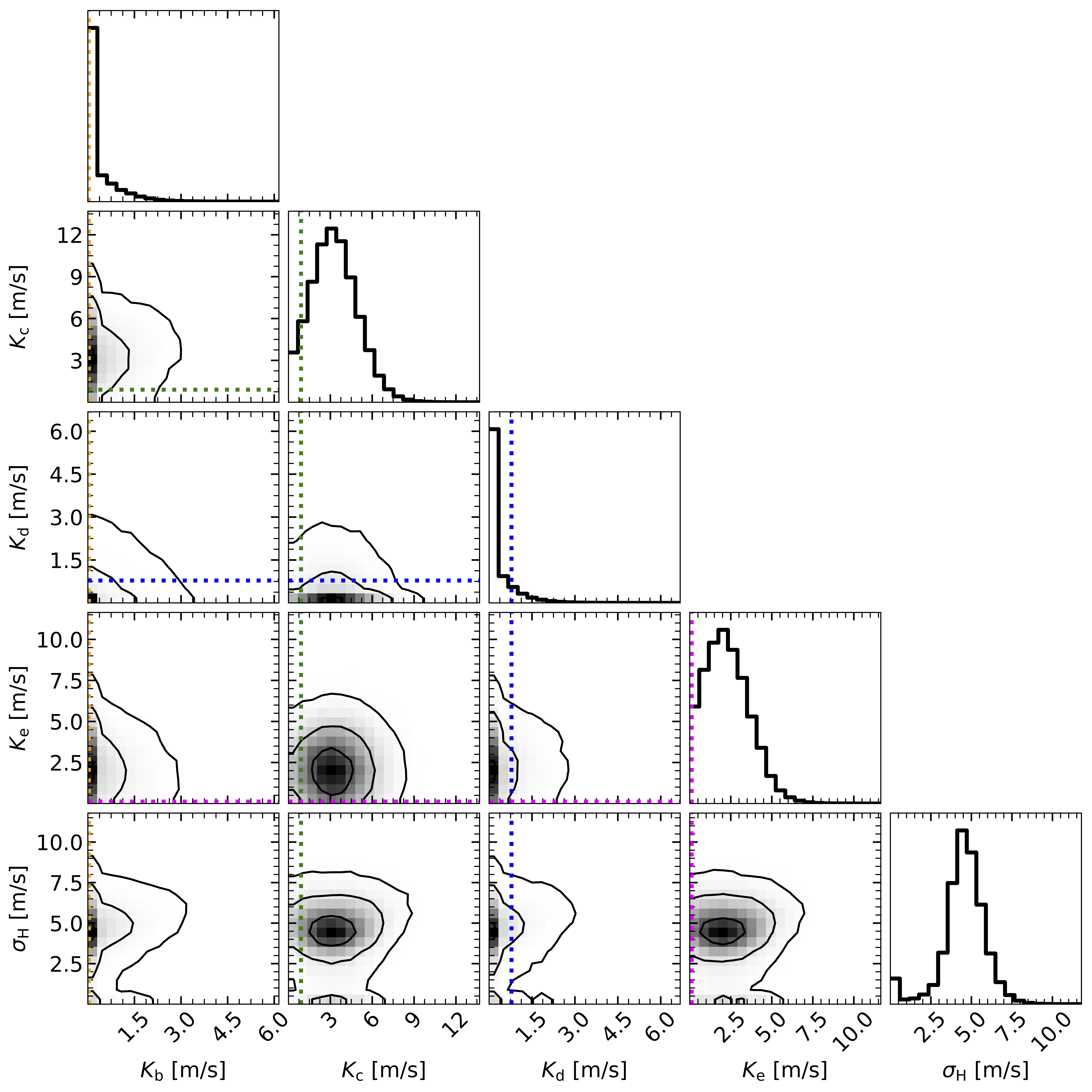}
\end{center}
\caption{\textbf{Joint and marginalized posterior distributions from the 4-planet fit to the \textit{Keck}/HIRES RVs.} We obtain upper limits on the masses of Kepler-138~b, c, d, and e. Contours highlight 1, 2 and 3$\sigma$ limits. The expected RV semi-amplitudes from the median parameters of the photodynamical fit are shown (dotted colored lines). The posterior distributions of the GP parameters are omitted: the  distributions of $\lambda$, $\Gamma$ and $P_\mathrm{GP}$ match their counterparts from the training step. }
\label{fig:rv_corner}
\end{figure}

\begin{figure}
\centering
   \includegraphics[width=0.7\linewidth]{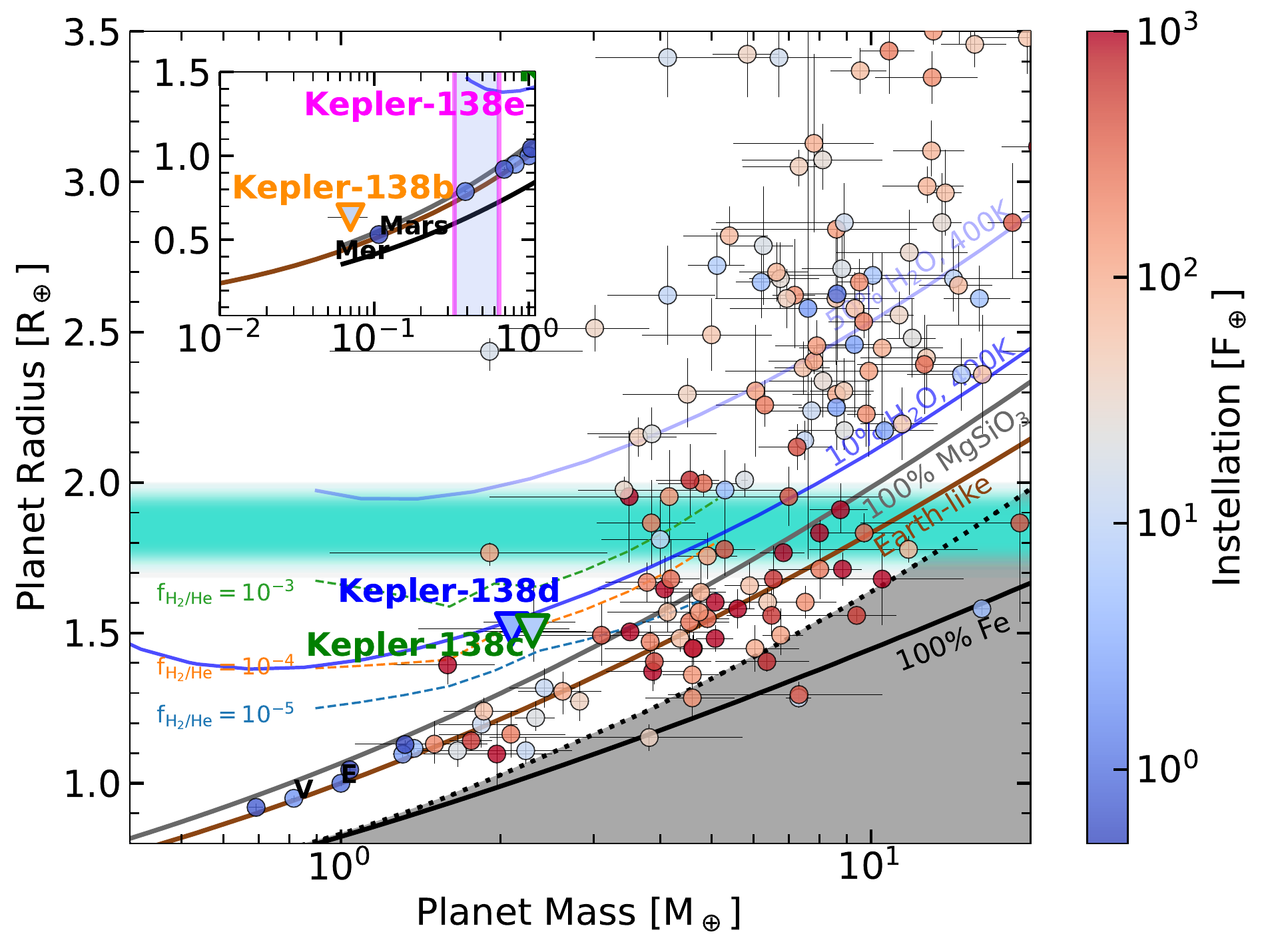}

\caption{\textbf{Mass-radius diagram of small planets.}
Kepler-138~b, c, and d (bold triangles) are shown along with the solar system planets (black letters) as well as small transiting exoplanets (\url{https://exoplanetarchive.ipac.caltech.edu}) with masses constrained to better than 50\% uncertainty, colored according to their instellation. The 68\% confidence constraint on the mass of Kepler-138e is shown in the inset as the vertical shaded region. Error bars correspond to the 68\% confidence region for the mass and radius of each planet. Planets are compared to model mass-radius curves for fixed rocky, volatile-rich and hydrogen-rich compositions (see Methods and Refs. \citep{zeng_mass-radius_2016,zeng_detailed_2013, aguichine_mass-radius_2021}). The radius valley is highlighted (transparent turquoise region). Kepler-138~b, c, and d all have low densities compared to a rocky composition.}
\label{fig:mass_radius}
\end{figure}

\begin{figure}[t!]
\begin{center}
\includegraphics[width=0.7\linewidth]{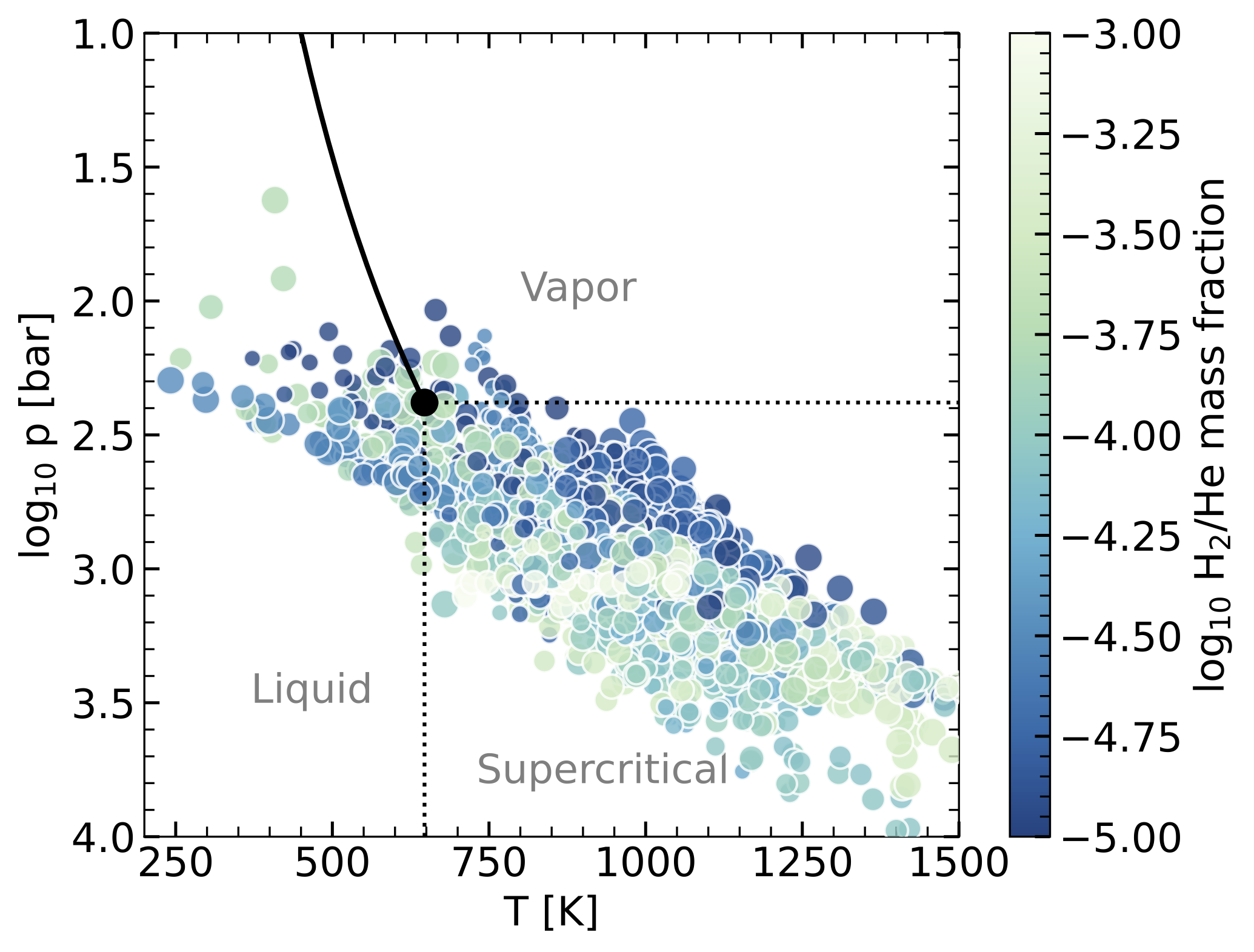}
\end{center}
\caption{\textbf{Water phases at the hydrogen-water boundary.} Phase of the water at the boundary between the hydrogen-rich layer and the underlying water layer (HHB), or at the RCB if it lies within the water layer. We draw 10,000 sample planet masses using the RV and photodynamical mass constraints for Kepler-138~d. The amount of water is varied between 10\% and 30\% (marker sizes) and the hydrogen mass fraction spans the range 0.001\% to 0.1\% (marker colors). The range of conditions is dictated by Kepler-138~d's energy budget (internal temperature, incident flux, Bond albedo), assuming a zero Bond albedo. Higher Bond albedos would result in lower temperatures in the water layer. Depending on the albedo and bulk composition of the planet, supercritical and even liquid water conditions are possible.}
\label{fig:HHB_conditions}
\end{figure}

\begin{figure}[t!]
\begin{center}
\includegraphics[width=0.99\linewidth]{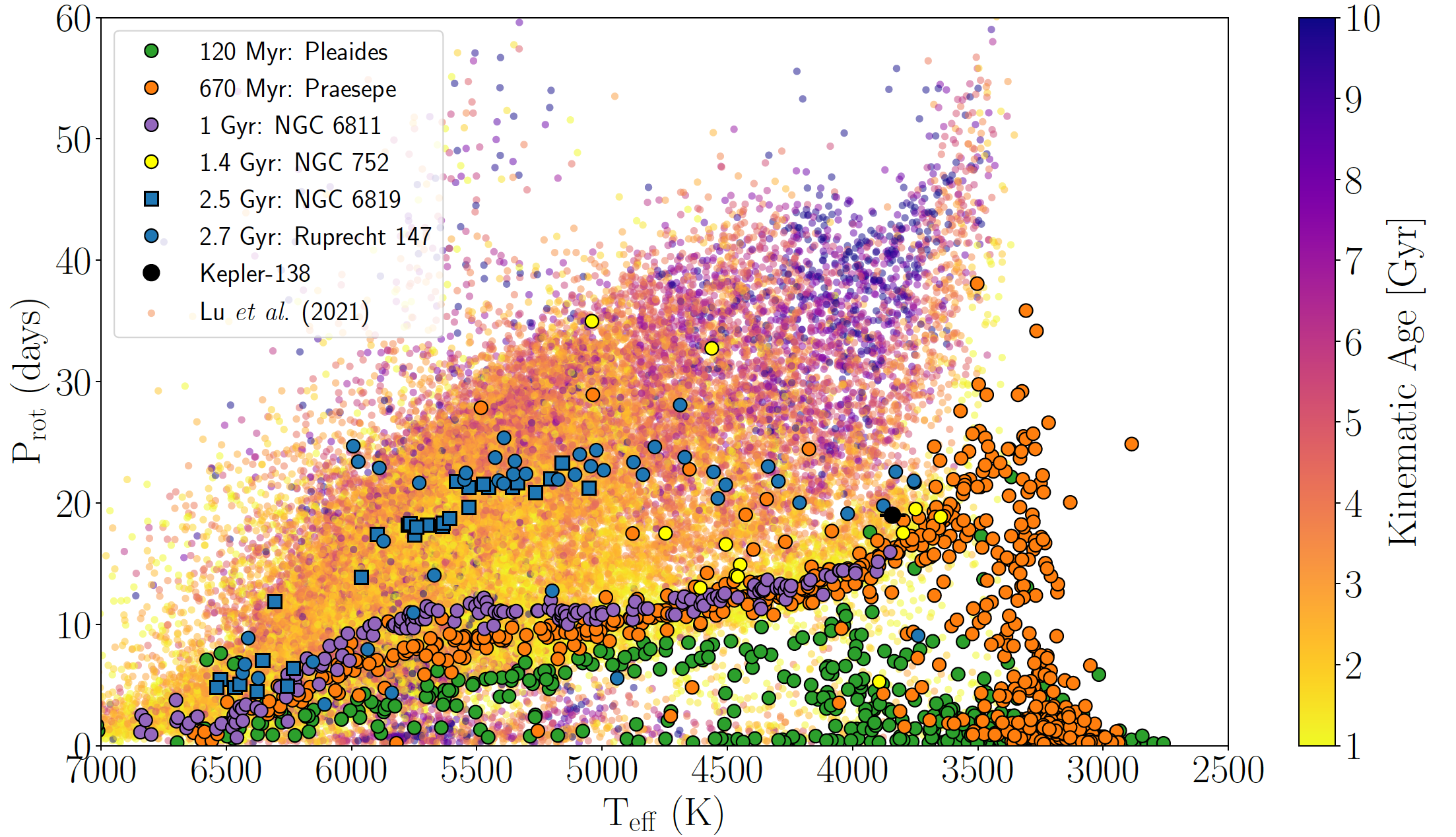}
\end{center}
\caption{\textbf{Age of Kepler-138 in the context of known open clusters.} Stars with known rotation period, effective temperature, and kinematic ages are shown. The effective temperature increases from right to left. Small points in the background are field stars with rotation periods measured from \textit{Kepler} light curves, with colors that correspond
to their kinematic age\citep{lu_gyro-kinematic_2021}. Stars that belong to known stellar clusters\citep{curtis_when_2020} are highlighted using the same
marker (see legend) and with black marker edges. The position of Kepler-138, with $T_\mathrm{eff} \sim 3841$~K (Extended Data Table \ref{tab:star_params}) and $P_\mathrm{rot} \sim 19$ days\citep{almenara_absolute_2018} is indicated with a black marker. Error bars correspond to the 68\% confidence interval on the parameters of Kepler-138. Kepler-138 lies above the tight 1 Gyr isochrone outlined by the NGC 6811 cluster, and below the 2.7 Gyr old Ruprecht 147 cluster, providing a model independent age of 1-2.7 Gyr.}
\label{fig:stellar_age}
\end{figure}

\begin{figure}[t!]
\begin{center}
\includegraphics[width=0.7\linewidth]{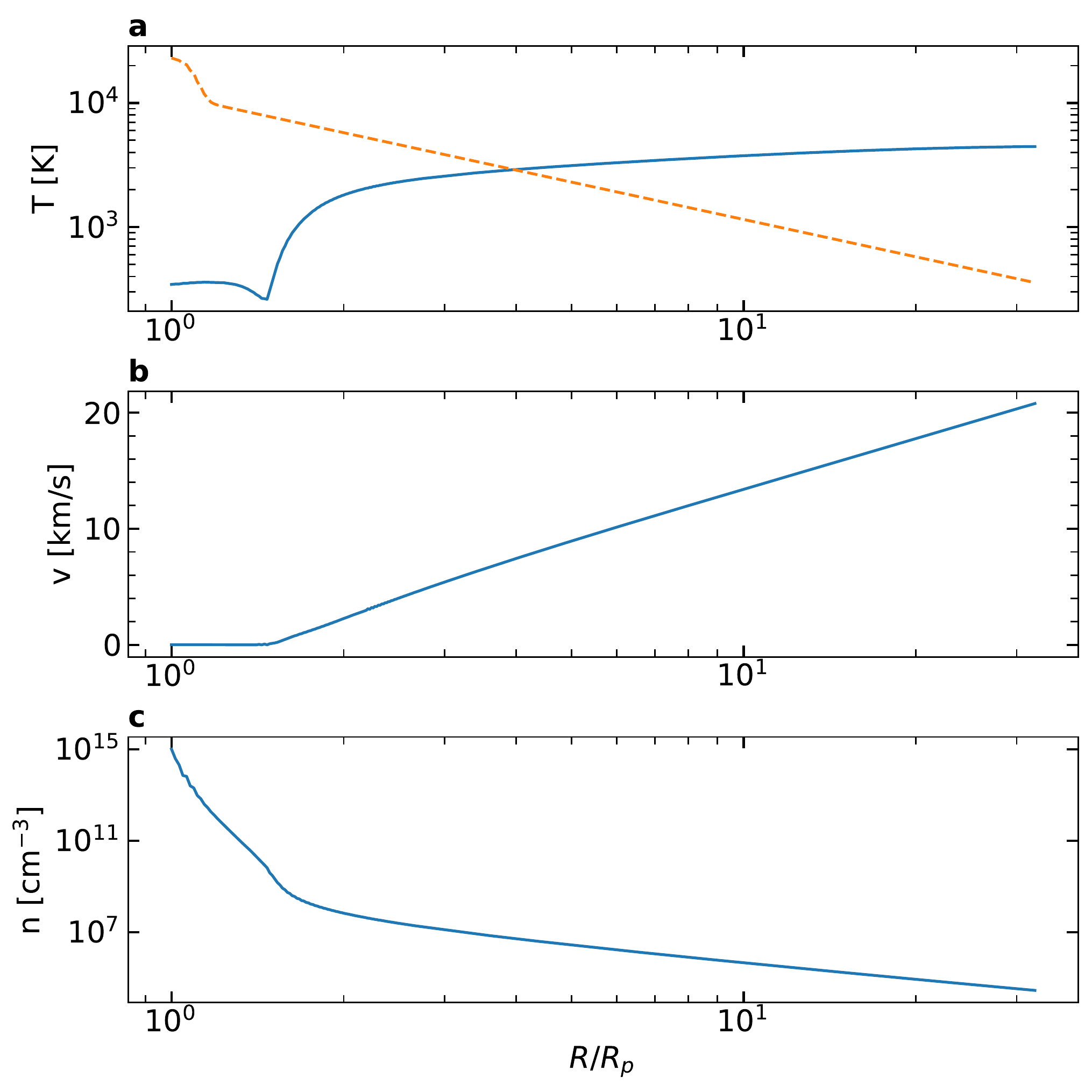}
\end{center}
\caption{\textbf{Upper atmosphere profiles for an escaping hydrogen-dominated atmosphere.} \textbf{a}, \textbf{b}, \textbf{c,} Temperature, velocity and number density profiles for a 1D hydrodynamic simulation of the upper atmosphere of Kepler-138~d assuming a hypothetical hydrogen-dominated composition (blue). Such an atmosphere lies in the blow-off hydrodynamic escape regime beyond $\approx 4R_p$ (temperature threshold shown in orange, dashed).}
\label{fig:upper_atmosphere}
\end{figure}

\begin{table*}
\begin{tabular}{lccc}
\hline
\hline
Parameter               &  Unit            &  Value  & Reference\\
\hline
Distance & pc & $66.86 \pm 0.11$ & Ref. \citep{bailer-jones_estimating_2018}\\
Effective temperature, $T_{\mathrm{eff}}$\ & K & $3841^{+50}_{-51}$ & Ref. \citep{muirhead_characterizing_2012}\\
Metallicity, $\left[\mathrm{Fe/H}\right]$ & dex & $-0.18 \pm 0.10$ & Ref. \citep{muirhead_characterizing_2012}\\
Surface gravity, $\log g_\mathrm{\star}$ & cgs & $4.71 \pm 0.03$ & This paper (derived) \\
Stellar radius, $R_\mathrm{\star}$ & $R_\odot$ & $0.535^{+0.013}_{-0.014}$ & Ref. \citep{berger_gaia-kepler_2020}\\
Stellar mass, $M_\mathrm{\star}$ & $M_\odot$ & $0.535 \pm 0.012$ & Ref. \citep{berger_gaia-kepler_2020}\\
Stellar mean density, $\rho_\star$ & g~cm$^{-3}$ & $4.9 \pm 0.4$ & This paper (fitted)\\
Stellar luminosity, $L_\star$ & $L_\odot$ & $0.056 \pm 0.004$ & This paper (derived) \\
\hline
\hline
\end{tabular}
\caption{\textbf{Kepler-138 stellar parameters.} Quoted error bars correspond to the 1$\sigma$ uncertainty on each parameter.\label{tab:star_params}}
\end{table*}

\begin{table*}
\begin{tabular}{lcc}
\hline
\hline
Parameter               &  Unit            &  Value \\
\hline
\textbf{\textit{GP hyperparameters}} &&\\
Log covariance amplitude, $\log_{10} a$  &  $\left(\mathrm{m~s^{-1}}\right)$   &  $-0.47^{+0.92}_{-1.03}$ \\
Log exponential timescale, $\log_{10} \lambda$  &  (days)   &  $1.34 \pm -0.02$ \\
Log coherence, $\log_{10} \Gamma$  &   &  $-0.51 \pm 0.01$ \\
Periodic timescale, $P_\mathrm{GP}$  & (days)  &  $19.52^{+0.15}_{-0.14}$ \\
Additive jitter, $\sigma_\mathrm{H}$  & $\left(\mathrm{m~s^{-1}}\right)$ &  $4.59^{+1.06}_{-1.03}$ \\
&&\\
\textbf{\textit{Planet parameters}} &&\\
\underline{\smash{Kepler-138 b}} && \\
RV semi-amplitude, $K_b$  &  $\left(\mathrm{m~s^{-1}}\right)$   &  $0.03^{+0.75}_{-0.03}$ \\
Mass, $M_b$  &  $\left(M_\oplus \right)$   &  $0.07^{+1.69}_{-0.07}$ \\
&&\\
\underline{\smash{Kepler-138 c}} && \\
RV semi-amplitude, $K_c$  &  $\left(\mathrm{m~s^{-1}}\right)$   &  $3.20^{+1.71}_{-1.58}$ \\
Mass, $M_c$  &  $\left(M_\oplus \right)$   &  $7.91^{+4.22}_{-3.90}$ \\
&&\\
\underline{\smash{Kepler-138 d}} && \\
RV semi-amplitude, $K_d$  &  $\left(\mathrm{m~s^{-1}}\right)$   &  $0.02^{+0.64}_{-0.02}$ \\
Mass, $M_d$  &  $\left(M_\oplus \right)$   &  $0.06^{+1.89}_{-0.06}$ \\
&&\\
\underline{\smash{Kepler-138 e}} && \\
RV semi-amplitude, $K_e$  &  $\left(\mathrm{m~s^{-1}}\right)$   &  $2.18^{+1.48}_{-1.28}$ \\
Mass, $M_e$  &  $\left(M_\oplus \right)$   &  $7.55^{+5.14}_{-4.43}$ \\
\hline
\hline
\end{tabular}
\caption{\textbf{Fitted and derived parameters from the final RV fit.}  We report the constraints on orbital and planetary properties using a trained GP activity model to account for the effect of stellar surface inhomogeneities. \label{tab:RV_params}}
\end{table*}

\begin{table}
\begin{tabular}{lcccc}
\hline
\hline
Instrument               &  Wavelength            &  Depth       & +1$\sigma$ &  -1$\sigma$ \\
                         &  [$\mu$m]              &  [ppm]       & [ppm]      &  [ppm] \\
\hline
\textit{Kepler}       &  0.43 -- 0.88  &  630 & 35 & 35 \\
\textit{HST}/WFC3 G141  & 1.11 -- 1.23 & 651.6 & 43.4 & 43.0 \\
& 1.23 -- 1.35 & 670.3 & 37.7 & 37.6 \\
 & 1.35 -- 1.47 & 650.3 & 36.4 & 36.8 \\
 & 1.47 -- 1.59 & 632.2 & 37.4 & 37.7 \\
\textit{Spitzer}/IRAC Ch1 &  3.05 -- 3.95 & 498.1 & 82.6 & 82.6 \\
\textit{Spitzer}/IRAC Ch2 &  4.05 -- 4.95 &  644.8  & 61.2 &  61.2 \\
\hline
\hline
\end{tabular}
\caption{\label{tab:spectrum} \textbf{Optical/IR transmission spectrum of Kepler-138~d (Updated HST and Spitzer results with the posterior distributions from new fits).}}
\end{table}

\end{supplementary}

\typeout{}
\clearpage
\bibliography{kepler-138d.bib,aux.bib, references.bib}
\newpage

\clearpage

\end{document}